\documentclass[conference]{IEEEtran}

\ifCLASSINFOpdf
\else
\fi
\usepackage{multirow}
\usepackage{mathtools}
\usepackage{xspace}
\usepackage{xcolor}
\usepackage{multirow}
\usepackage{graphicx}
\usepackage{booktabs}
\usepackage{tabularx}
\usepackage{array}
\usepackage{caption}
\usepackage{subcaption}
\usepackage{amsmath}
\usepackage{bm}
\usepackage{pifont}% http://ctan.org/pkg/pifont
\usepackage{url}
\usepackage{xurl}
\usepackage{makecell}
\usepackage{adjustbox}
\usepackage{enumitem}

\usepackage{caption}

\usepackage{algorithm,algpseudocode}
\algnewcommand{\To}{\textbf{To }}
\algnewcommand\Input{\item[\textbf{Input:}]}%
\algnewcommand\Output{\item[\textbf{Output:}]}%
\algrenewcommand{\algorithmiccomment}[1]{%
  \hfill\(\triangleright\)\textit{\tiny #1}%
}

\usepackage{booktabs} % Add this to your preamble
\usepackage{multirow}

\usepackage{mathtools}

\usepackage{hyperref}

\newcommand{\maria}[1]
{{\footnotesize\color{red}[MA: #1]}}

\newcommand{\generators}{SynNetGens\xspace}
\newcommand{\generator}{SynNetGen\xspace}
\newcommand{\sys}{TraceBleed\xspace}

\newcommand{\defense}{TracePatch\xspace}

\newcounter{packednmbr}

\newcommand{\mypara}[1]{\smallskip \noindent{\bf {#1}.}~}

\newcommand{\remove}[1]{}

\newcommand{\ie}{\emph{i.e.,}\@\xspace}
\newcommand{\eg}{\emph{e.g.,}\xspace}
\newcommand{\etc}{\emph{etc.}\xspace}

 \newcommand{\myitem}[1]{\vspace*{0.02in}\noindent\textbf{#1}}

\usepackage{ifthen}

\usepackage{listings}
\newcounter{insightlabel}
\newcounter{insightnmbr}
\renewcommand{\theinsightlabel}{\textbf{\theinsightnmbr}}

\newcounter{challengelabel}
\newcounter{challengenmbr}
\renewcommand{\thechallengelabel}{\textbf{\thechallengenmbr}}

\newcounter{researchquestionlabel}
\newcounter{researchquestionnmbr}
\renewcommand{\theresearchquestionlabel}{\textbf{\theresearchquestionnmbr}}

\usepackage{mdframed}

\newcounter{findinglabel}
\newcounter{findingnmbr}
\renewcommand{\thefindinglabel}{\textbf{\thefindingnmbr}}

\newenvironment{finding}{
  \refstepcounter{findinglabel}
  \refstepcounter{findingnmbr}
  \begin{mdframed}[
    linewidth=1pt,
    linecolor=black,
    leftline=true,
    rightline=false,
    topline=false,
    bottomline=false,
    innerleftmargin=6pt,
    innerrightmargin=0pt,
    innertopmargin=2pt,
    innerbottommargin=2pt,
    skipabove=6pt,
    skipbelow=6pt
  ]
  \textbf{Finding \thefindinglabel: }%
}{
  \end{mdframed}
}

\newcounter{limitationlabel}
\newcounter{limitationnmbr}
\renewcommand{\thelimitationlabel}{\textbf{\thelimitationnmbr}}

\newcommand{\commentout}[1]{}

\newcommand{\rom}[1]{\uppercase\expandafter{\romannumeral #1\relax}}

\begin{document}
\title{Cross-Flow Correlations Survive Synthesis: \\ \Large{Measuring Source-Level Privacy Leakage in Synthetic Network Traces}}
% \author{Paper \# 498, 13 pages + References + Appendix}
\author{
\IEEEauthorblockN{Minhao Jin}
\IEEEauthorblockA{Princeton University\\
minhaoj@princeton.edu}
\and
\IEEEauthorblockN{Hongyu H\`e}
\IEEEauthorblockA{Princeton University\\
hhy@g.princeton.edu}
\and
\IEEEauthorblockN{Maria Apostolaki}
\IEEEauthorblockA{Princeton University\\
apostolaki@princeton.edu}
}
	

%Telephone: (800) 555--1212, Fax: (888) 555--1212}


\IEEEoverridecommandlockouts
\makeatletter\def\@IEEEpubidpullup{6.5\baselineskip}\makeatother
\IEEEpubid{\parbox{\columnwidth}{
		Network and Distributed System Security (NDSS) Symposium 2027\\
		22 - 26 March 2027 , Seoul, Republic of Korea\\
		ISBN 979-8-9919276-8-0\\  
		https://dx.doi.org/10.14722/ndss.2026.[23$|$24]xxxx\\
		www.ndss-symposium.org
}
\hspace{\columnsep}\makebox[\columnwidth]{}}

\maketitle

\begin{abstract}

Synthetic network data generators (SynNetGens) are increasingly used to share realistic traffic traces without exposing sensitive raw data. While substantial effort has gone into improving fidelity, privacy is either assumed to be a built-in property of synthesis or addressed through differential privacy at the packet or flow level.

This paper uncovers a fundamental privacy vulnerability: SynNetGens preserve cross-flow behavioral correlations that expose source-level membership, allowing an attacker to determine whether traffic of a specific user, or service was included in the training data.
This leakage arises from a mismatch in abstraction: existing SynNetGens operate and are protected at the packet or flow level, while sensitive information is encoded in correlations across flows from the same source.
To demonstrate that this vulnerability is exploitable in practice, we develop TraceBleed, the first source-level membership inference attack against black-box SynNetGens.% and TracePatch,a generator-agnostic defense that mitigates leakage without harming fidelity.

Our evaluation spans five datasets and six SynNetGens, revealing that: (i) every generator leaks source-level information on at least some datasets; (ii) flow- or packet-level differential privacy fails to protect source privacy unless fidelity is degraded to unusable levels; and (iii) releasing 10× more synthetic data amplifies leakage by 130\% on average.
To support ongoing research in this area, we will maintain a public privacy–fidelity leaderboard so practitioners can choose generators that fit their needs and researchers can benchmark new designs faithfully.

\end{abstract}


% \fi
\IEEEpeerreviewmaketitle

\section{Introduction}

%\maria{Todos: threat model remove claims on obfuscating IPs; results on attacker as sensitivity analysis; reference table }

Synthetic Network Data Generation Pipelines (\generators) are generative systems trained on real network traces to produce synthetic traces that preserve key statistical properties of the originals. They are gaining traction in academia and industry as a practical way to support benchmarking, analytics, and collaboration without exposing raw sensitive traffic~\cite{han2019iemcon,han2019flowwgan,lin2020doppelganger,ring2019flowgan,wang2020packetcgan,xu2019ctgan,xu2020stan,yin2022netshare,sun2024netdpsyn,jiang2024netdiffusion}.

Considerable effort has gone into evaluating and improving their fidelity, namely, how useful the synthetic traces remain for testing, benchmarking, and other downstream tasks. Privacy, however, has received far less rigor despite being the primary motivation for deploying these systems. 
In some prior works, privacy is treated as an inherent byproduct of synthesis, implicitly assuming that synthetic traces are safe to release~\cite{han2019flowwgan,ring2019flowgan,xu2020stan}. In others, synthetic traces are considered sufficiently protected when the underlying \generator satisfies $(\epsilon,\delta)$-differential privacy (DP) at the packet or flow level~\cite{yin2022netshare,sun2024netdpsyn}.

This paper argues that protecting or evaluating synthetic traffic solely with packet- or flow-level privacy metrics is inadequate and may even be misleading to data owners.
In fact, we find that all state-of-the-art \generators leak training-set participation at the source level: from the released synthetic trace alone, an attacker can infer whether traffic from a particular source, such as a user, tenant, or service, was used to train the \generator.
%In fact, we find that all state-of-the-art \generators inadvertently leak source-level information: effectively allowing an attacker to confirm the inclusion of a source of traffic \eg an individual user or a service, in the training dataset of a \generator only by accessing synthetic traffic. 
This leakage is consequential. Confirming inclusion can also confirm an individual's location or condition, the violations of no-training agreements, or the load of a sensitive website. 
% To our knowledge, this is the first work to demonstrate that.

%Implicit in this practice is the notion that protecting flows/packets also protects the higher-level sources that generate them, such as users, services, or workloads. 

%This finding is non-trivial.
%While it is well-known that the traffic sources (\ e.g., users, services) expose their behavioral patterns (\ e.g., web browsing habits or communication patterns) in the traffic they create, no prior work has exposed that these behavioral patterns survive \generators' synthesis.
%The key insights here are that: \emph{(i)} behavioral patterns are manifested in \emph{cross-flow} correlations, not at individual packets or flows, which many \generators aim to obfuscate (\eg via $(\epsilon,\delta)$-DP privacy); and that \emph{(ii)} obfuscating individual flows does not automatically also obfuscate cross-flow correlations.  In effect, we find that even protected \generators leak source privacy.

This finding is non-obvious. While it is well-known that traffic sources (e.g., users, services) imprint unique behavioral patterns or fingerprints in the traffic they generate (e.g., due to their browsing habits or communication structure), prior work has not shown that such patterns survive synthesis: traffic generated by \generators often retains source-level patterns present in the training data.
This threat is distinct from traffic fingerprinting, which identifies which source produced an observed trace. Source-level membership inference asks a fundamentally different question: whether a source's traffic influenced the generator's training data. 
The key insight (missed by prior work) is that these patterns are expressed through \emph{cross-flow} correlations, rather than solely through individual packets, flows, or explicit identifiers. Hence, even \generators that protect individual flows or packets (e.g., via $(\epsilon,\delta)$-DP), do not eliminate these higher-order correlations. Compounding this issue, privacy is typically evaluated using packet- or flow-level membership inference attacks, leaving this form of leakage largely undetected.
In effect, current privacy evaluations provide a false assurance: a \generator may report near-random flow-level membership inference while exposing the majority of its training sources at the source level.

To expose the risk of the misalignment between the de facto privacy abstractions for synthetic network traces and the encoding of sensitive information to network traces, we design \sys, a source-level membership inference attack that provides an empirical lower bound on the source-level privacy metric an attacker can achieve against \generators. 
Given synthetic traces and a reference dataset of raw traffic, \sys predicts whether any source in the reference dataset contributed to the training data of the \generator.
%Critical for understanding \sys's practicality, is that it operates under strictly weaker assumptions than canonical MIAs in classical domains 
A key anchor of \sys’s practicality is that it operates under strictly weaker assumptions than canonical MIAs in classical domains
(e.g., images), which are widely recognized as a fundamental privacy threat in ML, motivating extensive research and regulatory attention~\cite{shokri2016membership,carlini2022membership}.
Concretely, \sys requires no query access to the generator, reflecting the fact that data holders release finite synthetic datasets, not queryable models. It requires no knowledge of the generator's architecture or training procedure, unlike state-of-the-art MIAs \cite{hayes2017logan,shokri2016membership,carlini2022membership,yeom2018privacy}. Also, \sys assumes no overlap between the reference and training datasets, unlike SOTA MIAs, which assume near identical copies of records shared between the datasets~\cite{shokri2016membership,van2023membership,Li2025enhanced}. Even MIAs in the networking context follow this approach: \eg DoppelGANger~\cite{lin2020doppelganger} reports near-random membership inference for exact flows.%, while source-level leakage remained entirely unexamined.
Importantly, \sys’s assumptions are not only weaker than canonical MIAs but also reflect realistic network visibility. In the networking context, an attacker can obtain a reference dataset through multiple practical channels:
%As with any MIA, \sys's effectiveness depends on the attacker's ability to acquire reference traffic. In the networking context, this is feasible through multiple channels: 
\emph{(i)} record her own traffic, as when a tenant or organization monitors her own traffic at the border; \emph{(ii)} independently reproduce target behavior, such as loading websites or running applications on premise; or \emph{(iii)} observe traffic on-path, leveraging the fact that each packet traverses multiple tens of routers manufactured by diverse vendors that are owned by diverse independently-managed autonomous systems, any of which may retain observations, without requiring privileged access to the generator or training data.
As we show in \S\ref{sec:eval}, \sys remains effective even under noisy and imperfect reference data.

We use \sys to empirically evaluate source-level privacy leakage across a diverse set of generators (GAN-, diffusion-, GPT-, and DP-based), ensuring that our findings are not an artifact of a particular \generator. We find that even DP-protected \generators leak source-level patterns, which an attacker can exploit for membership inference. \sys can confirm the inclusion of some sources with high certainty, which is particularly concerning because confirming the inclusion of even a single source can be impactful, \eg by revealing a tenant’s workloads were used in training despite contractual restrictions.
Moreover, releasing more synthetic data amplifies the threat: 10× more synthetic output increases the fraction of exposed sources by 130\%, because additional samples better cover each source's behavioral manifold in the synthetic output, giving the attacker more opportunities to match cross-flow patterns. This finding directly contradicts the assumption that sharing synthetic data is safer than sharing the generator itself.
To further rule out dataset-specific effects, we evaluate across multiple datasets capturing a wide range of cross-flow patterns (\eg user browsing and application-specific patterns). We observe that different \generators' source leakage varies across datasets, suggesting that each preserves different subsets of traffic features, some of which are more revealing for certain patterns (\eg inter-arrival times is more revealing of a website over a datacenter application). For example, RTF-Tab achieves higher privacy leakage (\eg AUC$\approx$75\%) for the WFP data set and lower privacy leakage (lower AUC score) for DC.
To investigate the impact of the reference' dataset, we augment our analysis with datasets that emulate realistic attacker capabilities. These include attackers that reproduce traffic patterns or/and  passively collect traffic from different locations, thereby introducing variability due to latency, routing, and load balancing in the reference data, as well as attackers that have access to only a limited number of sources. We find \sys consistently achieving high accuracy in identifying vulnerable sources under different methods of constructing the reference dataset.
Finally, our results suggest that source-level privacy is not an easy extension of existing packet- or flow-level protections: eliminating source-level leakage requires destroying fidelity, effectively rendering synthetic traffic useless.

To enable principled deployment decisions, we maintain a public privacy–fidelity leaderboard\url{https://tracebleed448.github.io} that lets practitioners compare generators under standardized source-level metrics and lets researchers benchmark new \generators and attacks against a common baseline. We invite the community to contribute additional methods, datasets, and defenses.

\section{Motivation }

We first explain why source-level membership inference attacks are a critical vulnerability worth investigating as part of \generator's privacy, even though they are not covered by existing packet- or flow-level metrics. Next, we define the threat model we consider for investigating it.

\subsection{Why confirming membership in training data matters?}\label{ssec:why}
%To understand why source-level privacy is critical, we need to consider the consequences of failing to protect it for various sources, be they individuals, websites, or services/workloads.

%To explain why source-level MIA on synthetic traffic is  critical, we need to understand when confirming membership in training data matters for a source whose traffic the attacker has already access to.

Next, we describe some representative attack scenarios to explain why confirming that a source’s traces contributed to the generation of a synthetic dataset is consequential, even when the attacker already has obtained traces from that source (reference). 
A cloud tenant can confirm that their provider trained on their workloads in violation of contractual terms by comparing the provider's synthetic release against their own recorded traffic. 
In this scenario, the ``attacker'' is the tenant who can obtain the reference simply by recording her own traffic \ie the attacker is the data subject. The threat extends to third parties who merely observe traffic in transit.
A state actor can confirm the loading of a particular website in the synthetic release of a campus network, thereby revealing that an individual accessed the site from that location and exposing them to surveillance or retaliation. In this scenario, the ``attacker'' is the state actor, and she can obtain the reference by loading the website and recording the traffic she receives. Unlike the state actor of the prior example, who must actively reproduce target behavior, an ISP already possesses reference traffic of its subscribers as a byproduct of routine operations, making the attack entirely passive.
%cloud provider can confirm that a specific service (\eg \todo{add maybe something ML related})  
The ISP can confirm that one of its subscribers often visit a specialized hospital after the hospital releases synthetic traffic from its network. %In this case, the ``attacker'' the ISP who has access to her customer's traffic when in network,  is turning routine observations into sensitive inferences.
Table~\ref{tab:threat-model} summarizes four realistic attack scenarios, each mapping to a dataset in our evaluation.\footnote{Datasets are not literal instantiations of the attacks — e.g., we lack access to production hospital networks but close approximations that capture the cross-flow correlation structure that makes the corresponding threat possible.}

%Observe that source-level membership inference in the networking context is meaningful, perhaps even more so, than packet- or flow-level membership inferences. 

These scenarios span a spectrum from active data subjects defending their own rights, to active adversaries reproducing behavior, to entirely passive observers exploiting routine logs. In none of them does the attacker inspect individual packets or flows. In none of them would existing flow- or packet-level privacy metrics detect the threat. And in each case, the entity releasing synthetic data — cloud provider, campus, hospital — would have no indication from current evaluation practices that source-level information remains exposed. DoppelGANger's reported near-random flow-level membership inference illustrates the danger~\cite{lin2020doppelganger}: practitioners may interpret such results as evidence of strong privacy, when in fact source-level leakage is entirely unaddressed.

\subsection{Threat Model for Source MIA }

We consider a data holder (\eg cloud provider, hospital) trains a synthetic generator (\generator) on an original packet trace $D$ and publicly shares the resulting synthetic trace $P(D)$.

\myitem{Adversary goal}: Determine whether a target source contributed traffic to the training dataset~$D$, given only the released synthetic trace~$\mathcal{P}(D)$.

\myitem{Adversary capabilities:} \emph{(i)}~black-box access meaning no knowledge of \generator internals, architecture, or training procedure, and no ability to query it; \emph{(ii)}~a reference trace~$R$ containing traffic from some sources present in~$D$, collected at a different time (and potentially different location) than~$D$, so that $R \cap D = \emptyset$ at the packet level; \emph{(iii)}~access to a \generator to augment~$R$.

%The adversary's goal is to confirm whether a target source (e.g., a tenant, individual, or service) contributed traffic to $D$ by inspecting $P(D)$.
%The adversary has black-box access to the generator, which means she does not know the \generator internals (training pipeline, architecture, loss function, etc.) and cannot query it or request unlimited synthetic samples. The attacker has access to a reference trace $R$ containing traffic from some sources present in $D$, but $R$ is collected at a different time than $D$, hence no packet is both $R$ and $D$. The attacker can use a \generator to augment $D$.

%Our black-box assumption matches today’s practice: data holders release finite synthetic datasets, not queryable models. More importantly, it allows the attack to reveal leakage of all \generators.

\myitem{Our assumption on R is weaker than in standard MIAs.} Classical MIAs assume the attacker holds data drawn from the same distribution as D and can query the target model or observe its internals~\cite{shokri2017membership,van2023membership,Li2025enhanced,hui2021practical}. In our setting, the attacker's capabilities are strictly more limited: \emph{(i)} $R$ and $D$ are collected at different times and potentially from different locations so no record in R appears in D -- exact or near-exact record matching are impossible in our setting: due to non-deterministic factors including CDN server selection, dynamic ad and tracker loading, TLS session parameters, congestion in network \etc and \emph{(ii)} the attacker cannot query the \generators or observe its internals —- only the released synthetic trace P(D) is available.
Importantly, these assumptions reflect common operational scenarios; and as we show in \S\ref{sec:eval}, \sys remains effective under these conditions.

\begin{table*}[t]
\centering
\caption{Threat model scenarios illustrating how an attacker obtains reference traffic $R$ to mount a source-level membership inference attack against synthetic traces $P(D)$. Each scenario maps to a dataset used in our evaluation.}
\label{tab:threat-model}
\footnotesize
\renewcommand{\arraystretch}{1.3}
\setlength{\tabcolsep}{4pt}
\begin{tabularx}{\textwidth}{p{5.0cm} p{2.8cm} p{2.8cm} p{3.2cm} X}
\toprule
\textbf{Scenario} & \textbf{Privacy consequence} & \textbf{How $R$ is obtained} & \textbf{Why $R \cap D = \emptyset$} & \textbf{Dataset} \\
\midrule
Tenant suspects cloud provider trained on its workloads
  & Violation of no-training agreement
  & Tenant records its own workload traffic
  & Different time window than provider's capture
  & DC \\
\addlinespace[2pt]
ISP confirms subscriber's traffic was used in hospital's synthetic release
  & ISP learns subscriber visited the hospital
  & Routine monitoring of subscribers
  & Different vantage point \& traffic
  & BFP \\
\addlinespace[2pt]
State actor confirms a user at a university accessed a sensitive website
  & Surveillance; retaliation against user
  & Loads target websites from own infrastructure
  & Different vantage point and time
  & WFP \\
\addlinespace[2pt]
Enterprise suspects an AS on the path trained on its traffic
  & Violation of data handling agreement
  & Records own traffic
  & Different time window; different vantage point
  & MAWI, CAIDA \\
\bottomrule
\end{tabularx}
\end{table*}
\section{\sys Overview}
\label{subsec:overview}
Having explained that confirming a source's presence in a training dataset (source MIA) is harmful, this section explains why source patterns are distinguishable through cross-flow correlations and introduces the main insights that make the inadvertent preservation of those correlations by \generators a privacy breach.

%\maria{a structural property into a privacy breach.}

\subsection{Vulnerability of Source Patterns }

We start by explaining where the pattern comes from, and then we show evidence that the pattern survives synthesis.

\myitem{Source-specific patterns (\ie cross-flow correlations) exist \& persist over time.} 
Flows from individual users reflect distinctive habits, \eg mobile-app usage, browsing behavior, and typing rhythm, which prior work has shown to be consistent over time and sufficient to de-anonymize users on Tor or VPNs~\cite{verde2014no,tu2018your,crichton2025rethinking,taylor2017robust}.
Flows generated as part of a website load or a service also reflect distinctive patterns which are consistent over time despite upgrades or software modifications~\cite{chen2025cd}, as shown by the extended website-fingerprinting literature~\cite{websitef,sirinam2018deepfp,deng2024robust}.
%Flows from services or servers also remain recognizable even after upgrades or configuration changes~\cite{chen2025cd}, e.g., Amazon’s traffic profile (loading ads and recommendations) will be consistently different from video-streaming.
Finally, flows generated by cloud workloads persist for years, e.g., the training of a specific model generates synchronized flows (bursts) of certain periodicity, while replication jobs create long and heavy flows. The consistency of such patterns allows operators to optimize congestion control and routing\cite{li2024understanding}.

\begin{figure}[t]
    \centering
    \begin{subfigure}{.49\linewidth}
        \centering
        \includegraphics[width=.99\linewidth]{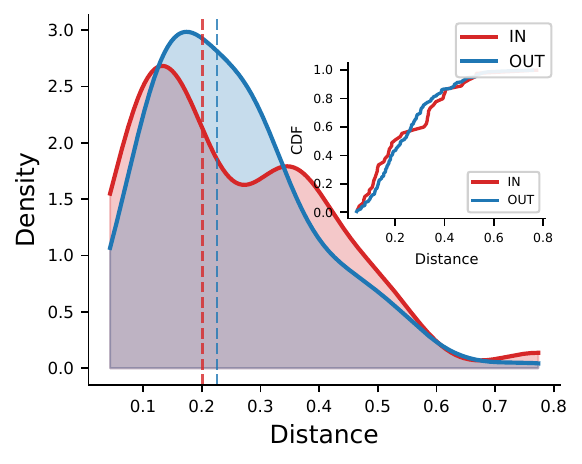}  
        \caption{}
        \label{fig:separable_a}
    \end{subfigure}
    \begin{subfigure}{.49\linewidth}
        \centering
        \includegraphics[width=.99\linewidth]{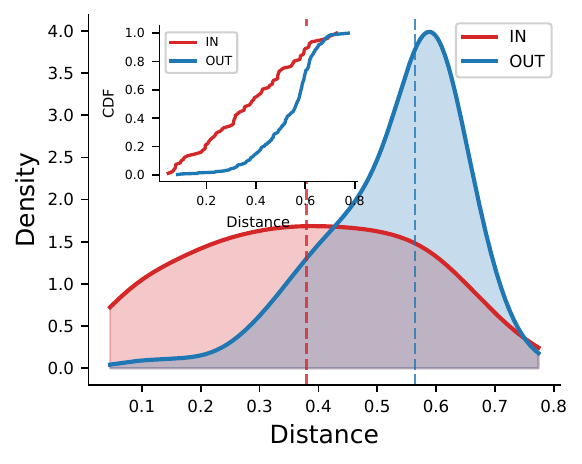}  
        \caption{}
        \label{fig:separable_b}
    \end{subfigure}
    \caption{(a) Source-level correlations survive NetDPSyn's synthesis: the distribution of distances of sources to their nearest synthetic counterpart is different for \texttt{IN} vs \texttt{OUT} sources (\ie those that have vs have not been used in the training of NetDPSyn). (b) \sys reduces the overlap between  \texttt{IN} vs \texttt{OUT} sources, hence facilitating a more accurate MIA attack.}
    \vspace{-0.3cm}
\end{figure}

While the existence of per-source patterns is well-known, they have been assumed not to survive synthesis. As a result, \generators privacy has been solely evaluated through per-packet or flow leakage. \textbf{ Our main insight is that this assumption is incorrect: source patterns manifested in cross-flow correlations survive synthesis.} To demonstrate this, we run a simple experiment. 
We start from a well-known raw trace, from the data center network which we split into $R$ and $D$.
We use this trace to train a DP-private \generator, namely NetDPSyn, which yields a synthetic trace $P(D)$. For each source in $R$ and  $P(D)$, we compute a feature vector of statistics including the mean packet size, the mean inter-arrival time, and the flow count.
For each source in $R$ we find the nearest neighbor source in $P(D)$ (measured as cosine similarity between pairs of feature vectors). We then plot the distribution of distances for sources in $R$ that were also (IN) $P(D)$ (hence seen by NetShare) and sources in $R$ that were (OUT) of  $P(D)$. If source patterns do not survive synthesis, the distances of IN and OUT raw sources from synthetic ones should be indistinguishable. Figure~\ref{fig:separable_a} shows this is not the case: IN sources exhibit higher similarity than OUT sources. 

%\myitem{Cross-flow correlation survive synthesis.}
%Fig.~\ref{} shows the distribution of similarities 

%evidence that source patterns survive synthesis through a simple experiment. 
%We compute per-source feature vectors (mean packet size, mean inter-arrival time, flow count) for both raw and synthetic traces generated by NetShare. We measure cosine similarity (distance) between each raw source and its closer neighbor in the synthetic dataset and plot the distribution of similarities. 

\subsection{Challenges of Source-Level MIA}

While Fig.~\ref{fig:separable_a} shows that the distributions are separable, many individual sources share similar marginal statistics, making per-source decisions unreliable. As a result, directly using the same feature vectors for per-source membership inference would achieve low precision. (as we show in Fig.~\ref{fig:baseline}). This distinction is critical: when confirming membership in a dataset, a high false-positive rate quickly undermines the credibility of the inference and can lead to misleading conclusions, including a false sense of privacy.

This is not an artifact of our example or the attack; user-level network MIA is fundamentally different from their predecessor, typical MIAs in domains like computer vision and natural language processing.
First, unlike inputs considered by typical MIAs (\eg images or text), which are static, user-level traces are highly volatile time-series data that change over time. A traffic source's presence in both the reference $R$ and training dataset $D$ does not imply the existence of an identical packet or sequence of packets from that source in both. In fact, in practice, this is never the case: even consecutive loads of the same website from the same location will yield different packet sequences. %Worse yet, \generators routinely obfuscate IPs and ports, which would have been consistent identifiers, unlike regular MIAs that leave such identifiers untouched. 
In effect, direct use of such MIAs  does not expose the real privacy leak as we show in \S\ref{sec:sys_effectiveness}.  Classification-based approaches face an additional limitation: they can only predict over source labels seen during training, and cannot generalize when source behavior evolves or when there is limited overlap between sources in R and D.
%We also observe that different \generators preserve different artifacts of the original traffic, making attacks based on hand-crafted features brittle and generator-specific. 
Further, many MIA approaches assume unrestricted access to the generator (active queries or access to internals) or synthetic data ~\cite{hayes2017logan, yeom2018privacy}, but this is not compatible with our threat model or networking practices regarding \generators.

%Consider, for instance, the MIA evaluation of DoppelGANger~\cite{lin2020doppelganger}, which reported that the attacker's probability of guessing whether a flow was used in training was near random even without using DP which its follow-NetShare~\cite{yin2022netshare}, a generalization of DoppelGANger, can leak user-level information (see Fig.~\ref{fig:fidelity_privacy} in \S\ref{subsec:privacyofsyngen}).

%Despite the evidence above, it is important to explore whether this is exploitable and whether it is true for all \generators. To this end, we need to design an attack under the threat model we described in \S\ref{subsec:MI}. This endeavor is hard despite the literature in  Membership Inference Attacks (MIAs) because networking has distinct characteristics.

%While Membership Inference Attacks (MIAs) have been extensively studied in domains like computer vision and natural language processing, the approaches are not directly applicable to networking.  

%Second, many existing MIA approaches assume unrestricted access to the generator (active queries or access to internals) or synthetic data ~\cite{hayes2017logan, yeom2018privacy}, yet in networking, only a limited volume of synthetic data is released. 
%Finally, unlike typical MIA, which do not explicitly remove identifiers from training data, \generators routinely obfuscate IPs and ports before training.
% This section gives an overview of \sys, the first membership inference attack against synthetic network data, and explain the key insights.  

\begin{figure}[t]
    \centering
    \includegraphics[width=0.99\linewidth]{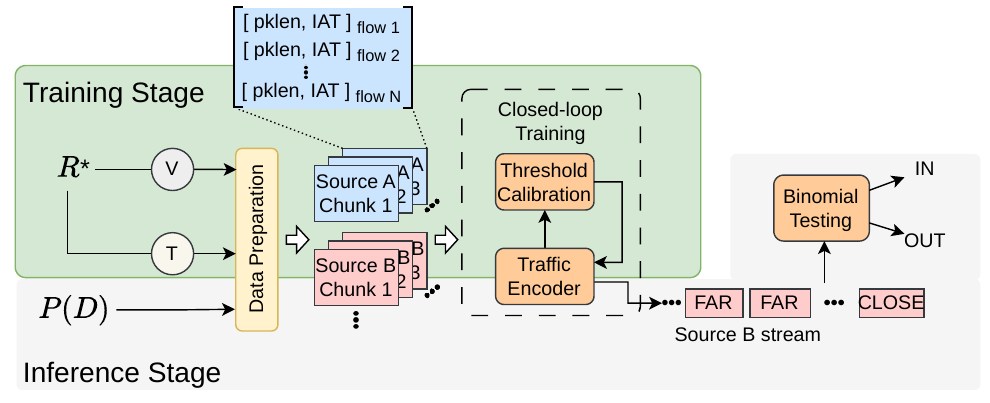}
    \caption{\sys trains a Traffic Encoder to capture per-chunk source-specific patterns using only the reference dataset $R$. 
    Given a synthetic trace $P(D)$, \sys uses the Encoder and Binomial Testing to determine if any synthetic source demonstrates a traffic pattern that resembles that of a source in the reference trace. 
    % \maria{Are colors users here? }
    % \maria{In the figure do trainING instead of train and put in the middle of each side} 
    }
    \label{fig:overview}
    \vspace{-0.3cm}
\end{figure}

% \subsection{\sys Overview}
% \label{subsec:overview}

\subsection{\sys key insights} 
Looking at source-level membership inference from first principles, a successful MIA must overcome three obstacles. First, it must recognize the same source despite natural behavioral drift over time and distortions introduced by synthesis. Second, it must operate across diverse \generators, each of which may obfuscate and preserve different artifacts of the source's pattern. Hence, the attack should not rely on fixed hand-crafted features which will work on some \generators but fail in others. Third, a source may appear only once in a trace, and only portions of its activity may be strongly identifying. An effective attacker, therefore, needs representations that are robust to drift, transferable across generators, and capable of exploiting informative temporal segments.

\sys is built around three corresponding insights. First, rather than relying on brittle features or fixed source labels, \sys uses contrastive learning to learn an embedding space in which traffic generated by the same source remains close while traffic from different sources remains separated. This allows the attack to tolerate temporal variation and adapt to heterogeneous leakage patterns across generators.  As an illustration, Fig.~\ref{fig:separable_b} shows that the overlap between IN and OUT sources is smaller when using \sys's embedding. Second, \sys applies sliding-window chunking to transform a single source trace into multiple overlapping views, enabling learning even when each source appears only once while localizing the most revealing segments of activity. Third, when reference observations are limited, \sys can augment them with generated traffic to create additional approximate views of the target source. These augmentations need not be exact replicas; they enrich the training signal and improve robustness to variability.
Fig.~\ref{fig:separable_b} highlights \sys's effectiveness to capture robust representation for better separability compared to simply using feature vectors.

Fig.~\ref{fig:overview} illustrates the end-to-end workflow of \sys, consisting of two stages: training and inference. 
In the training stage, \sys's  goal is to learn source-specific representations that remain stable despite traffic variability, solely using $R$. In the inference stage, \sys exploits these representations to determine whether each source in $R$ was \texttt{IN} or \texttt{OUT} of $P(D)$.
More specifically, during the training stage, \sys trains a Traffic Encoder on a subset ($T \subset R$) to map chunks of traffic from a given source to nearby points in the embedding space while separating chunks from different sources. Using the held-out set $(V = R \setminus T)$, \sys estimates a decision threshold of distance that when used to split chunks in $V$ distinguishes chunks of sources in $T$ (\texttt{CLOSE}) from chunks of sources not in $T$ (\texttt{FAR}).
During the inference stage, \sys converts embedding distances into membership decisions for the synthetic $P(D)$. Each chunk of each source in $P(D)$ receives an independent \texttt{CLOSE}/\texttt{FAR} prediction based on this threshold, and a binomial test aggregates these per-chunk decisions into a single per-source membership verdict with a controlled confidence level.

\section{ \sys Design}

% \maria{Let's discuss how to make an end-to-end figure containing all prep, training, and threshold determination (also including user-specified inputs e.g., certainty thresholds \etc SHould we not say in the begining raw is augmented? }

%Similar to traditional membership inference attacks on ML classifiers, \sys frames the problem as a binary classification task.
%That is, 
Given a source $src$, \sys predicts \texttt{IN}--if the data holder used $src$’s traffic to train \generators--or \texttt{OUT}--if $src$’s traffic was not used.
Given $R$, \sys first splits the data into a training set $T$ and a validation set $V$ by time and vantage points if multiple are available.
\sys uses  $T \in R$ to train the traffic encoder $M$.
% ( \S\ref{subsec:train}) and the validation $V \in R$ to determine a decision threshold above which it will predict the source as in(\S\ref{subsec:train}). %Details about them are discussed in \S\ref{subsec:train}.

\subsection{Training stage} \label{subsec:train}

In this section, we first introduce \sys's architecture of the traffic encoding and its training scheme. Then we explain how data augmentation improves \sys's performance. Finally, we discuss how \sys calibrates the decision threshold. 

\sys partitions both $T$ and $V$ into traffic chunks using a sliding window of size $W$ and stride $S$. 
For each chunk, \sys groups packets by source IP. Then, for each source, \sys further groups packets into flows (by source/destination IP, source/destination port, protocol) and  extracts the inter-arrival time and packet length as features. 
% For each packet in a flow, \sys extracts the inter-arrival time and packet length as features. 
% By design, \sys can use any packet-level fields, including stateful ones such as TCP flags, but for compatibility, we limit it to fields that most \generators generate.
 % \maria{in a flow? -- let;s include a clear chunk in the figure}

\begin{figure}[t]
    \centering
    \includegraphics[width=0.99\linewidth]{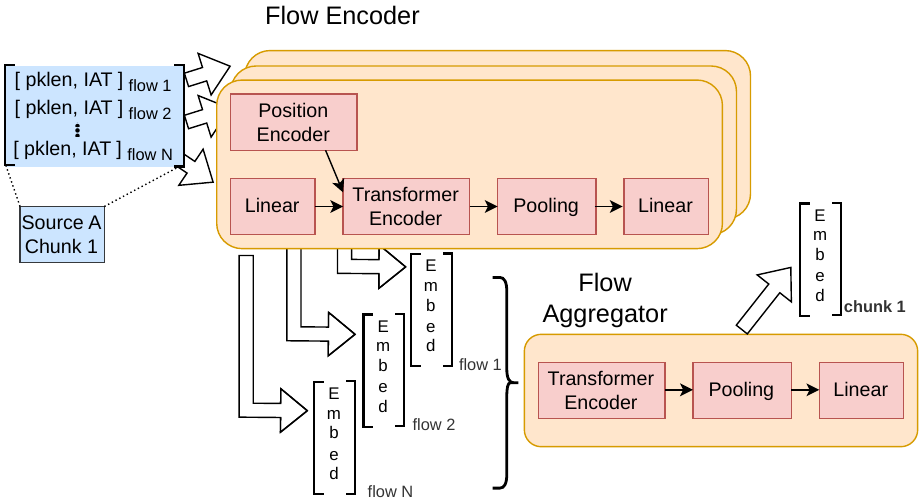}
    \caption{The traffic encoder $M$ map traffic chunks to embeddings by combining a flow encoder with an aggregator.
    % \maria{maybe we should show we turn the packet sequences into intervals and lengths }
    }
    \label{fig:arch}
    \vspace{-0.3cm}
\end{figure}

% The traffic encoder $M$ maps each traffic chunk to a traffic embedding.
% $M$ is an end-to-end differentiable model composed of \emph{(i)} a flow encoder, which maps flows into a flow embedding; and \emph{(ii)} a flow aggregator, which aggregates flow embeddings into a final traffic embedding per source. 
% Fig.~\ref{fig:arch} shows the architecture of $M$. 
% More specifically, the flow encoder first maps each flow (sequence of interarrival times and packet sizes) into a latent space using a linear layer. 
% Since each packet's position within a flow is critical (\eg The SYN packets appear first and are usually 40 bytes in a TCP session traffic), \sys uses a trainable encoder for positional encoding. 
% Then, the flow encoder aggregates the positional encodings to the latent space, passing through a stack of Transformer encoder layers that model the temporal dependencies for each flow. 
% After the flow encoder maps the flows to the flow embeddings, \sys uses the flow aggregator, which is a transformer encoder aggregating the flow embeddings to a single traffic embedding.

The traffic encoder $M$ maps each traffic chunk to a traffic embedding via two components (Fig.~\ref{fig:arch}): \emph{(i)} a flow encoder, which maps each flow's packet sequence which is represented as interarrival times and packet lengths into a flow embedding using positional encoding and stacked Transformer layers; and \emph{(ii)} a flow aggregator, a Transformer encoder that combines per-flow embeddings into a single source-level traffic embedding.

% \maria{so the input out of this part above is packet sequences into multiple embeddings for each source?}\minhao{It is already converted into a single embedding due to the flow aggregator}

\sys uses a few-shot contrastive learning scheme, prototype learning, to train $M$.
It is a self-supervised learning approach~\cite{tao2020self} and its goal is to ensure that traffic embeddings corresponding to the same source are similar \ie the distance across pairs of embeddings is minimized.
To achieve this, \sys computes a prototype embedding vector for each source, which is the mean embedding of all traffic chunks associated with the same source. During training, $M$ maps a batch of traffic chunks to traffic embeddings. A contrastive learning-specific loss function encourages embeddings to align with their corresponding prototypes while pushing them away from other sources.

The number of chunks per source varies in real-world traces. To prevent this from degrading accuracy for sources with few chunks
%For example, for the CAIDA dataset, the source with the most traffic chunks has 792 chunks, while the source with the least has only  10. The standard deviation is 233, highlighting a severe data imbalance problem.  
\sys employs a focal loss variant as a contrastive learning loss that dynamically emphasizes such underrepresented sources. This learning scheme generally improves the effectiveness of $M$, especially on underrepresented sources~\cite{wang2022calibrating,ghosh2022adafocal,song2022tam}.

% We further improve \sys for better membership inference accuracy by augmenting $T$ with synthetic data generated from $T$. 
\sys further employs synthetic data augmentation to improve its effectiveness.
For \sys, synthetic counterparts that preserve user-specific patterns increase chunk count for underrepresented sources and help the encoder learn transformation-invariant representations~\cite{jiang2024netdiffusion,Jin2023}.

Since modern SynNetGens avoid memorizing source-specific identifiers such as IPs~\cite{yin2022netshare}, we train a separate SynNetGen per source in $T$ to ensure that learned patterns remain source-specific. To avoid overfitting given limited per-source flows, we recommend packet-level SynNetGens such as NetDPSyn. Critically, the \generator used for augmentation is trained solely on $T$ and has no access to $D$ or $P(D)$, eliminating any circularity concern.

\begin{figure}[t]
\centering
\begin{subfigure}{.75\linewidth}
    \centering
    \includegraphics[width=.9\linewidth]{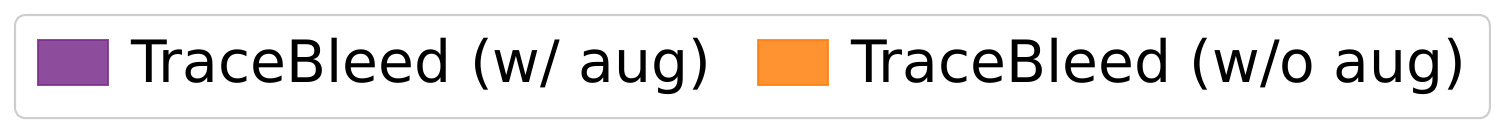}  
\end{subfigure}

\begin{subfigure}{.48\linewidth}
    \centering
    \includegraphics[width=.99\linewidth]{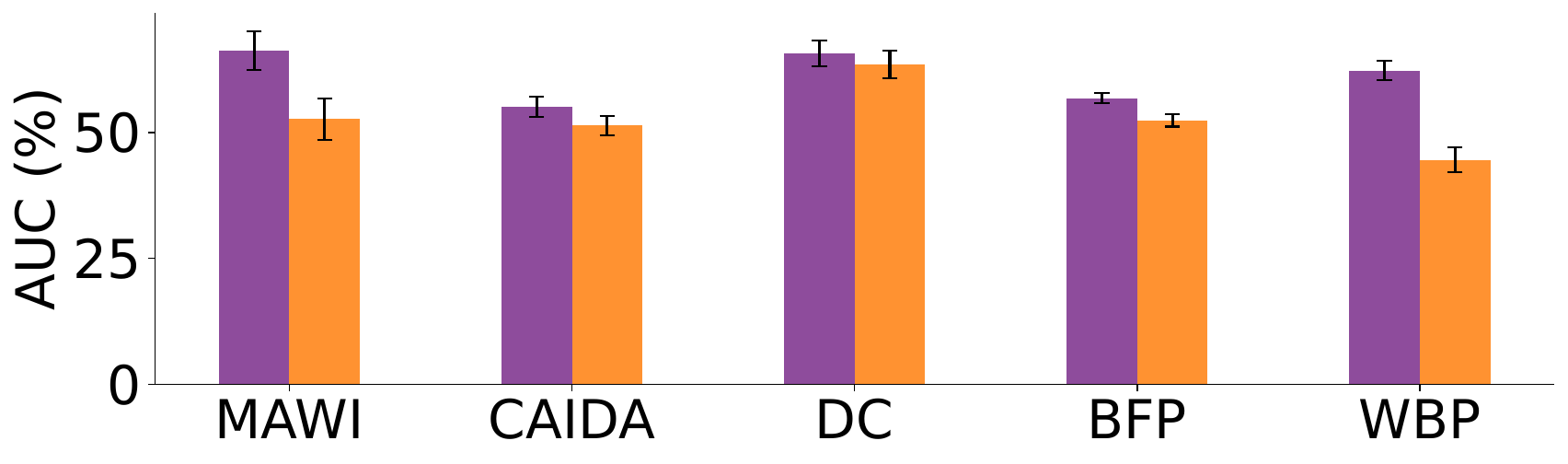}  
    \caption{AUC}
    \label{}
\end{subfigure}
\begin{subfigure}{.48\linewidth}
    \centering
    \includegraphics[width=.99\linewidth]{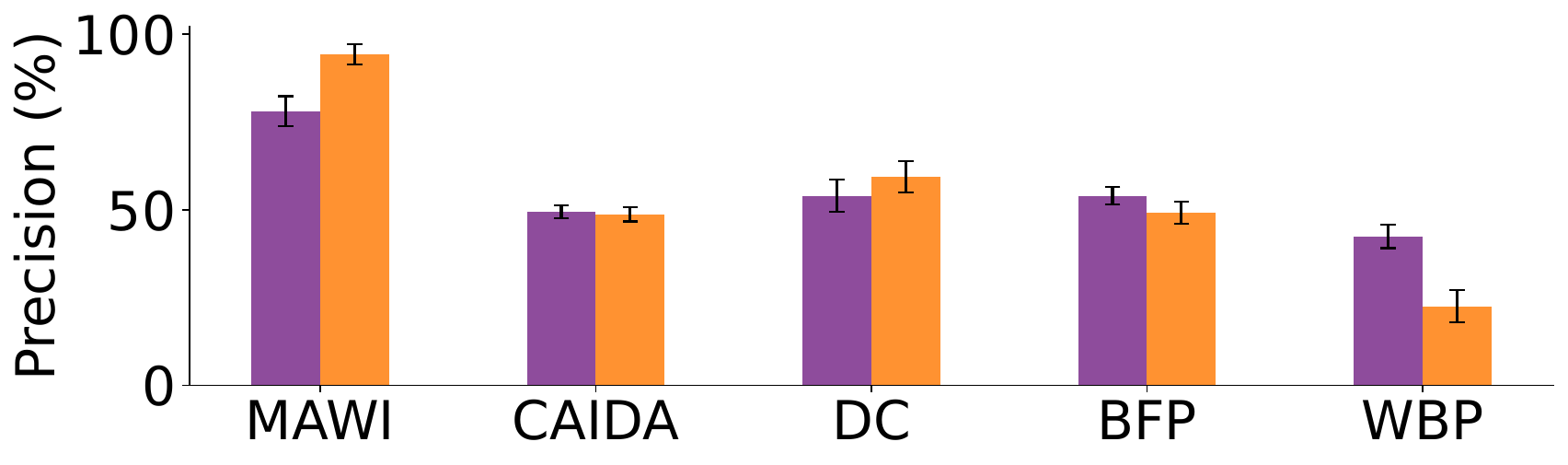}  
    \caption{Precision}
    \label{}
\end{subfigure}
\begin{subfigure}{.95\linewidth}
    \centering
    \includegraphics[width=.95\linewidth]{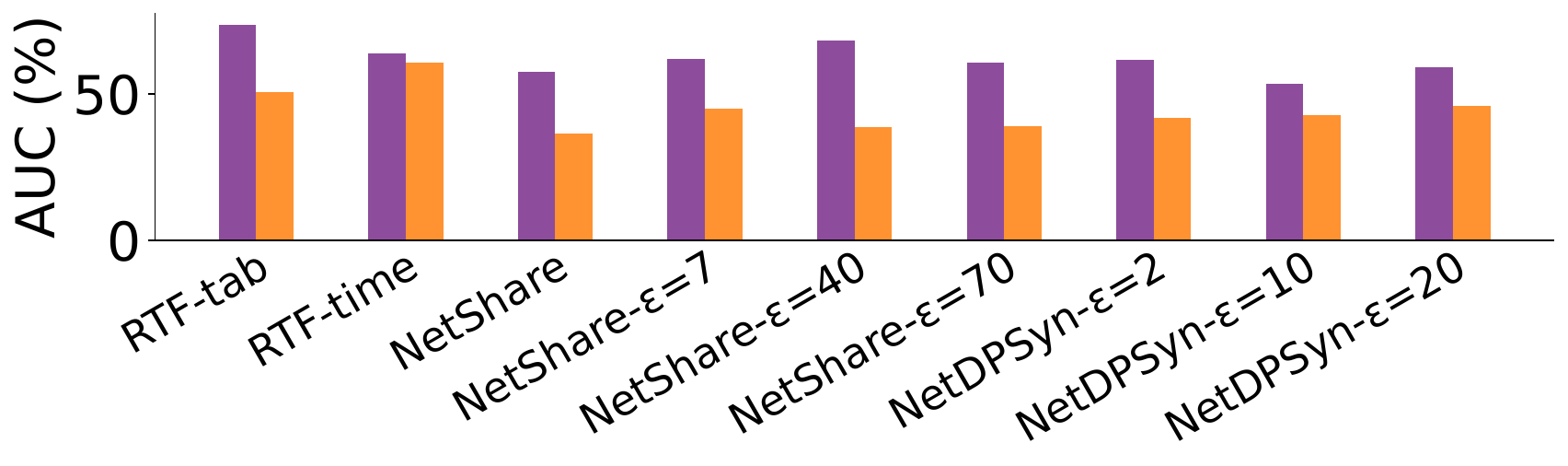}  
    \caption{AUC for all evaluated \generator (WFP dataset)}
    \label{fig:augmentation_ablation_3}
\end{subfigure}
\caption{With synthetic data augmentation, \sys can improve its membership inference performance by 16\%. The performance improvement is universal against $P(D)$ generated by any evaluated \generator even though \sys is only augmenting with one \generator (NetDPSyn $\epsilon=2.0$).} 
\vspace{-0.3cm}
\label{fig:augmentation_ablation}
\end{figure}

Fig.~\ref{fig:augmentation_ablation} reports the average AUC and Precision against 9 evaluated \generators across 5 datasets between \sys with and without augmentation (NetDPSyn $\epsilon =2.0$). Augmentation consistently improves by 16\% in AUC.
Please note that the augmented data is a synthetic versions of $T$, fully independent of $D$ and $P(D)$. Augmenting with any SynNetGen trained on $T$ does not violate our SynNetGen-agnostic property and doesn't cause augmentation circularity. Indeed, Fig.~\ref{fig:augmentation_ablation_3} shows that augmenting with a single \generator improves \sys's performance against $P(D)$ generated by all evaluated \generators, confirming that augmentation helps the encoder learn generalizable cross-flow behavioral patterns rather than \generator-specific artifacts.

% \maria{The entire pipeline, \ie flow encoder, flow aggregator, and prototype computation, trains end-to-end in a single optimization loop. Prototypes are recomputed at each training step as the mean embedding of all chunks belonging to each source in the current batch.}\maria{update figure to show the full training if possible?}

We now discuss how \sys calibrates the decision threshold.
Given a well-trained $M$, \sys can calculate the similarity of two traffic chunks T1, T2 by computing the distance of the two embeddings returned by $M$. \sys defines the distance as $||T1-T2 || = 1 - cosine(embed_{T1}, embed_{T2})$, which is in the range of [0, 2] with values closer to 0 indicating higher similarity.

To calibrate the threshold of similarity for inference, \sys uses $V$ and $T$, which together make $R$. 
Concretely, for any source $src$ and a random piece of $src$'s traffic chunk $i$ from the training set $T$, denoted as $T_{src,i}$, \sys computes the minimum distance with any of the traffic chunks in $V$, \ie 
$\text{dist}_{src,i} = min_x ||T_{src,i} - V_x||$, 
where $V_x$ represents any traffic chunk in $V$ regardless of their true source.
Next, for each $T_{src,i}$, \sys assigns a binary label: \texttt{CLOSE} if the source $src$ is close to any of the chunks in $V$, otherwise \texttt{FAR}.
If $src$ appears in $V$ (\texttt{IN}), then the $src$'s traffic chunks in $T$ should be \texttt{CLOSE} to the $src$'s traffic in $V$. If $src$ doesn't appear in $V$ (\texttt{OUT}), $src$'s traffic chunks should not be similar to any of the chunks in $V$, therefore \texttt{FAR}.

We define the threshold $th$ to predict the traffic chunk $T_{src,i}$ as \texttt{CLOSE} if $\text{dist}_{src,i} < th$; otherwise, \texttt{FAR}.
\sys will choose the optimal threshold that can achieve the best \texttt{CLOSE}-\texttt{FAR} prediction accuracy on $V$.
Note that the \texttt{CLOSE}-\texttt{FAR} prediction is at the chunk level, \ie  for each traffic chunk, we check whether it is \texttt{CLOSE} to any chunk in $V$. 
One source can have multiple chunks, therefore, a set of \{\texttt{CLOSE},\texttt{FAR}\} predictions. \sys leverages all of them to predict whether the source is \texttt{IN} or \texttt{OUT}, which will be explained in \S\ref{subsec:MI}.

\begin{figure}[t]
\centering
% \begin{subfigure}{.47\linewidth}
%     \centering
%     \includegraphics[width=.95\linewidth]{figs_ccs/kde_validate.pdf}  
%     \caption{Distance to the most similar traffic chunk in $V$}
%     \label{}
% \end{subfigure}
% \begin{subfigure}{.47\linewidth}
%     \centering
%     \includegraphics[width=.95\linewidth]{figs_ccs/kde_test.pdf}  
%     \caption{Distance to the most similar traffic chunk in $D$}
%     \label{}
% \end{subfigure}
\includegraphics[width=.55\linewidth]{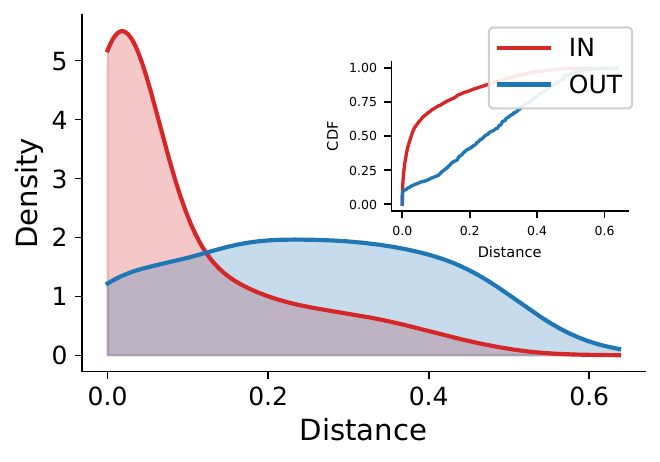} 

\caption{
%Over-time source patterns change less than cross-source patterns.
%\sys is able to capture the persistent and source-specific pattern.
While there is intra-source variability over time (distance is non-zero for IN), that variability is smaller than inter-source variability (distance of OUT), hence per-source characteristics persist over time and \sys can capture them. 
There is a clear distinction (threshold) in the embedding-space distance between sources which are \texttt{IN} and \texttt{OUT}.
% , which is consistent in both $D$ and $V$ (we can find it in $V$ and use it in $D$).  
% \maria{do we talk about this figure if not maybe we should reference it in 3 or even in 4}
}
\label{fig:gap}
% \vspace{-0.1cm}
\end{figure}

Fig.~\ref{fig:gap} shows the probability density for $\text{dist}_{src,i}$ when the source in $T$ is \texttt{IN}/\texttt{OUT} of $V$. The \texttt{IN} sources' traffic chunks have much \texttt{CLOSE}r distance to $V$ compared to \texttt{OUT} sources, which validates our intuition. 
% Meanwhile, the intra-source variability over time is smaller than inter-source variability. Hence, \sys can capture the per-source time-persistent characteristics, preserving the distinction when against $D$.

\subsection{Membership Inference Stage}
\label{subsec:MI}

After \sys trains $M$ and determines the threshold, she is ready to perform the attack. Given a source $src$, \sys first collects all the traffic chunks of $src$ in $T$. For each $T_{src,i}$, \sys calculates $\text{dist}_{src,i}$. During inference, instead of calculating the minimum distance from $V$, \sys computes minimum distance from $P(D)$ (the synthetic counterpart of $D$), \ie $\text{dist}_{src,i}=min_x ||T_{src,i} - P(D)_{x}||$. To do this, we use the pre-configured $W$ and $S$ to build traffic chunks for each source in $P(D)$. Then, by the determined threshold, \sys returns a set of predicted \{\texttt{CLOSE}, \texttt{FAR}\}. The size of the set depends on the number of traffic chunks of $src$ in $T$.

Given the output predictions, we use binomial testing under a null hypothesis to determine if $src$ is present in $D$. 
Each prediction is binary--either \texttt{CLOSE} or \texttt{FAR}, and the binomial test assesses whether the number of \texttt{CLOSE}/\texttt{FAR} predictions is significantly higher than expected. If the number of \texttt{CLOSE} is significantly higher than a prior probability, it suggests that many traffic chunks are similar to those in $D$, \sys infer that $src$ is present in $D$ with high confidence\footnote{We set confidence level to be 0.95 as default}. And \sys will return with the prediction \texttt{IN}. 
Similarly, we use the same way to test \texttt{FAR} and predict whether $src$ is not present in $D$ with high confidence. If with high confidence, \sys returns with prediction \texttt{OUT}.
If neither number of \texttt{CLOSE}/\texttt{FAR} is significantly high, \sys returns "unsure".
Users of \sys, such as data holders, can also override the returned "unsure" as \texttt{IN} or \texttt{OUT} using a majority vote.

Fig.~\ref{fig:prec_cov} highlights the coverage of sources and Precision, the fraction of predicted \texttt{IN} sources which are truly \texttt{IN}, when setting a higher confidence starting from 10\% to 99.9999\%. 
High precision at confidence levels such as 95\% and 99\% demonstrates TraceBleed's ability to capture time-persistent source patterns.
\section{Evaluation: \generators privacy}
\label{sec:eval}

In this section, we use \sys to empirically evaluate the source privacy leakage of the latest \generators.
We first describe our methodology, then \generators' properties and then \sys properties.
%\maria{we need a better structure for eval. 1: focused on \generators (privacy, 10x, confidence); 
%2: attack  is it more reliable than baselines? how does it behave with more/less sources? why does it work? which features does it use more?
%3:}

\subsection{Methodology} \label{sec:setup}
In this section, we explain our methodology. 
All our datasets and evaluation infrastructure is published on the leaderboard\cite{rbleaderboard}, where we will also welcome pull requests for contributing additional datasets. 

% \maria{Is it possible/easy to add the nearest neighbor we have in 3 to all dataset? in a table -- to emphasize leackage independently of \sys?}
% \minhao{Fig 1 in 3 is for NetShare in CAIDA datasets. Do you mean I fix NetSHare and plot for all datasets?}

\mypara{Datasets} 
We evaluate our approach on five datasets capturing diverse forms of source-level behavior,
spanning fully organic traffic (MAWI, CAIDA, DC) to controlled-but-conservative constructions (BFP, WFP). For the first three, which are publicly available, R is drawn from the same dataset as D with no overlap, providing a controlled setting to isolate source-level leakage under real-world traffic noise. The remaining two are generated through controlled interactions with live websites, with R collected at a different time and from a different location, introducing variability from routing, latency, and location differences.

 \myitem{MAWI \& CAIDA}: are packet traces collected from a Japanese Internet Service Provider (ISP)~\cite{tsareva2023mawi} and an Internet backbone link~\cite{CAIDA}, respectively. 
Hence, the source-level behavior stems from real users’ browsing patterns and real services’ traffic, preserving realistic cross-flow correlations. These datasets also capture real-world variability and noise, as collected traffic inherently reflects diverse applications, routing dynamics, and network conditions over time.

\myitem{Data Center (DC)} contains traffic captured from the ``UN1'' data center~\cite{benson2010dcdataset}. Hence, the source-level behavior stems from the traffic patterns of hosted applications (\eg MapReduce jobs), preserving realistic cross-flow correlations induced by application-level behavior. The dataset also captures real-world variability and noise, as traffic reflects dynamic workloads, scheduling decisions, and network conditions over time.%For all these datasets (\ie MAWI, CAIDA, DC), we do not control sources, hence $R$ and $D$ are drawn from the same dataset with 9:1 ratio and no overlap: no packet or flow was both in $R$ and $D$.

\myitem{Behavioral FingerPrinting (BFP) \& Website FingerPrinting (WFP)} are packet traces that we constructed with experiments and contain website loads.
For \textbf{BFP}, we emulate 50 distinct users with distinct browsing behaviors, each of which loads a randomly selected set of websites from three geographically distant locations and at distinct times: a campus on 11/22/2024, the CloudLab Wisconsin cluster on 6/19/2025, and the CloudLab Emulab cluster on 7/1/2025. $D$ is taken from Emulab while $R$ is from the campus network and the Wisconsin cluster.
%\maria{INTEGRATE this : While 30 websites is modest compared to the full web, prior website fingerprinting literature has shown that even small closed-world sets produce weaker attacker advantage than open-world settings with long-tail distributions [41], [15]. Our choice thus biases against the attacker.}

For \textbf{WFP} we load various website loads from the aforementioned vantage points. Hence, the source-level behavior stems from the websites, each producing multiple flows for loading different parts of a webpage, such as images and text ads. Similar to BFP, Emulab for $D$ and the other two for $R$.
While BFP and WFP are constructed, they contain realistic cross-flow correlations as they are derived from interactions with live websites. They also capture real-world variability and noise (exactly what makes the attack harder), since traffic is served from diverse locations, collected from different client vantage points, and observed at different times, preventing records from being identical.
 %We generated this dataset by loading websites from different locations.
%This setup models an attacker who visits a small set of sensitive websites from a remote vantage point relative to the victim's network.
%The attacker's reference is built from emulated users' traffic generated from locations distinct from those used to train the \generators.  Hence, this setup models an attacker who observes the user's behavior at a different location and time than the data holder. For instance, the attacker could capture the user’s traffic while at a café, whereas the data-holder who trains the \generator is the user's ISP.
We provide additional details in Appendix in \S\ref{apdx:detailsdatasets}.

For all selected datasets, we leverage NetDPSyn($\epsilon=20$) for synthetic data augmentation. We train a separate NetDPSyn model which generates 10X amount of packets for each source. We then filter out all the synthetic packets whose source IP doesn't belong to the target source, and augment the rest as a synthetic counterpart.
We define the window size $W$ as 1/10 of the duration of D, with a stride $S$ of $W/10$. We group user traffic based on the source IP of the packets and further segment into chunks using the configured $W$ and $S$. 
Recall that $M$ encodes user traffic behavior by accepting a set of flows, so we further group the user traffic chunks by their five-tuple (source/destination IP, source/destination Port, Protocol) into flows and remove any chunk if user traffic is inactive during that chunk period.\footnote{We consider a user to be inactive at a specific time window if there are fewer than 2 flows that have at least 5 packets.} Additionally, we drop users with fewer than 10 chunks, as their traffic is insufficient to model user behavior.   

\myitem{On the reality-to-evaluation gap}
We deliberately construct BFP and WFP from controlled website loads rather than organic browsing traces. Ethical and legal constraints prevent redistributing real users' browsing histories, but more importantly, our construction makes the attack strictly harder: real users exhibit richer behavioral fingerprints — habitual visit orderings, session durations, and long-tail site distributions — that would only strengthen cross-flow correlations. Our emulated users instead draw websites uniformly at random from a small pool of 30 sites, producing weaker and less distinctive signatures. Any leakage TraceBleed detects therefore represents a lower bound on what an attacker would achieve against organic traffic.
Crucially, three of our five datasets (MAWI, CAIDA, DC) contain fully organic traffic from real users and services. TraceBleed's effectiveness on these datasets demonstrates that source-level leakage is not an artifact of our controlled construction; BFP and WFP complement them by isolating specific threat vectors under conservative conditions.

%\mypara{Synthetic data augmentation}
%We augment synthetic data to evaluate the performance of upgraded \sys using WEB*. We select NetDPSyn as the candidate \generator as it generates traces at packet level and it provides privacy guarantee. We first group $T$ by the website (source). Then we train NetDPSyn (eps=20) for each websitew, generate equal number of synthetic data and augment into training set.

\mypara{Metrics} Given a user's traffic chunks, \sys predicts if this source is included in $D$ by inferring from $P(D)$. 
We assess the privacy of a given \generator by comparing \sys's prediction to ground truth using two primary metrics, namely AUC and Precision. 
\textbf{AUC} measures the probability that \sys produces a lower average embedding distance for a truly IN source than for an OUT source.%, which indicates stronger membership evidence. 
 AUC is a particular insightful metric because it is independent of any fixed decision threshold.
Any AUC above 50\% indicates the presence of exploitable source-level information in the synthetic data.
While AUC is very useful it measures average discriminative ability across all sources, but in  practice, an attacker does not need to identify every source; she needs to confidently identify specific sources. 
\textbf{Precision} measures the fraction of sources predicted as \texttt{IN} that are truly \texttt{IN}, directly capturing an attacker's confidence when committing to a specific accusation. 
For both AUC and Precision, lower values indicate higher data privacy.
We also report the fraction of vulnerable \texttt{IN} sources \sys can correctly identify with confidence $\geq95\%$, highlighting \sys's capability for accurate and confident membership inference.  
\begin{table}[t]
\centering
\small

\begin{tabular}{@{}lcccc@{}}
\toprule
\small
\multirow{2}{*}{\textbf{Dataset}} & \multirow{2}{*}{\textbf{IN vs. OUT}} & \multicolumn{2}{c}{\textbf{Random Guess}} & \\ \cmidrule(lr){3-4}
 &  & \textbf{AUC} & \textbf{Precision} &  \\ \midrule
MAWI  & 43 vs. 148 & 0.50 & 0.23  \\
CAIDA & 86 vs. 107 & 0.50 & 0.45  \\
DC    & 81 vs. 253 & 0.50 & 0.24  \\
BFP   & 25 vs. 38  & 0.50 & 0.40  \\
WFP   & 15 vs. 35  & 0.50 & 0.30  \\ \bottomrule
\end{tabular}
\caption{Our datasets are diverse in capture locations, and balanced. If the attacker does better than a random guess, then there is privacy leakage.}
\label{tab:setup}
\vspace{-0.3cm}
\end{table}

Table~\ref{tab:setup} presents the number of sources in $R$ which are included in $D$ (\texttt{IN}) versus those that are not (\texttt{OUT}), along with the probabilities of randomly guessing a user’s inclusion as a baseline. The datasets vary significantly in the number of sources and their inclusion ratios, reflecting diverse user traffic characteristics. If \sys prediction is better than random guessing (\eg higher AUC score), then the dataset and corresponding \generator leaks privacy. 
% Some datasets, such as MAWI and DC, exhibit a high imbalance between \texttt{IN} and \texttt{OUT} sources. Random guessing and naively predicting sources as \texttt{IN} lead to low precision scores.

To evaluate the fidelity of the \generator,  we follow and extend the practice of NetShare to calculate the fidelity (Appendix in \S\ref{apdx:fidelity}). In short, we present a summarized fidelity score, Avg(JSD+EMD), which aggregates the fidelity score such as packet length distribution similarity, flow length distribution similarity and etc. These fidelity score comprehensively describes the distributional similarity between the synthetic data and its raw counterpart.    

\mypara{Baselines}
To ensure the comparison is fair, we design three GAN-Leaks variants at increasing granularity. GAN-Leaks (packet) represents each source by the nearest-neighbor distance of its individual packets to the synthetic pool. GAN-Leaks (flow) uses 25 per-flow statistics drawn from the traffic classification literature~\cite{iscxvpn2016}. GAN-Leaks (source) uses 79 features specifically chosen to capture cross-flow source-level patterns, including statistics over flow IATs, per-flow packet counts, bytes/s, and flow counts, that prior work (and \sys) has identified as discriminative for source identification~\cite{tu2018your,verde2014no}. %These features directly encode the cross-flow correlations our paper argues are privacy-relevant. 
%We adapt GAN-Leaks~\cite{chen2020ganleaks}, a black-box nearest-neighbor MIA for generative models, to source-level membership inference at three granularities.
%\textbf{GAN-Leaks (packet)} features each packet as a relative timestamp to the first packet of that source and packet length, and scores a source by the mean nearest-neighbor distance from its packets to the synthetic packet pool. \textbf{GAN-Leaks (flow)} operates on 25-dimensional per-flow statistics. \textbf{GAN-Leaks (source)} aggregates each source into a single 79-dimensional vector of traffic statistics and computes its nearest-neighbor distance to synthetic sources directly, requiring no within-source aggregation. All three calibrate a decision threshold on the validation set and are evaluated under the same protocol as \sys. Details of the per-flow statistics and source aggregated statistics are in Appendix in \S\ref{apdx:detail-gan-leak}.
%Indeed, all these baselines are a natural extension of GAN-Leaks, the canonical black-box MIA attack on the networking traces. 
Observe that GAN-Leaks is the strongest applicable baseline. Recent MIA techniques from the ML privacy literature \eg LiRA, shadow-model attacks, and calibration-based methods~\cite{shokri2016membership,carlini2022membership,yeom2018privacy}, are incompatible with our threat model as they assume query access to the target model or the target model's loss or confidence scores on specific inputs. Training shadow \generators on subsets of R does not resolve this, as the attacker still lacks the per-sample loss scores from the target generator that these methods require for calibration. 
% \todo{check}

% \maria{could one say that these attacks are MIA or fingerprinting?} \minhao{we can say MIA I think}

\mypara{MIA \generators}
We evaluate a wide spectrum of \generators, including GAN-based~\cite{xu2019ctgan} \textbf{NetShare}~\cite{yin2022netshare}; Diffusion-based~\cite{ho2020diffusion} \textbf{NetDiffusion+}~\cite{jiang2024netdiffusion}, GPT-based~\cite{brown2020gpt} \textbf{RealTABFormer(RTF)}~\cite{solatorio2023realtabformer} and marginal distribution-based synthesizers~\cite{zhang2021privsyn,sarela2003marginal} \textbf{NetDPSyn}~\cite{sun2024netdpsyn}. For RTF, we use both \textbf{RTF-Tab} and \textbf{RTF-Time}, which are different by generation granularity (packet versus flow).
These \generators differ in model architectures, training logic, generation granularity (per flow vs. per packet), and approach to privacy. Appendix in \S\ref{apdx:detailsgen} includes more details.
For each \generator, we generate a sufficient amount of synthetic data (10X of the original training size\footnote{For \generators with packet generation granularity, 10X means 10X packets. For those with flow granularity, it means 10X flows. For NetDiffusion+, we fail to generate 10X data within reasonable time budget (e.g., 6h budget for generation only.)}).

For all DP-protected models, we use $(\epsilon, \delta)$ DP with $\delta=1e-5$.
We evaluate three different $\epsilon$ values for each DP-protected \generator (NetShare and NetDPSyn). 
Note that $\epsilon$ values between NetShare and NetDPSyn are not directly comparable, as they apply DP at different granularities, hence offering different protections, \ie NetShare protects flows while NetDPSyn protects packets. 

To ensure that they are well trained, we evaluate the fidelity of generated data and summarize our results in Fig.~\ref{fig:apdx:general_fidelity} in Appendix \S\ref{apdx:fidelity} (lower value indicates better fidelity).

\mypara{Testbed} Appendix~\S\ref{apdx:testbed} details the used CPU/GPUs.

\subsection{Privacy of \generators}
\label{subsec:privacyofsyngen}

% To evaluate the privacy of \sys we focus on three questions: \emph{(i)} Do state-of-the-art \generators protect user privacy? \emph{(ii)} How do the common practices of incorporating differential privacy on \generators affect source-level privacy and fidelity? and \emph{(iii)} Is sharing the \generator model safe? \maria{this is not uptodate, we might prefer to remove it not a priority but now it is wrong}

\begin{figure*}[t]
\centering
\begin{subfigure}{.45\linewidth}
    \centering
    \includegraphics[width=.99\linewidth]{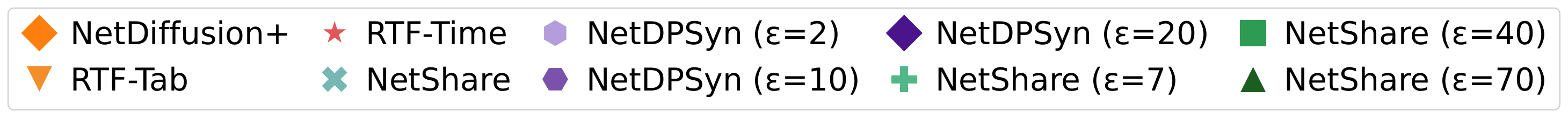}  
    \label{}
\end{subfigure}

\begin{subfigure}{.195\linewidth}
    \centering
    \includegraphics[width=.99\linewidth]{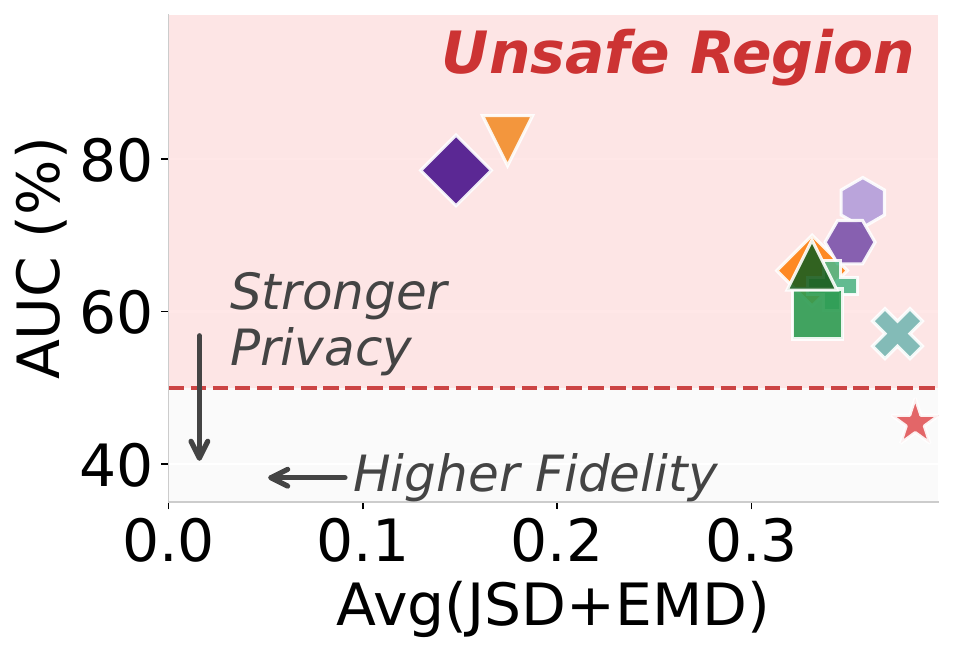}  
    \caption{MAWI, AUC}
    \label{fig:fidelity_privacy_e}
\end{subfigure}
\begin{subfigure}{.195\linewidth}
    \centering
    \includegraphics[width=.99\linewidth]{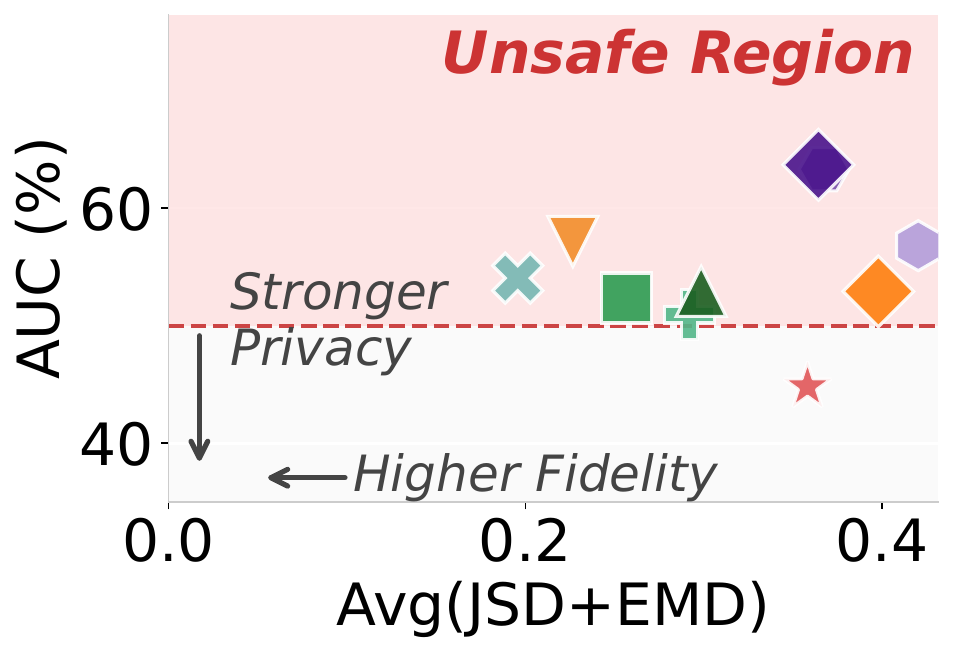}  
    \caption{CAIDA, AUC}
    \label{fig:fidelity_privacy_a}
\end{subfigure}
\begin{subfigure}{.195\linewidth}
    \centering
    \includegraphics[width=.99\linewidth]{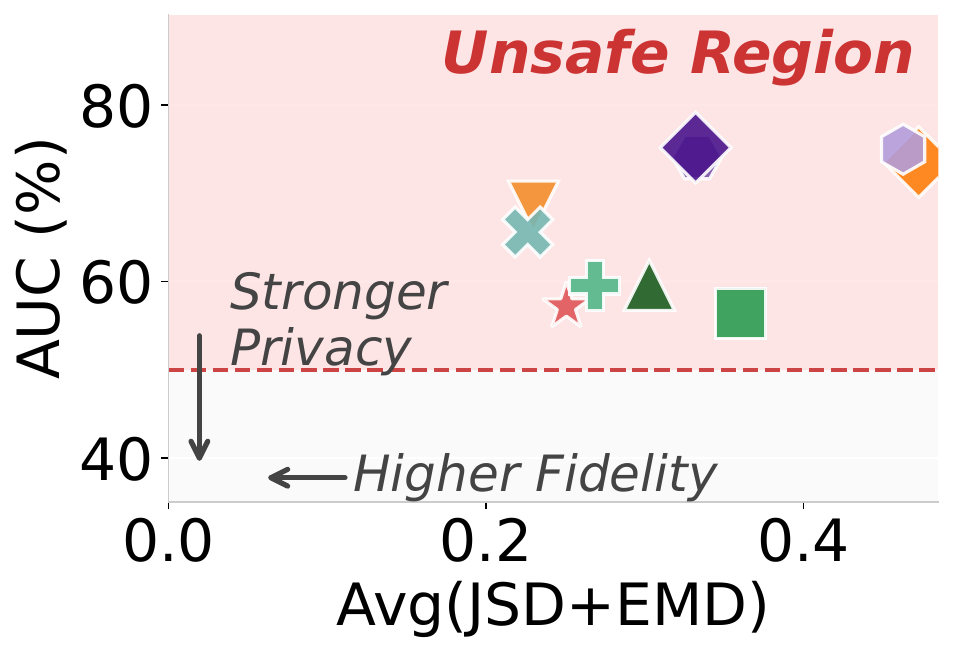}  
    \caption{DC, AUC}
    \label{fig:fidelity_privacy_c}
\end{subfigure}
\begin{subfigure}{.195\linewidth}
    \centering
    \includegraphics[width=.99\linewidth]{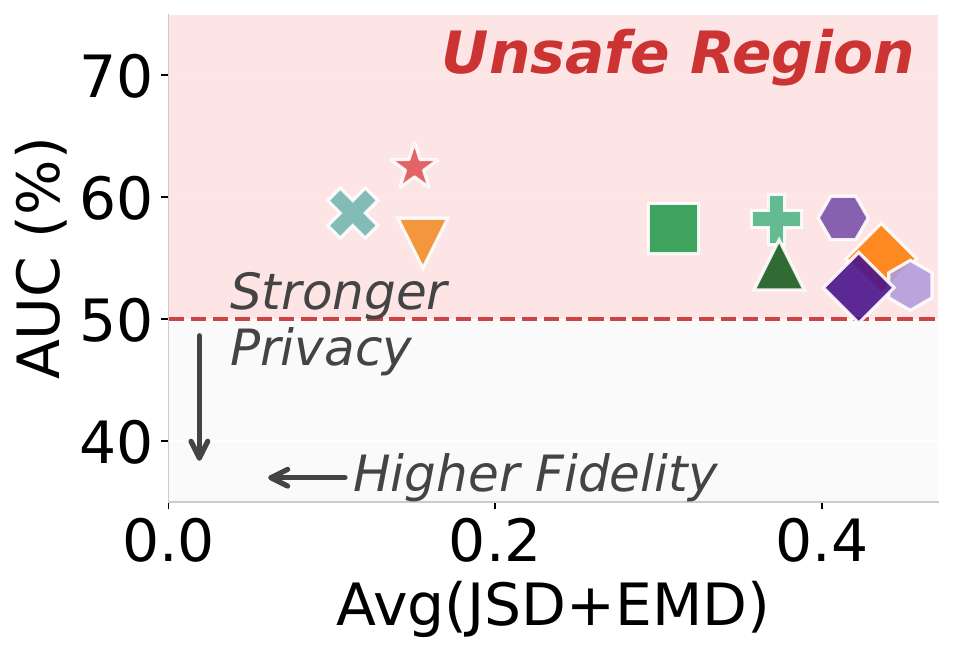}  
    \caption{BFP, AUC}
    \label{fig:fidelity_privacy_e}
\end{subfigure}
\begin{subfigure}{.195\linewidth}
    \centering
    \includegraphics[width=.99\linewidth]{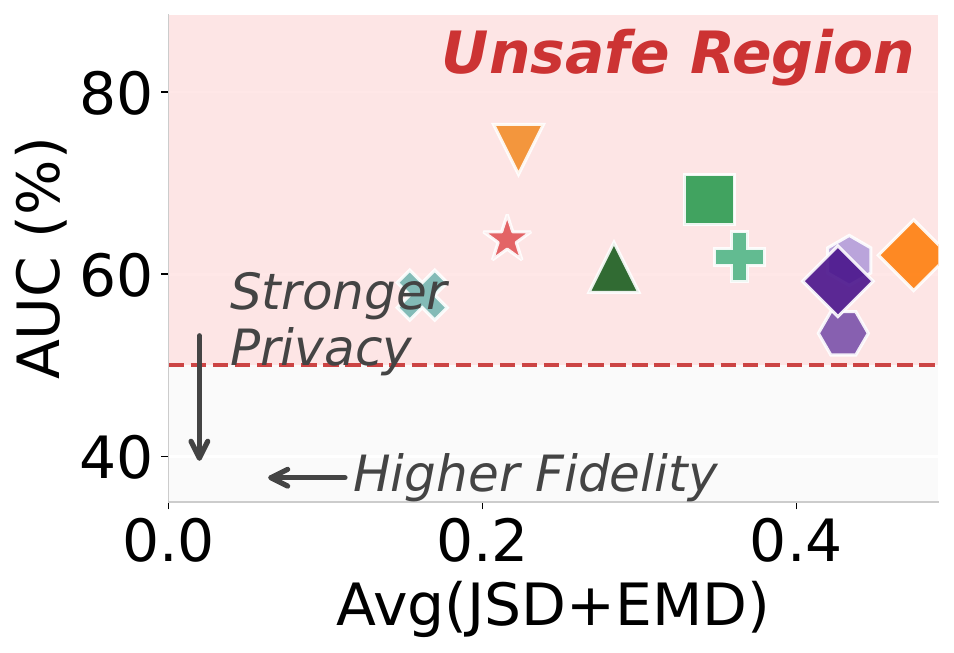}  
    \caption{WFP, AUC}
    \label{fig:fidelity_privacy_g}
\end{subfigure}

\begin{subfigure}{.195\linewidth}
    \centering
    \includegraphics[width=.99\linewidth]{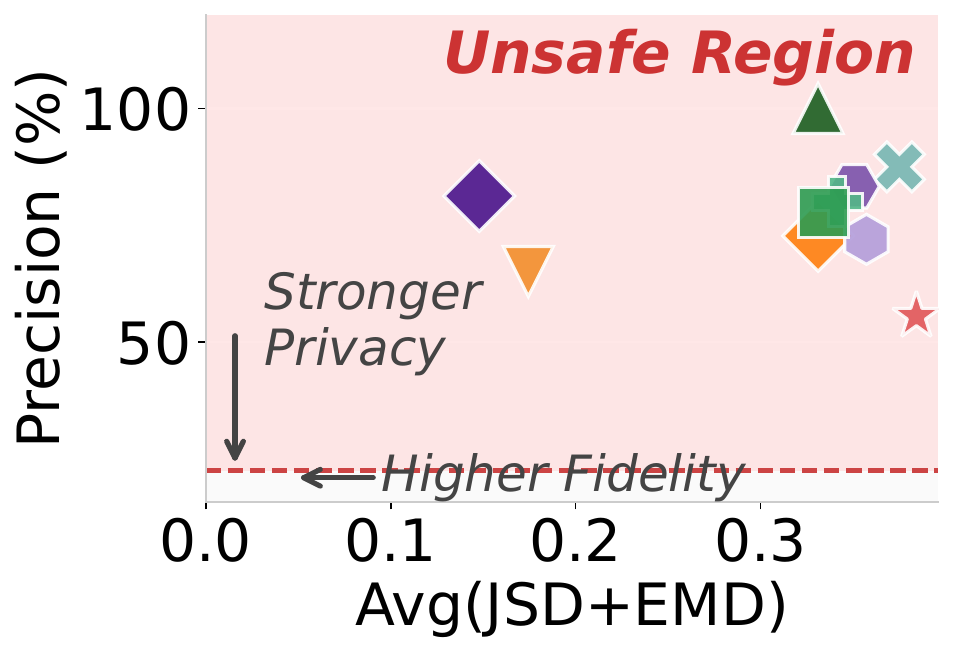}  
    \caption{MAWI, Precision}
    \label{fig:fidelity_privacy_f}
\end{subfigure}
\begin{subfigure}{.195\linewidth}
    \centering
    \includegraphics[width=.99\linewidth]{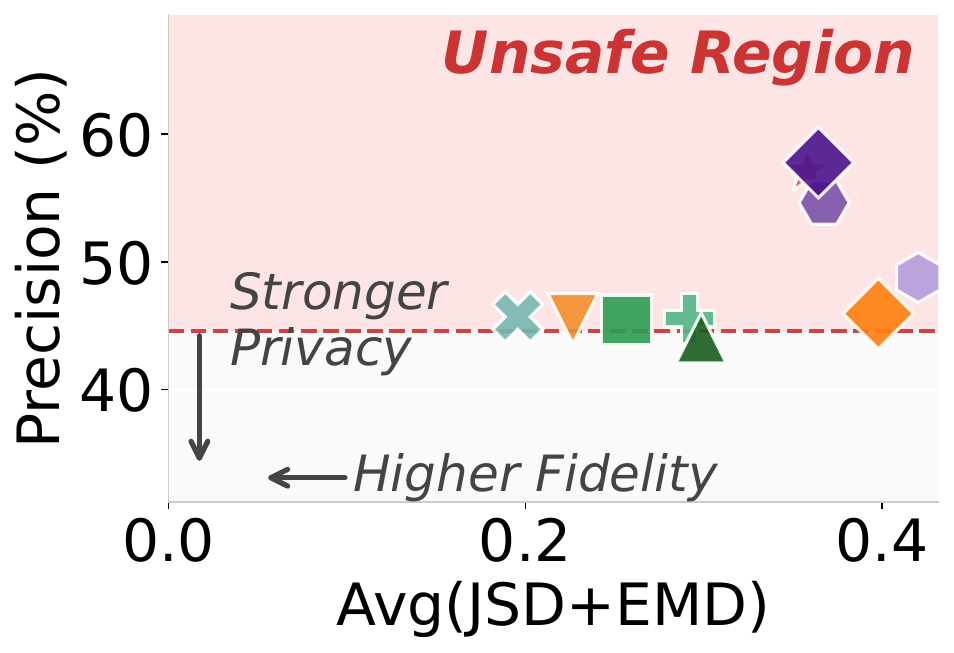}  
    \caption{CAIDA, Precision}
    \label{fig:fidelity_privacy_b}
\end{subfigure}
\begin{subfigure}{.195\linewidth}
    \centering
    \includegraphics[width=.99\linewidth]{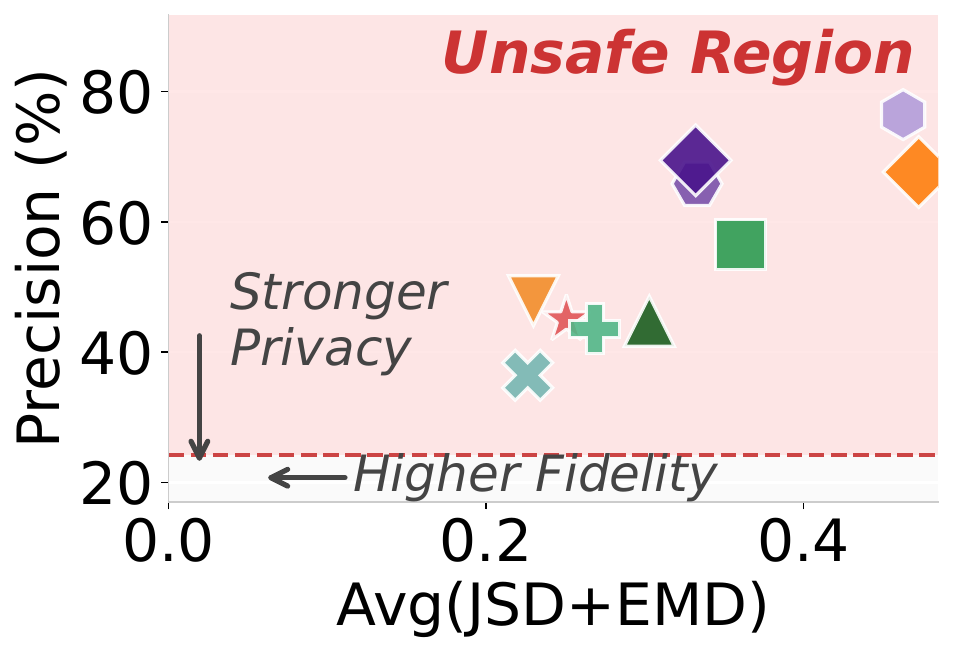}  
    \caption{DC, Precision}
    \label{fig:fidelity_privacy_d}
\end{subfigure}
\begin{subfigure}{.195\linewidth}
    \centering
    \includegraphics[width=.99\linewidth]{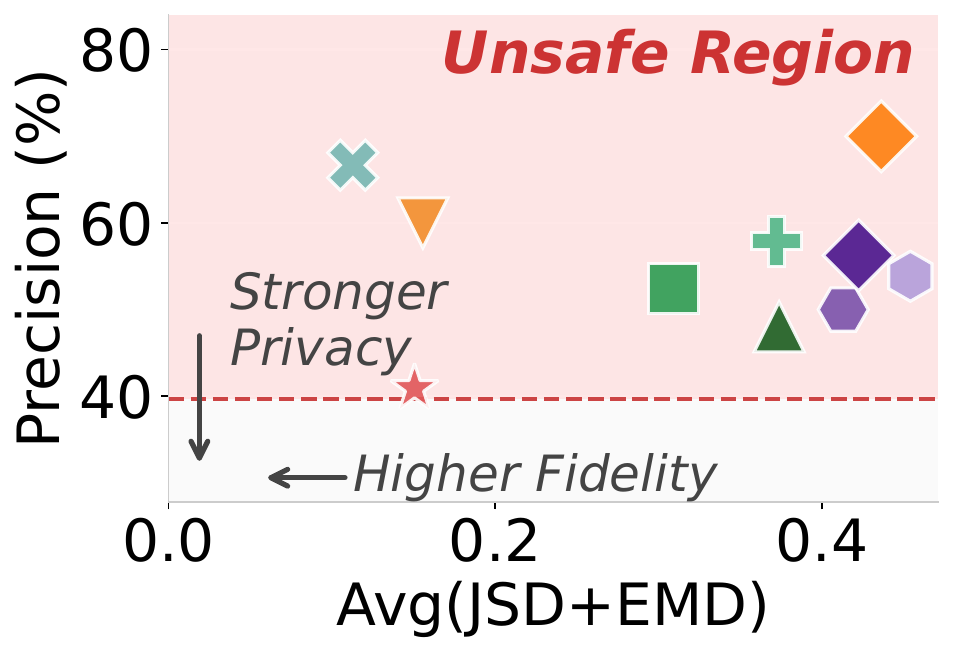}  
    \caption{BFP, Precision}
    \label{fig:fidelity_privacy_f}
\end{subfigure}
\begin{subfigure}{.195\linewidth}
    \centering
    \includegraphics[width=.99\linewidth]{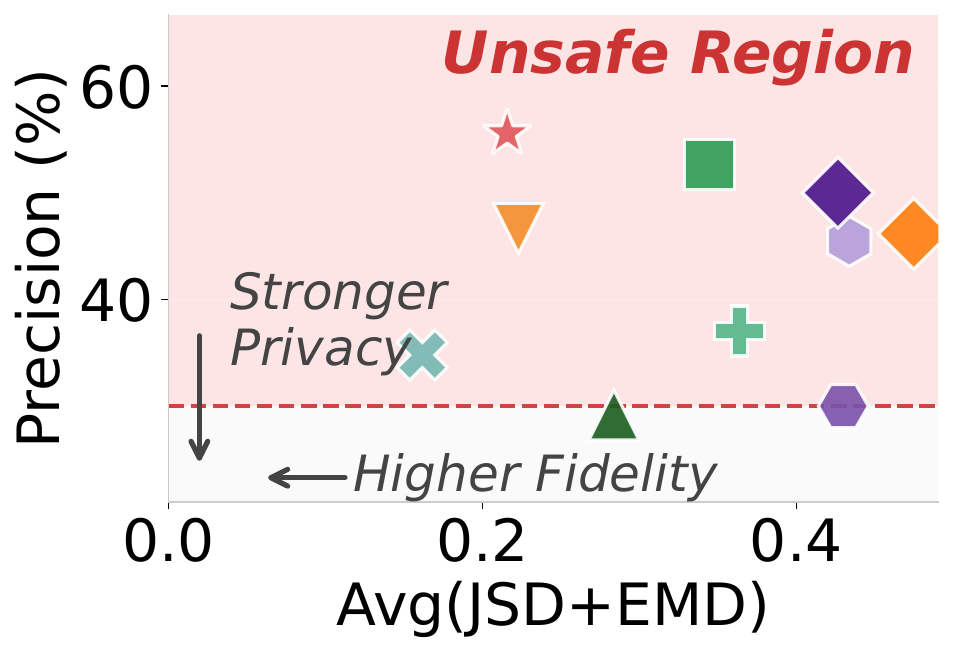}  
    \caption{WFP, Precision}
    \label{fig:fidelity_privacy_h}
\end{subfigure}

\caption{
% Fidelity versus F1 score against \sys for synthetic traces. 
The mere use of \generators does not automatically guarantee privacy; neither does the use of DP. Synthetically generated traces leak user privacy (in shaded regions) or severely degrade fidelity (higher values).}
\label{fig:fidelity_privacy}
% \vspace{-0.1cm}
\end{figure*}

\myitem{Using \generators to share network traces does not automatically guarantee source privacy, even when using flow- or packet-level DP.}
We first train \generators on $D$ and generate a sufficient amount of synthetic data. Fig.~\ref{fig:fidelity_privacy} reports both the fidelity and privacy leakage of the synthetic data generated by \generators across datasets. 
To make the graph more interpretable, we shade in red the region where AUC and Precision exceed random-guess performance, indicating \generators that leak source-level information in Fig.~\ref{fig:fidelity_privacy}. All \generators (including those with privacy guarantees at the packet or flow level) leak source-level privacy.   
For example, \sys achieves 64\% higher AUC and 183\% higher Precision than random guess on traffic generated by RTF-Tab under MAWI dataset, indicating that \generators leak privacy.

Fig.~\ref{fig:fidelity_privacy} also shows that DP-protected \generators fail to improve privacy while suffering from additional fidelity degradation. 
In fact, the use of DP causes 31\% fidelity degradation and the privacy is on-par with non-DP-protected \generators.
Notably, our findings hold regardless of the DP granularity. Both flow-level DP (NetShare) and packet-level DP (NetDPSyn) fail to disrupt the cross-flow behavioral correlations, suggesting the vulnerability is fundamental to the abstraction mismatch rather than an artifact of any specific DP implementation.
% The packet- or flow-level DP fails to disrupt the cross-flow behavioral correlations, confirming that source-level privacy requires protection at the source granularity.

While \sys's AUC scores already show there is privacy leakage (any value above 0.5 represents exploitable information~\cite {chen2020ganleaks}, a data holder should care about), the privacy threat is in practice more prominent. 
We stress that AUC measures average discriminative ability and therefore understates the practical threat. An attacker need not identify every source, confirming even a single source with high confidence can be consequential (e.g., proving a contractual violation or confirming a user visited a sensitive site). The relevant question is not "how well does the attack perform on average" but "how many sources can be identified with near-certainty." We address this next.

\myitem{By tuning the binomial test confidence threshold, \sys allows an attacker to trade source coverage for precision, identifying a smaller subset of sources with near-certainty.}
As described in \S\ref{subsec:MI}, \sys's binomial test significance level controls the confidence of each membership prediction. 
By tightening this threshold, \sys commits only to predictions with strong statistical evidence, trading coverage for precision.
Fig.~\ref{fig:prec_cov} (and Fig.~\ref{apdx:fig:prec_cov} in Appendix in \S\ref{apdx:sec:prec_cov}) show precision as a function of source coverage as the confidence level increases from 10\% to 99.9999\%. The x-axis is reversed; moving right indicates higher confidence and lower coverage.
% We also mark circle, diamands and squares for confidence at 90\%, 99\% and 99.99\%.

As confidence increases, \sys covers (\ie attempts membership inference on) fewer sources but achieves higher precision. For example, the Precision increases from 55\% at conf. 90\% to 80\% at conf. 99.99\%, 2X improvement over random guess for CAIDA dataset. 
This demonstrates that \sys can identify some sources with certainty, raising more serious privacy concerns.
At this level of precision, it is enough for an organization to substantiate a contractual violation claim against the network operator.
% At 99\% confidence, \sys identifies N sources in the CAIDA with zero false positives out of M\todo{ADD} predictions; enough for an organization to say substantiate a contractual violation claim against the Japanese Internet Service Provider\todo{find name?}.

While coverage at high confidence levels is naturally smaller, the absolute count remains operationally significant. For instance, in the CAIDA dataset, \sys identifies 76\% sources at 99\% confidence. In the surveillance scenario, confirming even one user's access to a sensitive website constitutes a meaningful privacy violation; in the cloud-tenant scenario, a single confirmed source suffices to establish a contractual breach.

Meanwhile, the shape of the precision-coverage curve is dataset and \generator dependent. For example, CAIDA and DC show steep increases in precision as coverage drops, indicating that some sources are much more vulnerable (recognizable) than others. MAWI and BFP are flatter, describing a more uniform vulnerability across sources. WFP shows \generator dependent behavior, indicating the vulnerability is sensitive to a specific \generator.

\begin{figure}[t]
\centering
\begin{subfigure}{.99\linewidth}
    \centering
    \includegraphics[width=.99\linewidth]{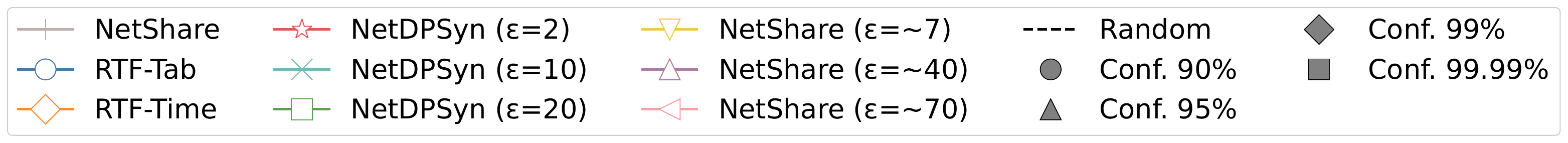}  
    \label{}
\end{subfigure}

\begin{subfigure}{.49\linewidth}
    \centering
    \includegraphics[width=.99\linewidth]{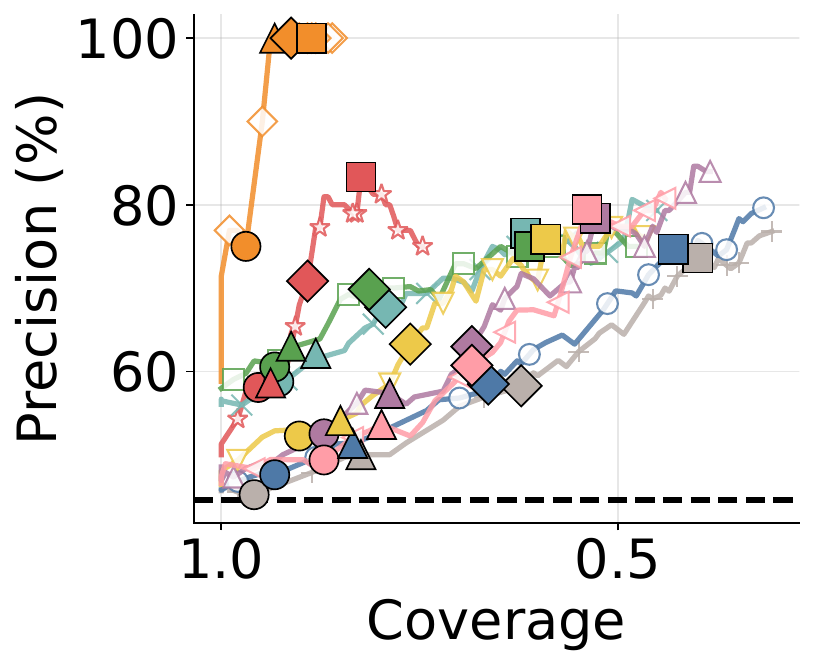}  
    \caption{CAIDA}
    \label{}
\end{subfigure}
\begin{subfigure}{.49\linewidth}
    \centering
    \includegraphics[width=.99\linewidth]{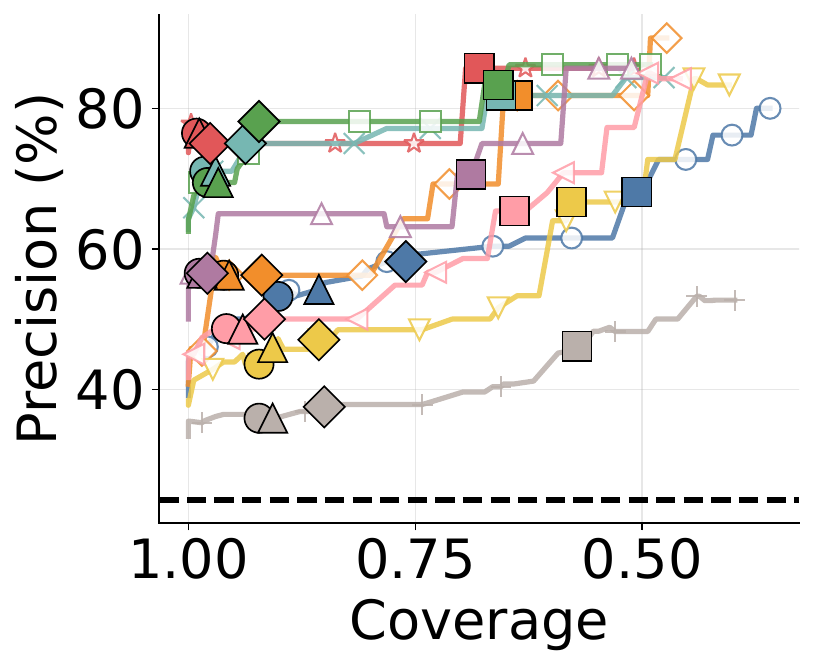}  
    \caption{DC}
    \label{}
\end{subfigure}

\caption{Precision and source coverage at different confidence levels. When setting a higher confidence level (x-axis moving from left to right), \sys infers the membership of fewer sources but with 75\% higher precision over random guess. \sys true privacy leakage is hence more worrying than what AUC scores alone imply.
%This demonstrates that \sys is a targeted attacker, identifying high-confidence sources with higher or even 100\% Precision.
}
\label{fig:prec_cov}
\vspace{-0.5cm}
\end{figure}

%By focusing only at sources for which she is very confident 
%\sys allows the attacker to trade coverage for precision by raising the confidence threshold. Next, we describe the associated experiments and findings.
%, 
%Indeed, as we demonstrate in Fig. 9. At 99.99\% confidence, TraceBleed achieves up to 100% precision on the sources it commits to — meaning every accusation it makes is correct. A data holder cannot dismiss this by observing that average AUC is "only" 0.65, because the targeted sources face certain exposure.

%We further investigate the privacy-fidelity trade-off when DP is used during training. 

\begin{figure}[t]
\centering
\begin{subfigure}{.99\linewidth}
    \centering
    \includegraphics[width=.99\linewidth]{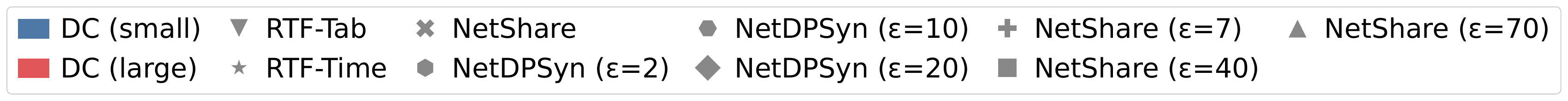}  
    \label{}
\end{subfigure}

\begin{subfigure}{.49\linewidth}
    \centering
    \includegraphics[width=.99\linewidth]{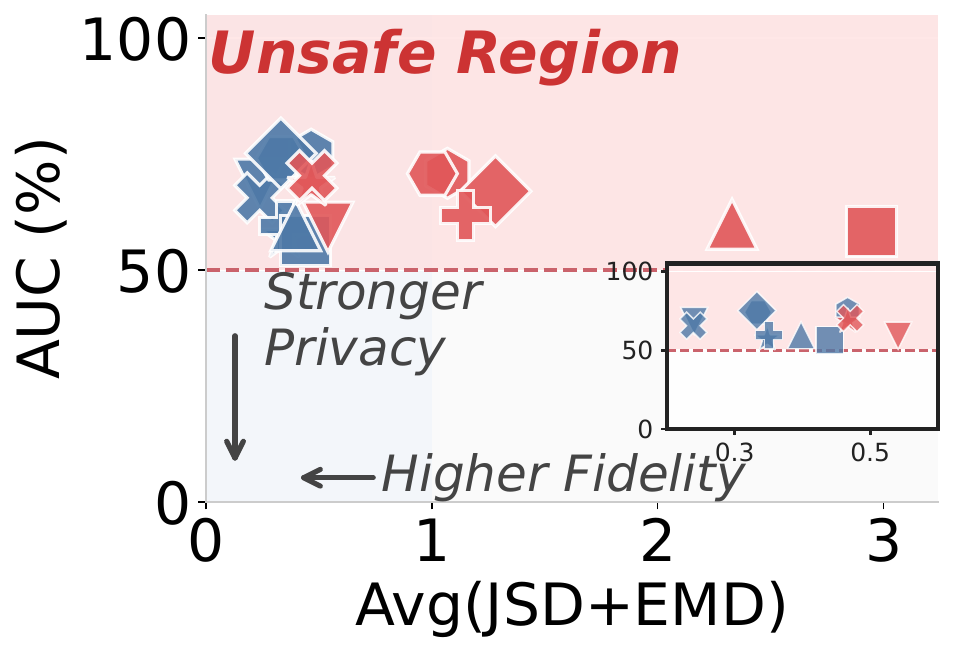}  
    \caption{AUC}
    \label{}
\end{subfigure}
\begin{subfigure}{.49\linewidth}
    \centering
    \includegraphics[width=.99\linewidth]{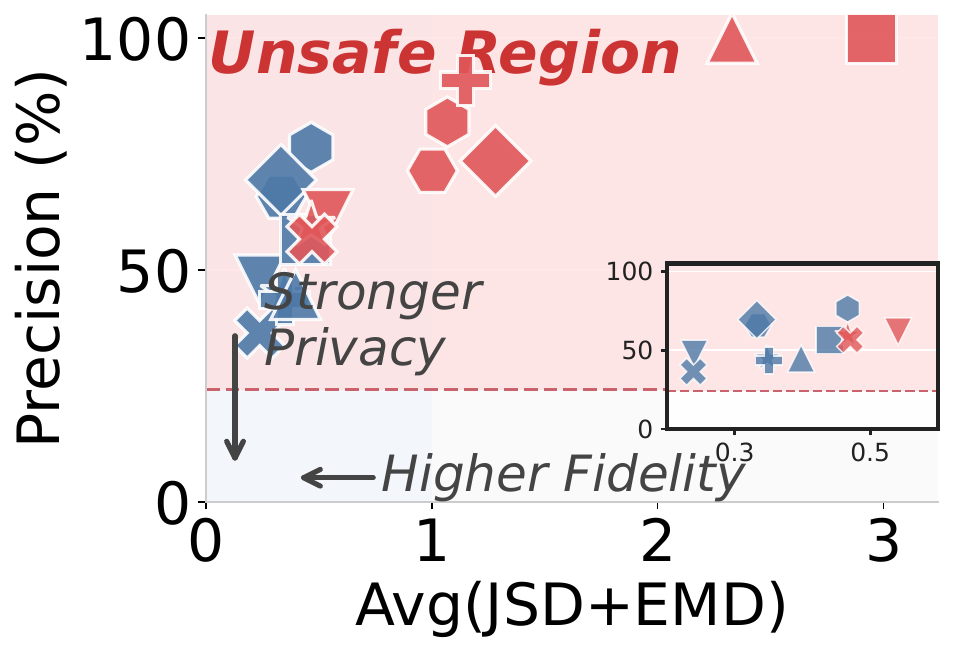}  
    \caption{Precision}
    \label{}
\end{subfigure}

\caption{\generators training with 10X more data fail to improve data privacy while suffering from an average of 2X fidelity degradation.}
\label{fig:more_users_dc}
\end{figure}

\myitem{\generators training on more data (sources) degrades fidelity by 2X on average, with variable privacy gains.}
We investigate whether increasing training data, specifically increasing the number of sources, improves privacy of \generators. Intuitively, this should hold as the \generator's tendency to memorize sources should lessen with their number. To investigate this hypothesis, we aggressively increase the size of $D$ for the DC dataset called DC-large by 10X more packets to train \generators. 
We keep $R$ the same to ensure the same capability of \sys for fair comparison.
Surprisingly, Fig.~\ref{fig:more_users_dc} highlights that when using a larger dataset, \generators cannot improve their privacy but their fidelity drops by 2X. All the \generators suffer fidelity degradation with its degree being  \generator-dependent. Worse yet, none of the \generators is private. \generator such as RTF-Tab and NetDPSyn (eps=20) still leaks privacy, albeit leaving the attacker with lower AUC. \generators like NetShare ($\epsilon=40/70$) preserve the same level of AUC.
Moreover, the enlarging of the training data can cause most of the \generators (\eg NetShare ($\epsilon=40/70$))  a higher Precision.
Therefore, we ask for careful control of the volume of training data and the corresponding number of sources to balance data fidelity and privacy when training \generators.

\myitem{Sharing 10x more data increases the fraction of exposed sources by 130\%. DP-protected \generators are more vulnerable to extended data releases.}
We evaluate how privacy leakage evolves with the amount of synthetic data shared. 
Fig.~\ref{fig:more_data} (and Fig.~\ref{apdx:fig:more_data} in Appendix~\S\ref{apdx:sec:more_data}) show the fraction of vulnerable sources ( \texttt{IN} sources confidently identified by \sys with confidence $\geq95\%$) when \generators generating data 1X(4X, 7X, 10X) the size of original $D$. 
Some data points are missing because, in certain cases, the synthetic data do not contain enough packets from the same source to form valid source chunks. These instances reflect poor fidelity.

Our results indicate that privacy leakage grows with the amount of data shared. 
Specifically, the fraction of exposed sources increases by 130\% when the synthetic data size increases from 1X to 10X. Notably, this effect is even more pronounced for DP-protected \generators, which rises by 158\%, a rate 2X faster than that of the non-DP models.  
These findings underscore the importance of carefully controlling the volume of synthetic data released and caution against the idea of sharing the \generator itself instead of the synthetic output. 

\begin{figure}[t]
\centering
\begin{subfigure}{.99\linewidth}
    \centering
    \includegraphics[width=.99\linewidth]{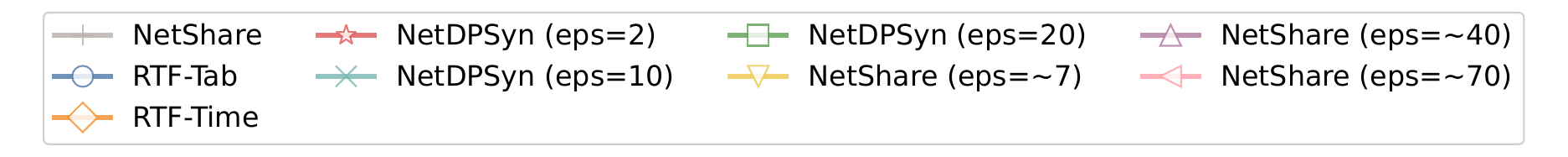}  
    \label{}
\end{subfigure}

\begin{subfigure}{.325\linewidth}
    \centering
    \includegraphics[width=.95\linewidth]{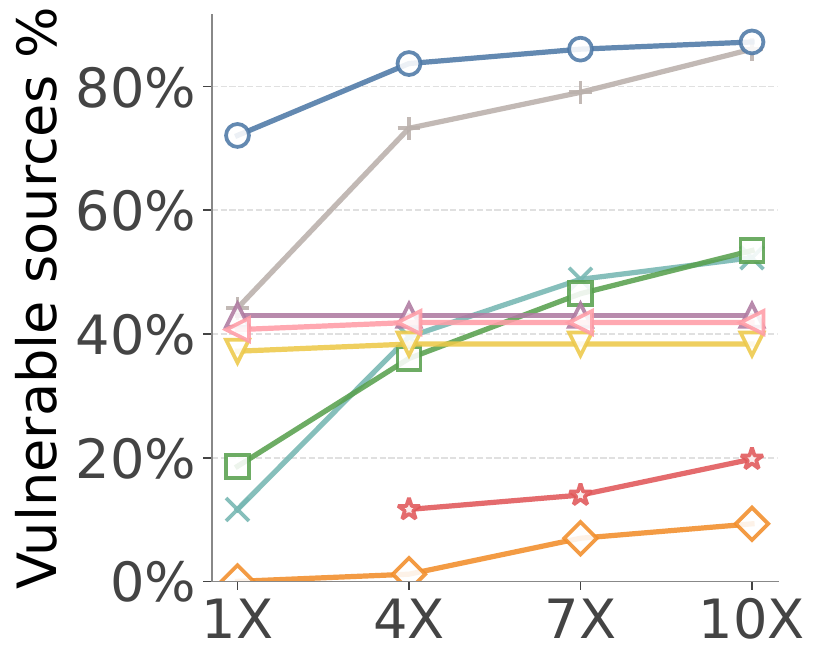}  
    \caption{CAIDA}
    \label{}
\end{subfigure}
\begin{subfigure}{.325\linewidth}
    \centering
    \includegraphics[width=.95\linewidth]{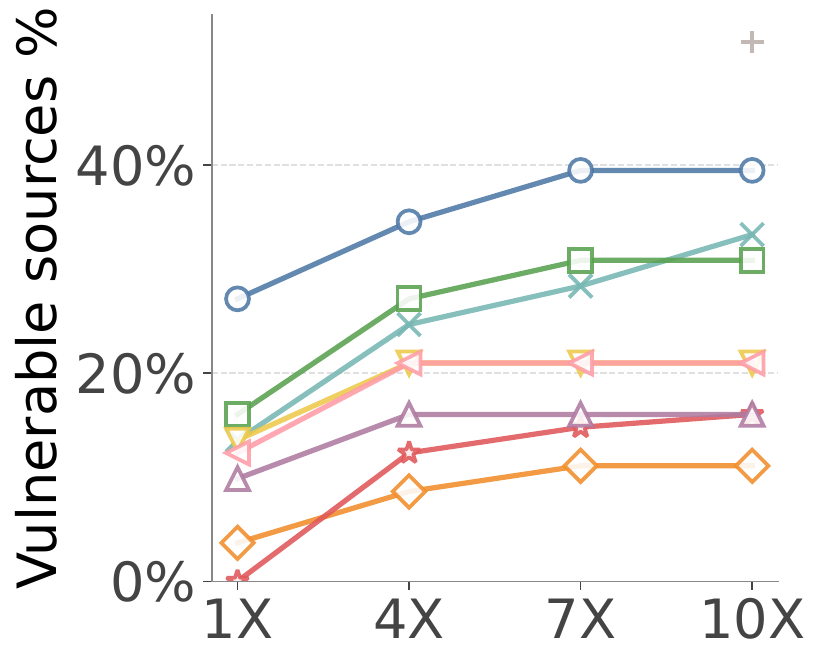}  
    \caption{DC}
    \label{}
\end{subfigure}
\begin{subfigure}{.325\linewidth}
    \centering
    \includegraphics[width=.95\linewidth]{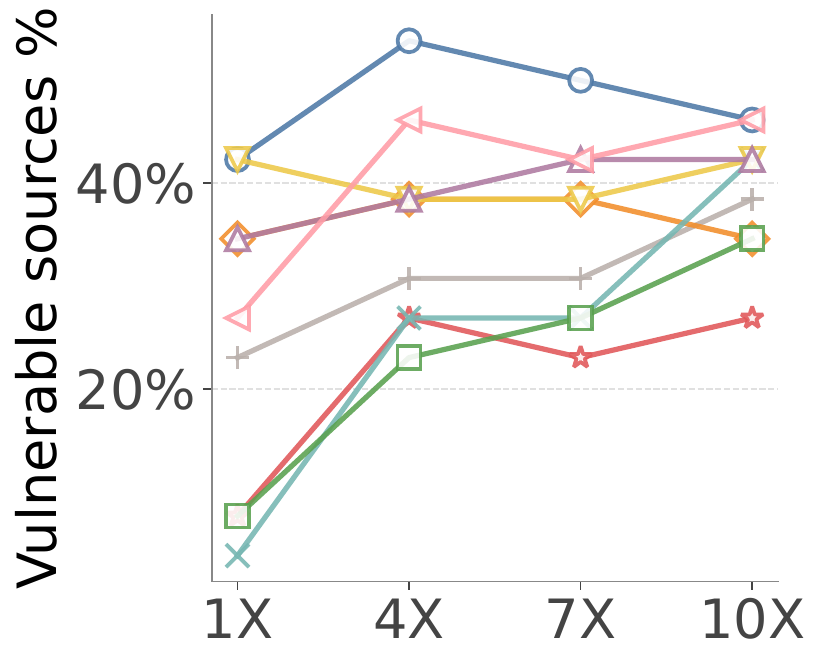}  
    \caption{BFP}
    \label{}
\end{subfigure}
\caption{%The F1 score increases by \change{59\%} with 10× more data, showing that data holders should limit synthetic data volume 
While prior work suggested sharing the generator is as a safe alternative to synthetic data, we find that increasing the amount of shared synthetic data (x-axis) substantially increase the number of vulnerable sources , with a 130\% increase at 10× the original dataset size.
}
\vspace{-0.4cm}
\label{fig:more_data}
\end{figure}

\subsection{Effectiveness of \sys}
\label{sec:sys_effectiveness}
\myitem{\sys can achieve 41\% more effective in source-level MIA on average than baselines. Its effectiveness is not sensitive to the number of sources in $R$ as well as their groundtruth \#\texttt{IN}-\#\texttt{OUT} ratio.}
\begin{figure}[t]
\centering
\begin{subfigure}{.75\linewidth}
    \centering
    \includegraphics[width=.99\linewidth]{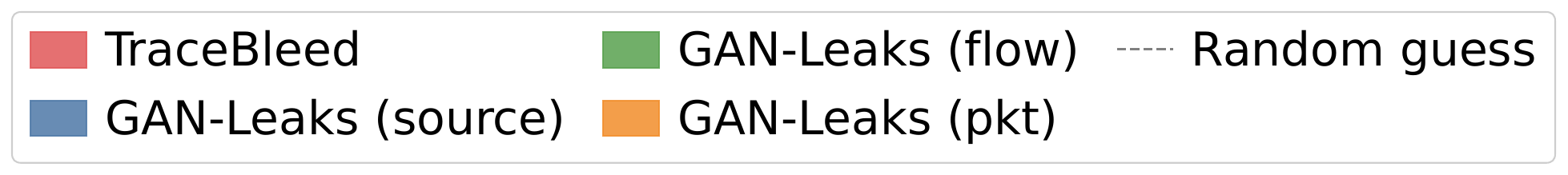}  
    \label{}
\end{subfigure}

\begin{subfigure}{.49\linewidth}
    \centering
    \includegraphics[width=.99\linewidth]{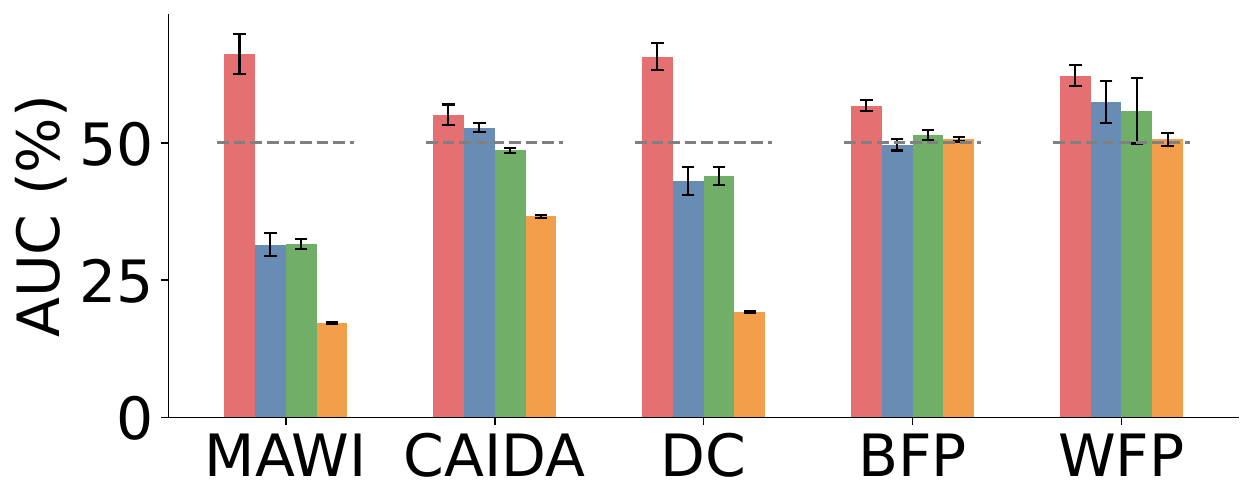}  
    \caption{AUC}
    \label{}
\end{subfigure}
\begin{subfigure}{.49\linewidth}
    \centering
    \includegraphics[width=.99\linewidth]{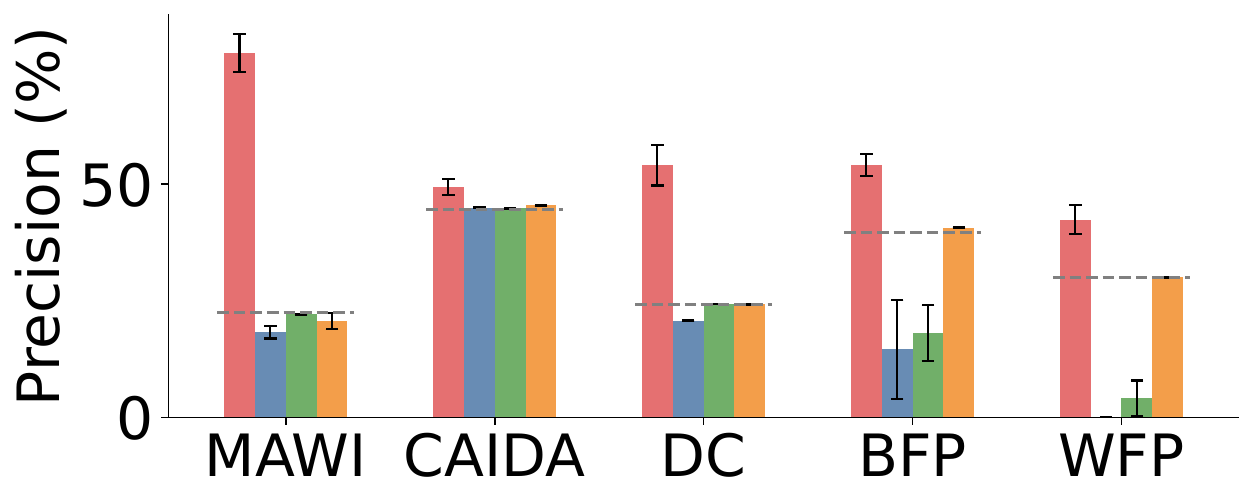}  
    \caption{Precision}
    \label{}
\end{subfigure}

\caption{\sys more reliably exposes privacy leakage, achieving 41\% higher AUC and Precision than GAN-Leaks (the most reliable baseline) across all five datasets.} 
\label{fig:baseline}
\end{figure}
We evaluate \sys's and GAN-Leaks variants' membership performance against \generators. GAN-Leaks is the canonical black-box MIA which determines membership based on the intuition that a sample in the training set should be closer to its nearest neighbor in the synthetic pool than a sample that is not.  Our goal here is to understand which attack would be more reliable in exposing privacy leakage.
Fig.~\ref{fig:baseline} reports AUC and Precision averaged across \generators for each dataset.
\sys outperforms GAN-leaks (source/flow/pkt) by 22\%/23\%/41\% in AUC and 64\%/59\%/38\% in Precision.
\sys exceeds random guess on all five datasets on both metrics, while no GAN-Leaks variant consistently exceeds random guess across all datasets, including GAN-Leaks (source), which directly extracts the per-source statistics-based traffic pattern. 
While \sys is not optimal, this empirical evaluation highlights the benefit of \sys's contrastive learning design for extracting source-specific behavior embeddings rather than nearest-neighbor-like matching in raw or aggregate feature space. 

\begin{figure}[t]
\centering
\begin{subfigure}{.55\linewidth}
    \centering
    \includegraphics[width=.95\linewidth]{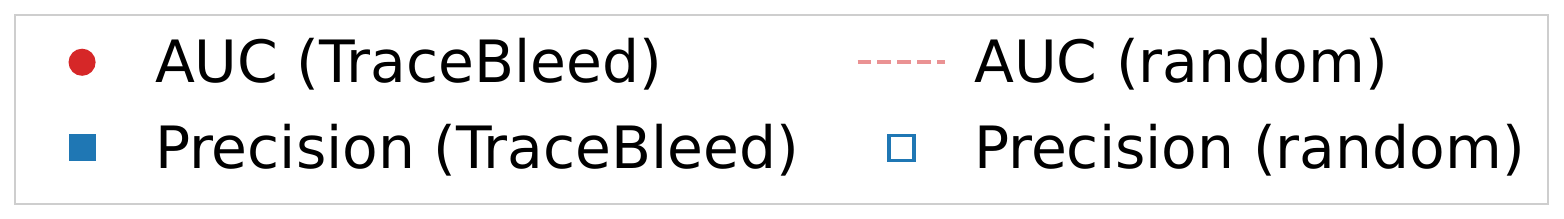}  
\end{subfigure}

\begin{subfigure}{.47\linewidth}
    \centering
    \includegraphics[width=.95\linewidth]{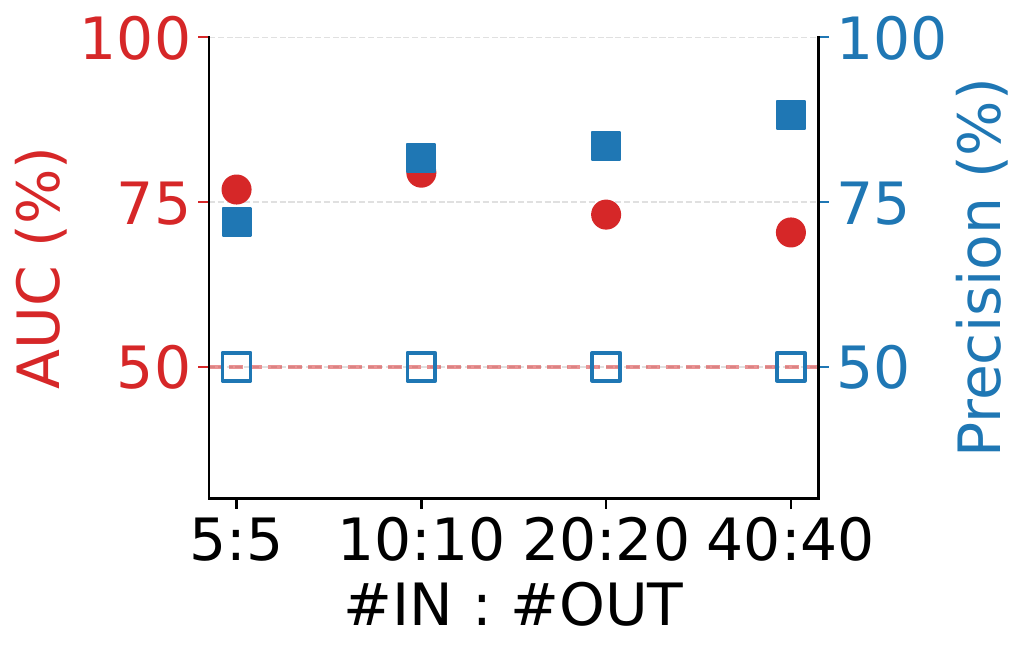}  
    \caption{Few sources}
    \label{fig:mawi_few_sources}
\end{subfigure}
\begin{subfigure}{.47\linewidth}
    \centering
    \includegraphics[width=.95\linewidth]{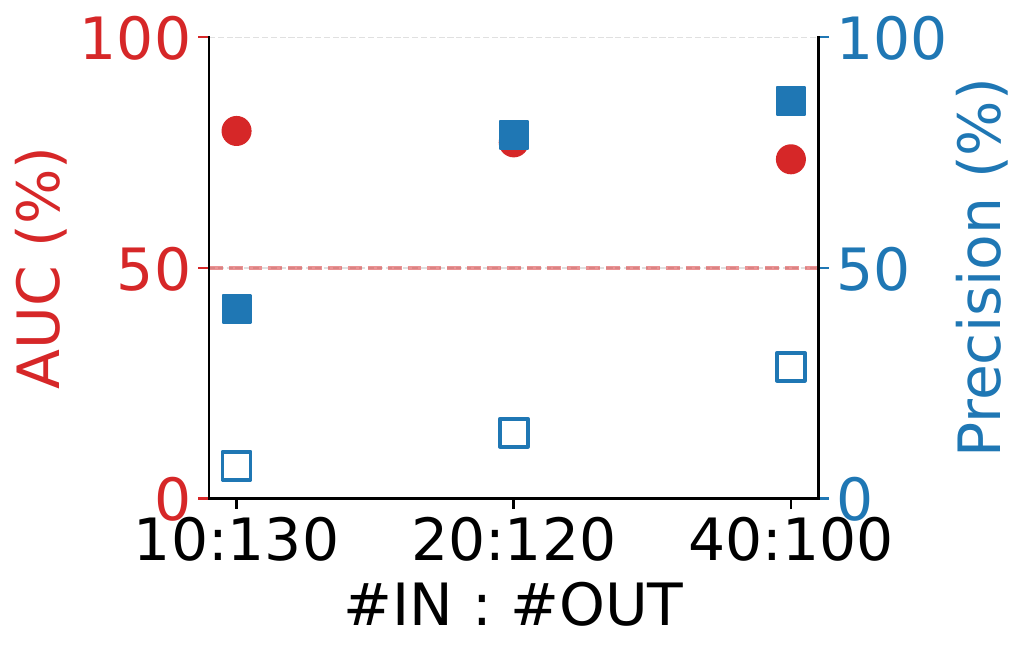}  
    \caption{Unbalanced sources}
    \label{fig:mawi_unbalanced_sources}
\end{subfigure}

\caption{
\sys can achieve 50\% higher AUC and at least 72\% higher Precision than random guess, even when $R$ includes very few or highly unbalanced sources for the MAWI dataset. In contrast, GAN-Leaks baselines cannot outperform random guess even when $R$ has a sufficient number of sources.  
}
\label{fig:mawi_few_and_unbalanced}
\vspace{-0.2cm}
\end{figure}

As in all MIAs, $R$ plays a significant role in \sys effectiveness. Next, we investigate \sys's effectiveness when the collected $R$ includes very few sources or the number of \texttt{IN} and \texttt{OUT} sources is highly unbalanced. 
To do so, we use the MAWI dataset and sample a different number of \texttt{IN} and \texttt{OUT} sources to construct $R$. 
These scenarios correspond to realistic attackers: who have a small $R$ and/or is less likely to have \texttt{IN} sources. 
We do not investigate larger number of sources in $R$ as increasing $R$ increases the attackers' power on capturing more fine-grained source-specific pattern. Meanwhile, if the size is too large for a single \sys model to differentiate sources, attacker can either use multiple ones.
% but also because an attacker could either multiple models \maria{say this better}.

Fig.~\ref{fig:mawi_few_sources} summarizes our results. \sys can achieve 50\% higher AUC and Precision compared to random guess even when $R$ has only 10 sources (\#\texttt{IN}:\#\texttt{OUT} = 5:5). As the number of sources increases from 5:5 to 40:40, the AUC is stable at 0.75, and the precision is increasing from 72\% to 88\%, achieving 22\% relative improvement. 
Meanwhile, \sys is also effective when the \#\texttt{IN} and \#\texttt{OUT} are highly unbalanced. Fig.\ref{fig:mawi_unbalanced_sources} shows that even when the $R$ has 10 \texttt{IN} and 130 \texttt{OUT}, \sys can still achieve 60\% higher AUC and 5X higher Precision than a random guess. When the unbalance changes from 10:130 to 40:100, the Precision increases by 1X, and the AUC remains around 0.75.
Therefore, \sys retains its high effectiveness on membership inference even when $R$ has few or highly unbalanced sources. In contrast, even when $R$ has a sufficient number of sources (Fig.~\ref{fig:baseline}), GAN-Leak variants cannot outperform random guess.

\myitem{Sources with longer and temporally diverse inter-arrival patterns are more likely to leak privacy.}
\begin{figure}[t]
\centering
\begin{subfigure}{.99\linewidth}
    \centering
    \includegraphics[width=.99\linewidth]{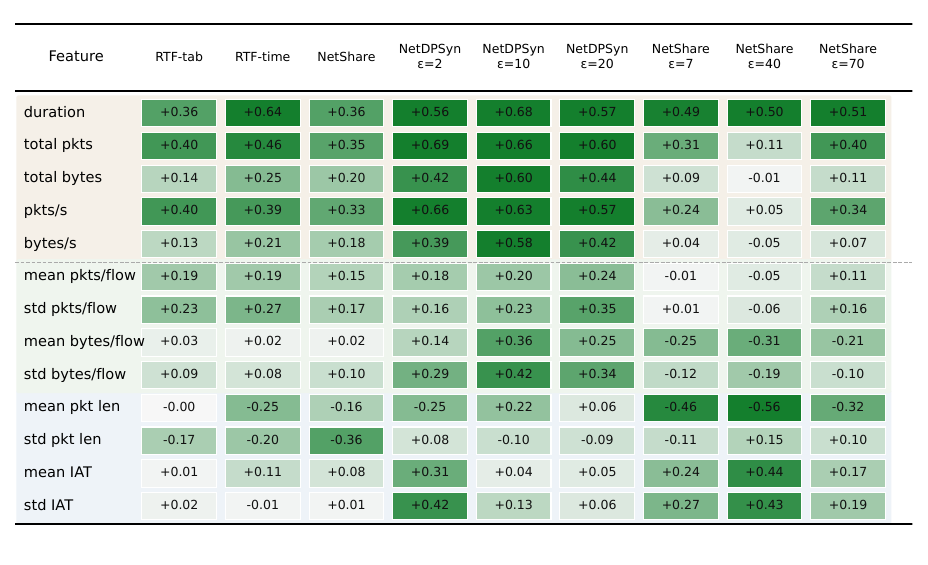}  
    \label{}
\end{subfigure}
\vspace{-3em}
\caption{Sources with longer (high \texttt{duration}, \texttt{total  pkts} and \texttt{pkts/s}) and diverse (high \texttt{std IAT}) traffic behavior are more likely to expose membership to \sys.
For the comparison of each feature, we mark $HV > LV$ as green and $HV < LV$ as red.}
\label{fig:feature_comparison}
\end{figure}
To better understand what makes some sources more vulnerable compared to the rest, we further analyze \texttt{IN} sources of each dataset.
We partition them into three groups. We label them as HV (high vulnerability) if \sys correctly identifies with confidence $\ge 95\%$; LV(low vulnerability) if \sys predicts them incorrectly, and we exclude the remainder from this analysis.
To investigate what is special in these sources we extract some features such as \texttt{duration}, packets per second (\texttt{pkts/sec}), bytes per second (\texttt{bytes/sec}), flow-level features such as the number of packets/bytes per flow (\texttt{pkts/flow}, \texttt{pkts/flow}), and packet level statistics statistics such as interarrival time (\texttt{IAT}) and packet length (\texttt{pkt len}).

We evaluate these features on HV and LV sources across all \generators. Fig.~\ref{fig:feature_comparison} describes the feature value comparison for HV and LV sources and reports the signed Cohen's d for each feature-generator pair. Signed Cohen's d shows the separability of HV and LV given the feature. Concretely, $d = (\text{mean}\text{HV} - \text{mean}\text{LV}) / \sigma_\text{pooled}$ where $\sigma_\text{pooled}$ is the pooled standard deviation for the two groups HV and LV~\cite{Colend}. A positive $d$ indicates $meanHV > meanLV$; a negative indicates $meanHV < meanLV$. The larger $|d|$, the more separable the feature. 
% \maria{The colors need to be for the absolute values, then not?}

Our results show that HV sources have longer duration, more total packets, and higher pkts/s and bytes/s, suggesting that high-volume ("elephant") sources are more susceptible to membership inference. HV sources also exhibit significantly higher std IAT for some \generators, indicating more temporally diverse traffic patterns that serve as richer fingerprints for \sys.
We note that HV sources have more chunks, which increases the binomial test's statistical power. 
However, chunk count alone does not explain vulnerability. A source with many chunks but uniform, featureless traffic would remain indistinguishable from \texttt{OUT} sources in embedding space. Indeed, it is behavioral properties such as high std IAT and longer duration that create the distinctive embeddings \sys exploits.

\remove{
\myitem{Source-level leakage carries semantic content: \sys can infer which websites a user visited from synthetic traces.}
We have so far shown that \generators leak source-level membership which is though a binary signal. A natural follow-up question is: does this leakage carry semantic content? Specifically, can an attacker go beyond confirming a source's presence and infer what that source was doing? We show that \sys learned representations are rich enough to enable exactly this, turning source-level membership leakage into user activity inference.

To answer this question we need to ... and we report prediction error rather than AUC because the task is multi-class.

Fig.~\ref{fig:adf} summarizes our result. We find that \sys can find the websites a given source is visiting 126\% more accurately compared to a random guess, on average across \generators.  \maria{Does this include the error of both finding the user and finding the websites?}
Although not definitive, our results indicate that \generators could reveal even more sensitive information, motivating deeper investigation and the development of refined privacy metrics.}
\section{Evaluation: Defense}

In this section, we investigate the effectiveness of a wide range of canonical defenses in avoiding \generators leaking source-level privacy.
We show that none of the off-the-shelf approaches can protect privacy while preserving fidelity. However, our fine-grained analysis reveals that different defenses destroy different feature distributions, suggesting that task-specific obfuscation is a more promising path than uniform perturbation.

%While we show that none of the off-the-shelf approaches can protect privacy while preserving fidelity of the synthetic traces, we identify an opportunity for task-specific approaches. We first summarize the evaluated defenses before we explain our results. 

 \subsection{Methodology:} 
 We consider both pre- and post-generation defenses. We limit the source-level DP evaluation to NetShare because it is the only evaluated SynNetGen with native DP support, making it the most natural candidate for this extension and our conclusion less implementation-dependent. More importantly, the result already establishes the fundamental tension: source-level DP must model the joint distribution of all flows from a source simultaneously, a much higher-dimensional object than individual flows or packets. This dimensionality increase is inherent to any generator, not specific to NetShare's architecture, so we expect similar or worse fidelity degradation from any \generators operating at source granularity.
For post-generation defenses, we implement eight obfuscation methods inspired by the website fingerprinting defense literature~\cite {rbp,juarez2016toward} and evaluate their privacy gain and fidelity degradation against all 5 datasets. These can be broadly split into packet-level and distribution-aware. 
\textbf{Fixed Padding (FP)} pads all packets to 1500 bytes; \textbf{Random Padding (RP)} adds a random number of bytes to each packet up to 1500 bytes; \textbf{Timing Obfuscation (TO)} advances or delays each packet's arrival time by a random offset; \textbf{FP+TO} and \textbf{RP+TO} combine the aforementioned. 
 \textbf{Traffic Morphing (Morph)}~\cite{Wright2009TrafficMA}  pads packets toward the population packet-size distribution, targeting to reduce the inter-source distinctiveness on packet size distribution;  \textbf{Burst Shaping (Burst)}~\cite{rbp} enforces a maximum packet rate per flow equal to the population median, smoothening the pkts/s and duration features and \textbf{Adaptive Padding (Adaptive)}~\cite{juarez2016toward} inserts dummy packets during idle periods exceeding a threshold, reducing IAT variance. 
We do not evaluate against traditional anonymization techniques, which preserve source affinity through one-to-one deterministic mappings, hence cannot protect against \sys or similar source-level MIAs.
Indeed, procedures such as CryptoPAn~\cite{cryptopan} and tcpmkpub~\cite{tcpmkpub} focus on value transformation from the sensitive fields(\eg IP) without altering the traffic pattern which \sys relies on (\ie interarrival times and packet lengths).

\begin{figure}[t]
\centering
\begin{subfigure}{.6\linewidth}
    \centering
    \includegraphics[width=.9\linewidth]{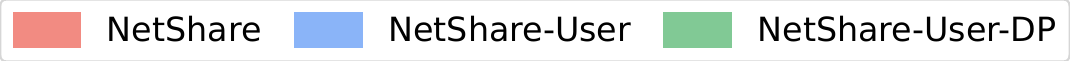}  
\end{subfigure}
\begin{subfigure}{.49\linewidth}
    \centering
    \includegraphics[width=.95\linewidth]{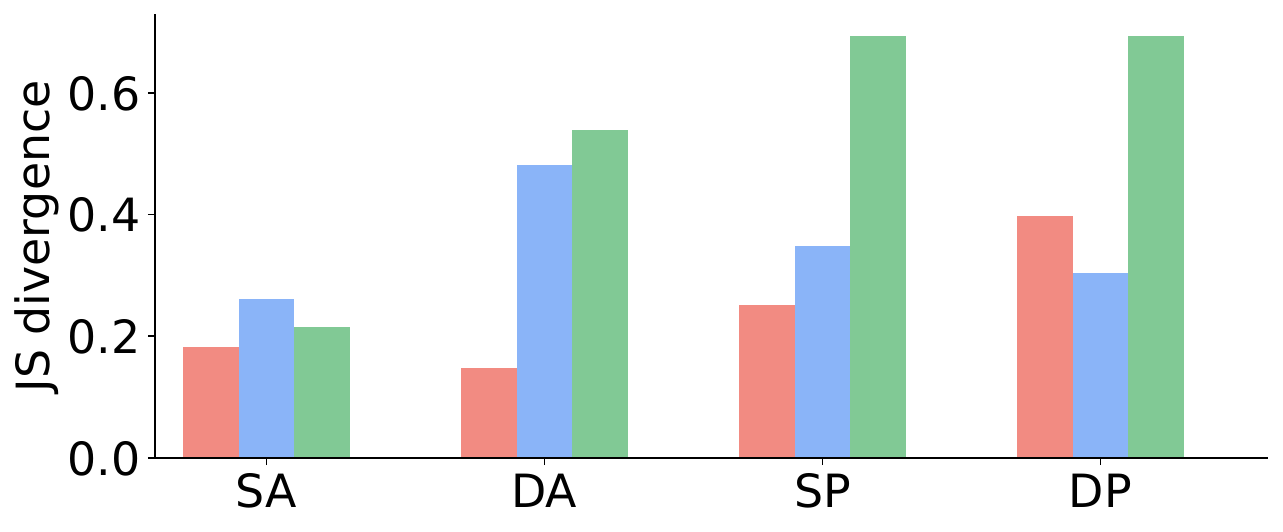}  
    \caption{JSD}
    \label{}
\end{subfigure}
\begin{subfigure}{.49\linewidth}
    \centering
    \includegraphics[width=.95\linewidth]{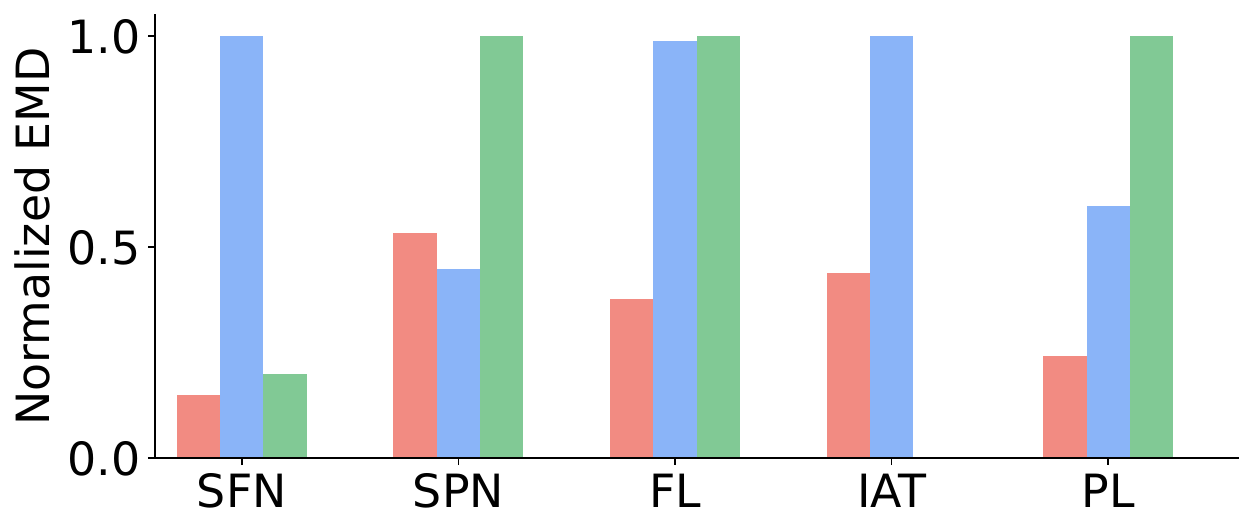}  
    \caption{EMD}
    \label{}
\end{subfigure}
\caption{
% Fidelity metrics of NetShare generating per-flow versus per-source under CAIDA dataset. 
The fidelity of synthetic traces drops by 81\% even when modifying NetShare to work at per-source granularity. We omit NetShare-DP (source, eps=7) from this graph because it fails to generate flows with more than one packet.}
\label{fig:netshare_user_dp}
\end{figure}

\myitem{Source-level DP significantly degrades data fidelity.}
Source-level DP is guaranteed to not leak source-level privacy, hence we only evaluate fidelity. 
Fig.~\ref{fig:netshare_user_dp} shows the fidelity metrics of three versions of NetShare: NetShare (\ie the vanilla flow-level non-protected version) and NetShare-User (which we train by modifying the NetShare encoding from flow to source granularity) and NetShare-User-DP (which we fine-tune the NetShare-User with DP). 
We use Avg(JSD+EMD) as a single summary fidelity score for readability; since JSD is bounded in $[0,1]$ and we normalize EMD to $[0,1]$, the two terms are on comparable scales. We show more detailed results later.
Turning NetShare into source-level granularity (\ie NetShare-User) already causes  81\% worse fidelity than its vanilla counterpart. Notably, NetShare-User fails to preserve the flow inter-arrival time and flow length distributions. 
Table.~\ref{tab:netshare_user_dp} in Appendix in \S\ref{apdx:sec:source-dp} shows the flow-level statistics for  NetShare-User-DP. NetShare-User even fails to generate flows that have more than one packet. 
Such loss of fidelity renders the synthetic data unusable.

\myitem{Post-generation defenses improve privacy only by 2\% for 14\% fidelity degradation.} We evaluate the privacy and fidelity before and after applying the obfuscation approaches to the synthetic data. 
Fig.~\ref{fig:trace_obfuscation} (and Fig.~\ref{apdx:fig:trace_obfuscation} in Appendix in \S\ref{apdx:sec:trace_obfuscation}) summarize our findings.
%Unfortunately, naively using obfuscation only improves privacy by 2\% but suffer from 14\% fidelity drop. \maria{find another number to report not the same as myitem}
Most obfuscation approaches only marginally improve privacy up to 8\%. 
Worse yet, Morph and Burst cause the AUC measured for Privacy to be even higher, highlighting more severe separability between \texttt{IN} and \texttt{OUT} sources. 
The reason can be its design on reducing the distribution difference for packet-size, pkt/s and duration after obfuscating $P(D)$. It  preserves the aggregated cross-flow pattern memorized by \generators and fails to disrupt the separability of \texttt{IN} and \texttt{OUT}.
% make privacy leakage even worse in some cases due to its special design on reducing the distribution difference for packet-size, pkt/s and duration after obfuscation, suggesting that perturbations may inadvertently create new distinguishable pattern which can be exploited by the attacker. \maria{that was weird to me. is the obfuscation applied in P(D) or in R as well?}
This highlights the risk of applying obfuscation methods without understanding their effect on source-level behavioral patterns.
Even for methods that reduce leakage, \eg FP+TO, the fidelity of the synthetic data degrades disproportionately by 29.53\%. 
% \maria{add number diff from myitem}.

\begin{figure}[t]
\centering
\begin{subfigure}{.8\linewidth}
    \centering
    \includegraphics[width=.7\linewidth]{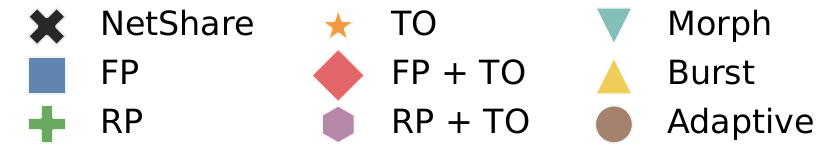}  
\end{subfigure}
\begin{subfigure}{.49\linewidth}
    \centering
    \includegraphics[width=.95\linewidth]{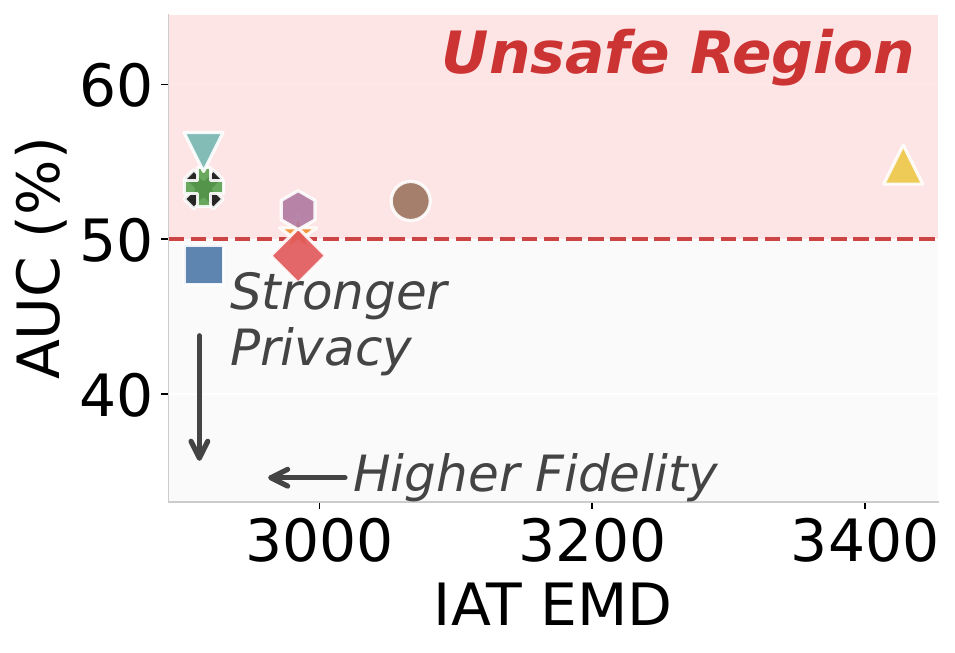}  
    \caption{IAT distribution, AUC}
    \label{}
\end{subfigure}
\begin{subfigure}{.49\linewidth}
    \centering
    \includegraphics[width=.95\linewidth]{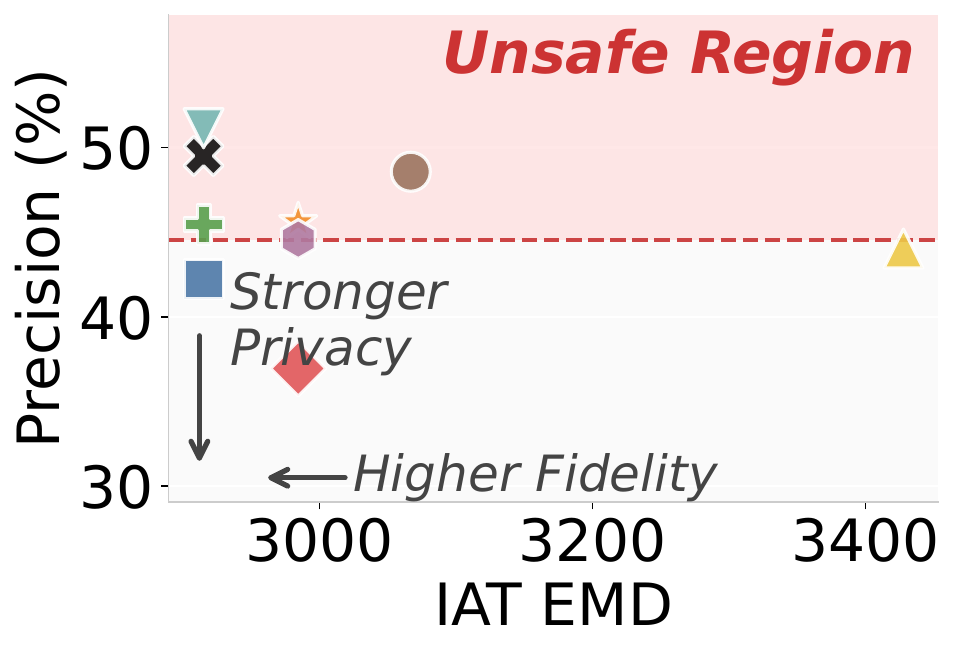}  
    \caption{IAT distribution, Precision}
    \label{}
\end{subfigure}
\begin{subfigure}{.49\linewidth}
    \centering
    \includegraphics[width=.95\linewidth]{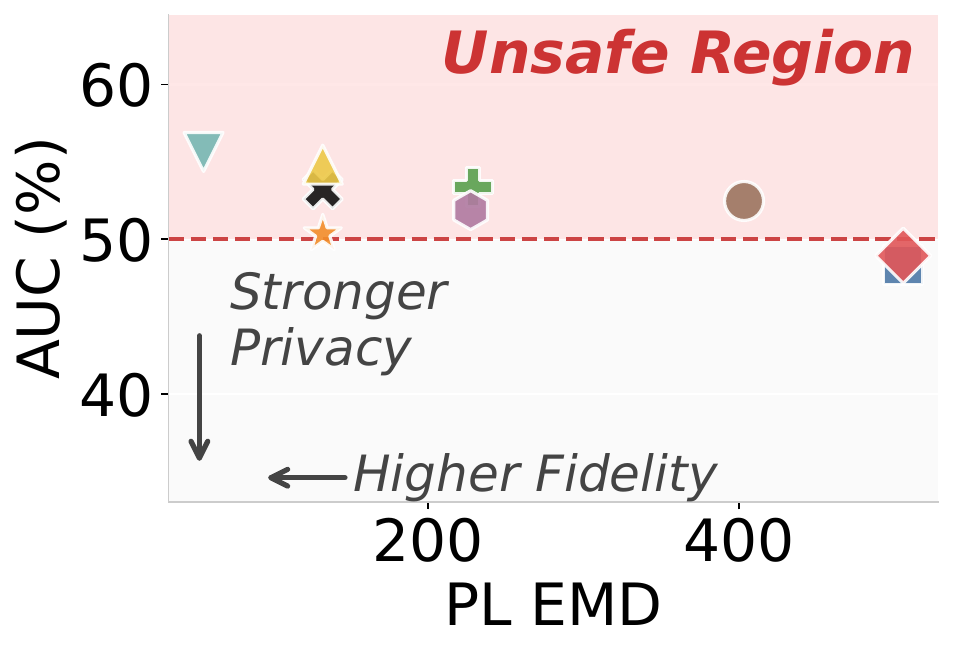}  
    \caption{PL distribution, AUC}
    \label{}
\end{subfigure}
\begin{subfigure}{.49\linewidth}
    \centering
    \includegraphics[width=.95\linewidth]{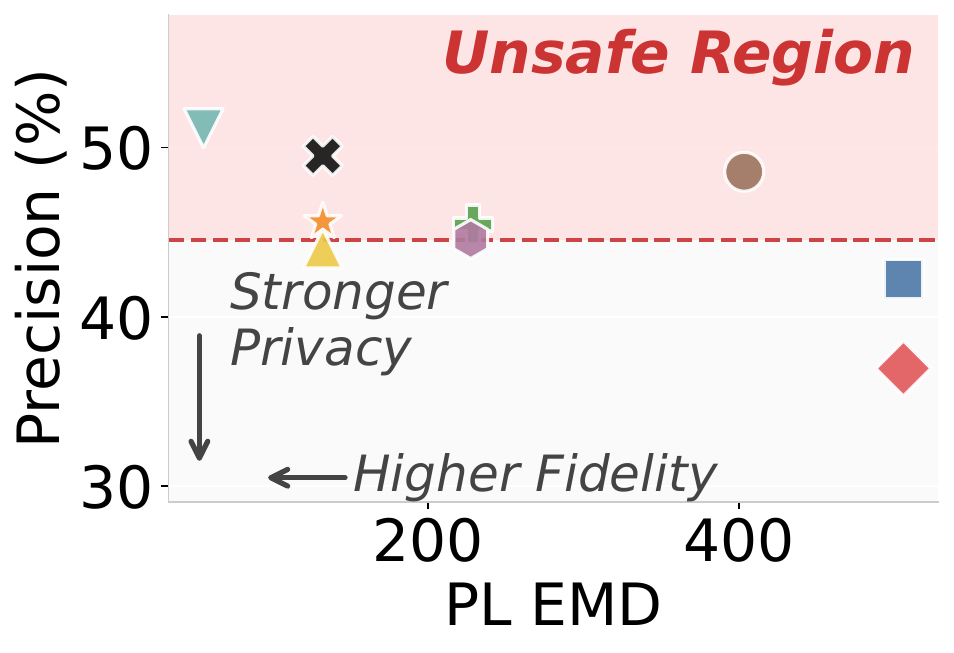}  
    \caption{PL distribution, Precision}
    \label{}
\end{subfigure}
\caption{Obfuscation methods face poor privacy-fidelity trade-off. Burst, FP and FP+TO improve privacy, but at the unbearable expense of fidelity.}
\label{fig:trace_obfuscation_netshare_caida}
% \vspace{-0.5cm}
\end{figure}

\myitem{Post-generation obfuscations affect fidelity in a non-uniform way, revealing an opportunity for task-specific protection.}
We further investigate the fidelity degradation through more fine-grained metrics. %Fig.~\ref{fig:trace_obfuscation_netshare_caida} shows the privacy (measured as AUC and Precision) as well as the data fidelity when obfuscating the synthetic NetShare data for CAIDA. 
%Since the obfuscation approach mainly perturbs the packet inter-arrival time as well as the packet length, we focus on 
Concretely, we measure the packet length (PL) and inter-arrival (IAT) distribution similarity individually, instead of leveraging Avg(JSD+EMD) for a summarized fidelity score. 
Fig.~\ref{fig:trace_obfuscation_netshare_caida} reports that FP, FP+TO and Burst can achieve a score lower than random guess for both AUC and Precision while the others are not private. 
However, FP and FP+TO suffer from 3X worse packet length distribution and Burst suffers from 18\% worse inter-arrival time distribution compared to the vanilla NetShare trace. 
Practitioners must therefore carefully consider which downstream applications their synthetic data will support before applying obfuscation
For example, FP or FP+TO may be acceptable when downstream applications depend only on the temporal pattern of the traffic and not packet sizes.

To validate this intuition concretely, we evaluate utility under anomaly detection using NetML\cite{yang2020comparative} on synthetic vs raw data.
% Our goal is to show that privacy-preserving obfuscation can retain high utility when it does not substantially perturb the features used by the downstream task.
 NetML is a commonly used downstream application for \generators\cite{yin2022netshare} and leverages an one-class SVM to determine the ratio of the traffic which is anomalous. 
 % To evaluate the data utility, for each synthetic or obfuscated dataset, we train the OCSVM on that and evaluate it against a raw CAIDA trace. NetML returns with the anomaly ratio which is the fraction of traffic considered as anomalous. 
NetML supports different modes that extract different flow features. If an obfuscation doesn't modify the features a mode relies on, the anomaly ratio should remain close to the unobfuscated baseline ($\approx 0$ relative error).
Table~\ref{tab:netml} shows the relative error for obfuscation under different NetML modes. We list the distribution each mode primarily relies on, as well as the relative error when selecting the best obfuscation approach (among FP, FP+TO and Burst) compared to a fixed approach as reference (FP + TO). When selecting the obfuscation which has the minimal effect on the relied distribution, such as FP when inter-arrival, the relative error is 0.34\%, 90\% lower than always selecting a fixed reference, \eg FP+TO.

% \maria{can you improve the readability of this? Can we say this is a downstream task used by some generator? also add some number? why do we say outlier and not IDS?}

Our results reveal a consistent pattern: defenses that reduce source-level leakage do so by destroying fidelity, effectively making synthetic data useless, albeit private. Declaring a defense effective therefore requires evaluating both privacy and task-specific fidelity jointly, not in isolation. A practitioner who knows which distributions the downstream application relies on can concentrate perturbations on the remaining features, preserving utility where it matters while disrupting source-identifying correlations elsewhere. %We hope that \sys and the accompanying leaderboard provide the tools for this evaluation as new defenses are proposed.

\begin{figure}[t]
\centering
\begin{subfigure}{.4\linewidth}
    \centering
    \includegraphics[width=.7\linewidth]{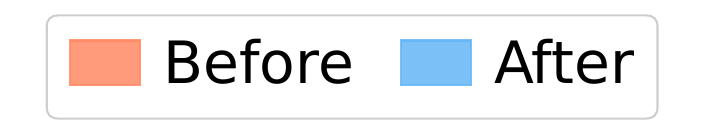}  
\end{subfigure}
\begin{subfigure}{.99\linewidth}
    \centering
    \includegraphics[width=.9\linewidth]{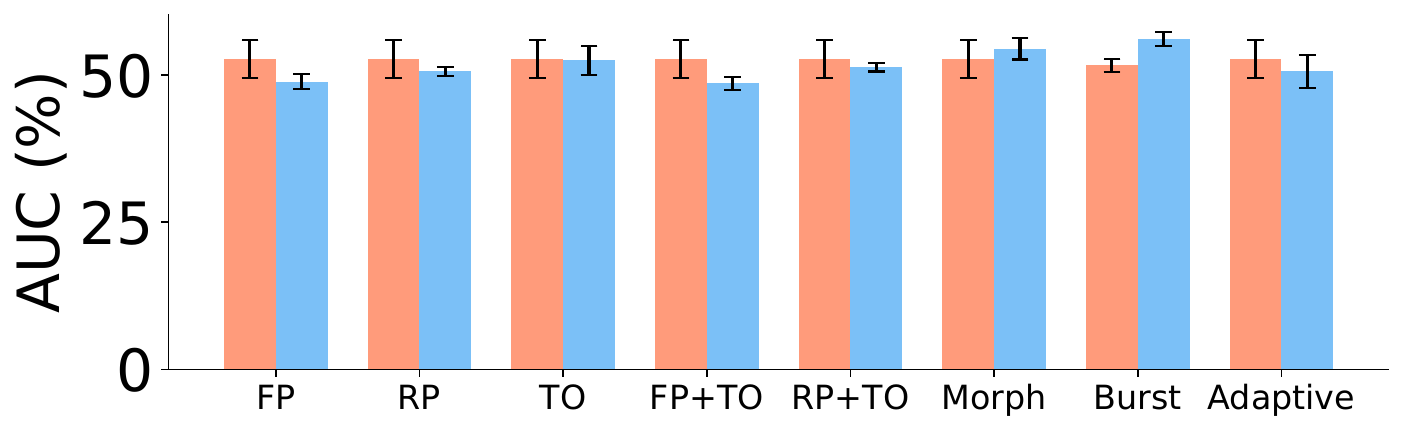}  
    \caption{AUC (Privacy)}
    \label{}
\end{subfigure}
\begin{subfigure}{.99\linewidth}
    \centering
    \includegraphics[width=.9\linewidth]{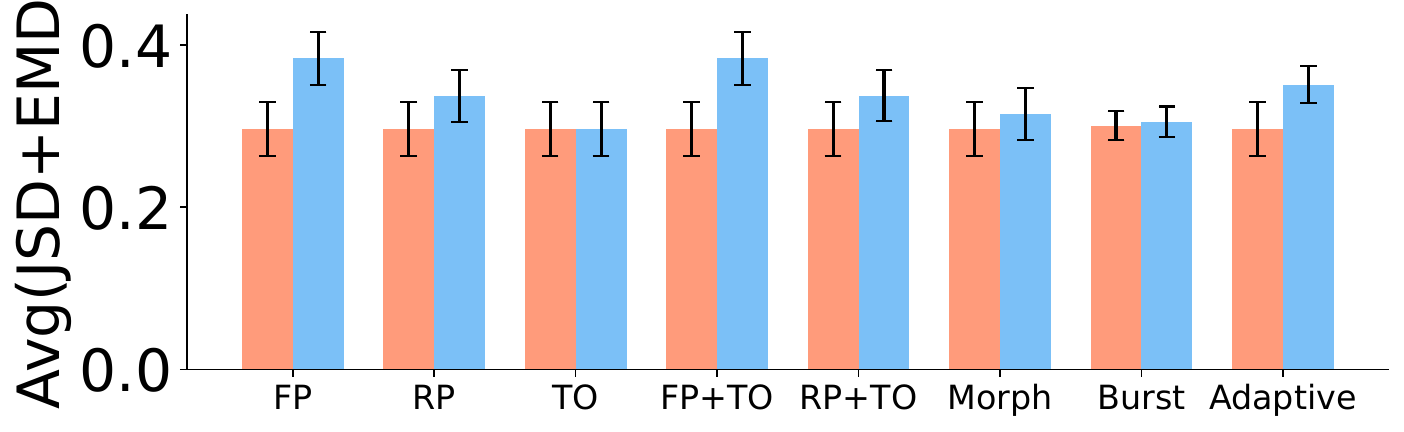}  
    \caption{Avg(JSD+EMD) (Fidelity)}
    \label{}
\end{subfigure}
\caption{Privacy and fidelity change after applying obfuscation for CAIDA dataset. The trace obfuscation approaches have minor privacy improvement while suffering 14\% fidelity drop. Morph and Burst unexpectedly increase privacy leakage in some cases, suggesting that distribution-aware obfuscation can create new distinguishable pattern.}
\label{fig:trace_obfuscation}
% \vspace{-0.3cm}
\end{figure}

\begin{table}[t]
\centering
\begin{tabular}{lrrrr}
\toprule
Mode & Relied Dist. & Best Ob & \makecell{Relative Err. \\ \scriptsize{(Best Ob)}} & \makecell{Relative Err. \\ \scriptsize{(Ref.:FP+TO)}}\\
\midrule
IAT & Inter-arrival & FP & 0.00\%  & 0.63\%\\
STATS & Mixed & All & 0.00\%  & 0.00\%\\
SIZE & Packet length & Burst & 0.32\%  & 4.37\%\\
SAMP\_NUM & Inter-arrival & FP & 0.00\%  & 10.10\%\\
SAMP\_SIZE & Packet length & Burst &  1.37\% & 2.39\%\\
\bottomrule
\end{tabular}
\caption{Selecting the obfuscation approach which doesn't change the distribution the downstream application relies on can achieve 90\% lower relative error compared to selecting a fixed obfuscation.  }
\vspace{-0.3cm}
\label{tab:netml}
\end{table}

\section{Related Works}
Having discussed the connection of \sys to MIA and existing \generators, we discuss attacks that seem relevant.

\remove{
\myitem{MIA against Generative models}
% \maria {Logan}
The MIA against generative models generally falls in two categories. The first category requires auxiliary information from the generative model.
For example, LOGAN~\cite{hayes2017logan} targets GANs by leveraging the prediction confidence of the discriminator--which distinguishes real from generated samples--to infer whether a target data point was in the training set.
Similarly, \cite{carlini2021extracting} demonstrates an MIA against GPT2 model using perplexity, \ie how well the model predicts the tokens in the sequence.

The second category generally assumes a full black-box setup, \ie, attackers cannot get any additional information except the synthetic data. 
These attacks typically exploit the frequency of the generated samples: generators tends to replicate or produce samples closer to the training data more frequently. Notable examples are GAN-leaks~\cite{chen2020ganleaks} against GAN and \cite{carlini2023extracting} against the Diffusion model.
}

%\myitem{Other Privacy metrics}
%\cite{alabdulwahab2024privacy} compares the privacy and utility of synthetic data generators. To test privacy, the paper uses singing-out, likability and inference. For the latter two, the attributes used were source and destination ports, protocol, and mean and standard deviation of packet length. While these are said to expose user identity, this was not shown in practice. 

\myitem{Privacy attacks on real packet traces}
Packet traces enable several types of privacy attacks. Website fingerprinting~\cite{sirinam2018deepfp,websitef,deng2024robust} exploits differences in packet size and timing patterns—revealing visited sites even over Tor or VPN. Similarly, applications and devices exhibit predictable traffic behaviors, such as periodic updates or characteristic packet mixes~\cite{Yan2015EnablingQL,iscxvpn2016,sharma2023vca,utmobile}.
\sys builds on the idea that services have distinct patterns, but its key innovation lies in identifying the unique combination of these patterns to distinguish users over time. 
%Other attacks leverage the fact that the traffic patterns remain correlated from end to end \ie, from the sender to the receiver, and while the flow lasts, but not over time for the same sender or receiver as \sys.
Cross-flow correlations can be used to identify pairs that communicate over Tor. The attacker investigates the correlation between two traffic chunks observed at the Tor entry and exit. Even if seemingly independent, such attacks match flows that correspond to the same stream~\cite{nasr2018deepcorr,lopes2024sumo}, not just to the same source as \sys. 
Active Watermarking (Injection)~\cite{iacovazzi2019duster}, leverages the same insight but also assumes an attacker that can inject some uniqueness on the stream (\eg creating bursts or causing throughput oscillations), hence is beyond the scope of this paper. 

\myitem{Privacy attacks on anonymized packet traces} such as the so-called fingerprinting attack and injection attack~\cite{brekne2005anonymization,brekne2005circumventing,yen2009browser} are effective against anonymized traces (\eg CryptoPAn). In those attacks, adversaries either already know some network flows in the original traces (by observing the network or from other relevant sources, e.g., DNS and WHOIS databases) \cite{burkhart2008risk}, or have deliberately injected some forged flows into such traces. 
Unlike \sys, the attacker in such attacks recognizes unique combinations of the \emph{unchanged fields} of known flows in the anonymized traces, namely, fingerprints (e.g., timestamps, IPs and protocols)~\cite{brekne2005anonymization}.

\section{Conclusion}

%This paper is the first to show that \generators leak training-set participation at the source level even when satisfying packet/flow privacy guarantees.
%We show that this vulnerability is rooted on that cross-flow correlations survive synthesis.
%We also present \sys the first source-level MIA against \generators that demonstrates these correlations can ne exploited under realistic constraints \ie black-box access and a reference dataset with overlap with the training. 
%We also show that the privacy-fidelity tradeoff fundamentally changes at the source level making mitigating source-level MIAs harder. 

This paper is the first to show that \generators leak training-set participation at the source level, even when satisfying packet- or flow-level privacy guarantees. We demonstrate that this leakage arises because cross-flow correlations (\ie fingerprints on the training dataset of a \generator) survive synthesis (\ie are distinguishable in synthetic traces). We then present \sys, the first source-level membership inference attack against \generators, showing that these correlations can be exploited under realistic constraints, including black-box access and a reference dataset with no overlap with the training data. Finally, we show that mitigating source-level leakage fundamentally reshapes the privacy–fidelity tradeoff, making source-level privacy substantially harder to achieve than existing packet- or flow-level notions suggest.

\bibliographystyle{abbrv} 
\newpage
\begin{small}
\bibliography{base}
\end{small}

\newpage

% \section{Ethics}
% % Note from the CFP that this section must include a statement about
% % ethical issues; papers that do not include such a statement may be
% % rejected.

% This paper doesn't raise any ethical issue.

% \section{Appendix}

\section{Details of BFP datasets}
\label{apdx:detailsdatasets}
We simulate users with unique website-access pattern by sampling up to 6 websites from the top 30 websites per user based on a user-specific website preference probability function. Next, we collect traffic from 50 loads of these 30 different websites (\eg google.com, bing.com) and use them as building blocks to emulate users. Concretely, we modify the packets' source IP and timestamp to construct a user's traffic with a unique access pattern to the websites multiple times.

% As mentioned, the SIM dataset is collected at the campus entity and the Multi-VA one is collected at three different entity at different times: a campus entity on 11/22/2024, the CloudLab Wisconsin cluster on 6/19/2025, and the CloudLab Emulab cluster on 7/1/2025, which are geographically far away. 

\section{Details of the features extracted for GAN-Leaks}
\label{apdx:detail-gan-leak}
For GAN-Leaks(flow), we extract 25 features including the statistics of the flow inter-arrival time and packet length as well as bytes/s, packets/s, fraction of small(large) packets, etc.

For GAN-Leaks(source), we extract 79 features such as packet-level statistics for packet length and source inter-arrival time, flow-level statistics such as the statistics of the mean/max of the flow inter-arrival time for all the flows given a source, source-level statistics such as number of flows, bytes/s, packets/s, etc.

\section{Details of \generators}
\label{apdx:detailsgen}

\textbf{NetShare}:~\cite{yin2022netshare} is a GAN-based IP header trace generator that can generate high-fidelity, privacy-preserving synthetic traces. It constitutes an extension of the time-series data generator DoppelGANger and incorporates network-specific domain knowledge to preserve data fidelity. 
In addition, NetShare allows users to optionally enable a privacy-preserving mode that uses Differential Privacy (DP), offering a formal, theoretically grounded privacy guarantee\footnote{The latest NetShare codebase no longer supports DP. We replicate the DP mode based on the original paper's description.}. When DP is on, NetShare first pre-trains on a public dataset, which is disjoint with the original training set. Next, it is fine-tuning on the original set using DP-SGD to provide DP guarantee during model training. In our evaluation for DP-proteced NetShare, we select another dataset which is disjoint with $D$ and $R$ to pretrain. NetShare's generation granularity is per network flow and its use of DP is also defined per flow.

\textbf{NetDiffusion+}: 
 NetDiffusion~\cite{jiang2024netdiffusion} is a Diffusion-based generator that leverages a fine-tuned Stable Diffusion model to generate network traces. During its training phase, NetDiffusion first converts network flows into image-like representations and then uses them to fine-tune an image-data pre-trained diffusion model. Leveraging the powerful generative capabilities of diffusion models in the image domain, NetDiffusion is able to synthesize high-fidelity network traces that closely resemble real traffic. However, NetDiffusion does not incorporate any explicit privacy mechanisms—such as Differential Privacy (DP)—during training or generation.

NetDiffusion is designed with connection-level (bi-directional flow) granularity, which is neither flow or packet-level. 
When extending it to be flow-level or packet-level, each translated image only capture very limited information, making the fidelity even worse.
To address this, we developed NetDiffusion+, a variant that generates traffic per source IP. 
We first group the training data by source IP, resulting in per-user records containing multiple flows. 
We then apply the same training and fine-tuning procedures as specified by the authors.

In NetDiffusion’s generation pipeline, the final step is post-processing, which enforces consistency constraints with network protocols by correcting errors made during generation. 
This process significantly alters the generated packets, modifying fields such as IP addresses, flags, and sequence numbers, and converting unidirectional flows into bidirectional ones.
To maintain single-source, unidirectional flows, we change the post-processing step by disabling the rewriting of source IP addresses and adjusting the rewriting process to sample multiple destination IPs.

For generation, we fail to generate 10X sources within a reasonable timeframe (e.g., 6h). Hence, we provide a hard stop once the generation time achieves the budget, and evaluate the fidelity and privacy of it as reference. Fig.~\ref{fig:fidelity_privacy} reports the fidelity and privacy of NetDiffusion+ as well as other \generators.

\textbf{RealTabFormer (RTF)}:~\cite{solatorio2023realtabformer} is a GPT-2 based synthetic tabular data generator. Although it was originally designed for general purposes, related works~\cite{Yin2024thesis,Yin2024CANShare} show that RTF is applicable to network traces synthesis. RTF offers two modes—RTF-Tab and RTF-Time—allowing data generation at the packet and flow levels, respectively.  Similar to NetDiffusion, its original design doesn't include privacy protection techniques.

\textbf{NetDPSyn}:~\cite{sun2024netdpsyn} is a marginal distribution-based networking trace synthesizer. Unlike the pipelines introduced above, NetDPSyn does not train a generative model to learn the original data's distribution. Instead, it leverages multiple marginal distributions to capture the overall data distribution. NetDPSym integrates Differential Privacy (DP) directly into its framework design, in contrast to NetShare where DP is optional and can be toggled on or off. In NetDPSym, both the data generation and privacy protection operate at the packet level.

\section{Fidelity metrics of \generators}
\label{apdx:fidelity}

We report JS Divergence for the distribution of \textbf{SA} (source IP address), \textbf{DA} (destination IP address), \textbf{SP} (source port), \textbf{DP} (destination port). We omit the distribution of protocol as all the packets are TCP. We also report the normalized Earth Mover's Distance (EMD) for the distribution of \textbf{SFN} (flow per source), \textbf{SPN} (packets per source),  \textbf{PL} (packet length), \textbf{IAT} (packets inter-arrival times), \textbf{FL} (flow length). Since the value of EMD is unbounded, we use min-max normalization to normalize EMD (min is 0 and max is the maximum EMD any \generator achieves). We note that data fidelity varies based on use case. For example, using synthetic networking traces to train a more robust ML-based networking application has different fidelity requirements from those to reproduce real-world behaviors. Discussion of the selection of fidelity metrics is out of the scope of this paper.

\begin{figure*}[htb]
\centering
\begin{subfigure}{.6\linewidth}
    \centering
    \includegraphics[width=.95\linewidth]{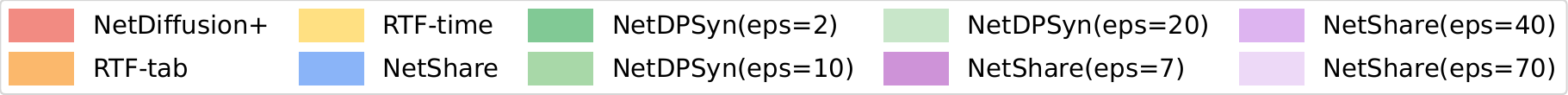}  
    \label{}
\end{subfigure}

\begin{subfigure}{.49\linewidth}
    \centering
    \includegraphics[width=.95\linewidth]{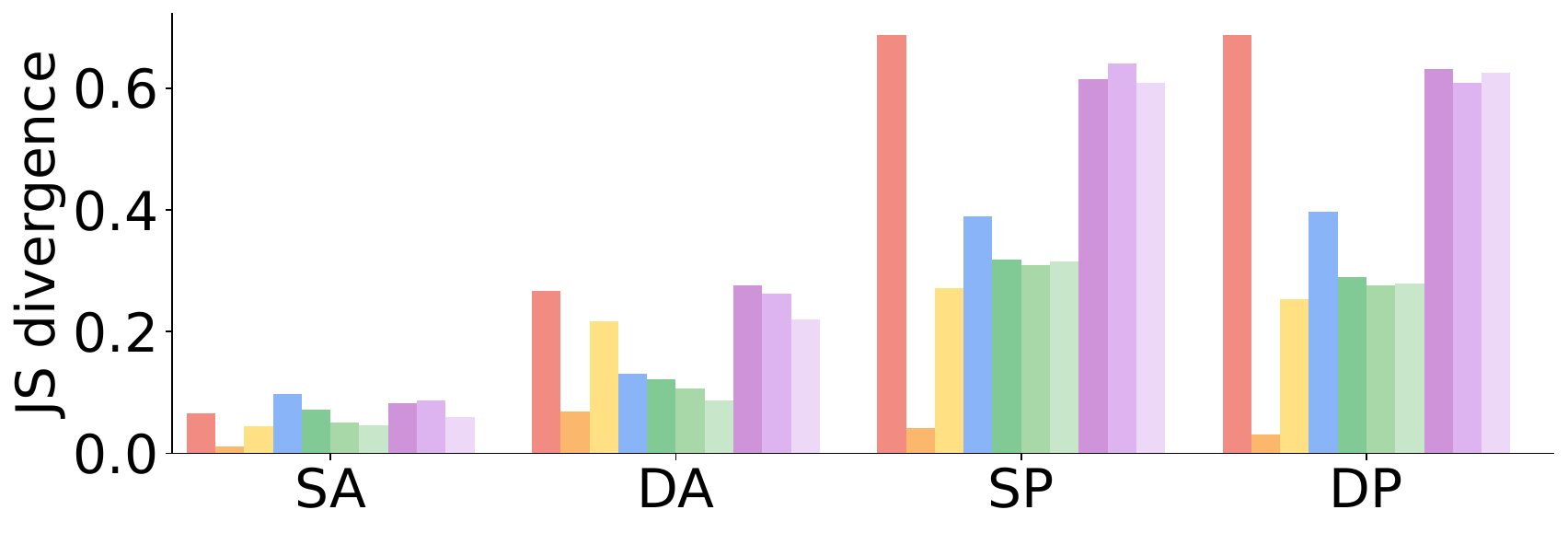}  
    \caption{JSD, MAWI}
    \label{}
\end{subfigure}
\begin{subfigure}{.49\linewidth}
    \centering
    \includegraphics[width=.95\linewidth]{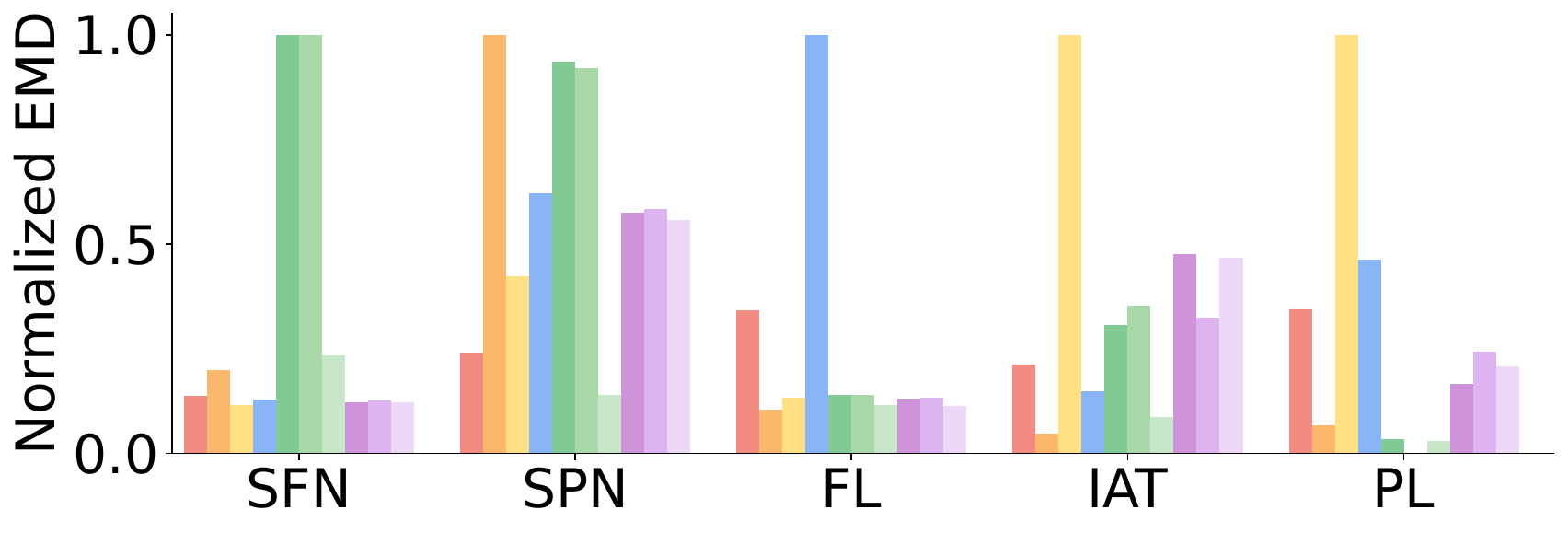}  
    \caption{EMD, MAWI}
    \label{}
\end{subfigure}

\begin{subfigure}{.49\linewidth}
    \centering
    \includegraphics[width=.95\linewidth]{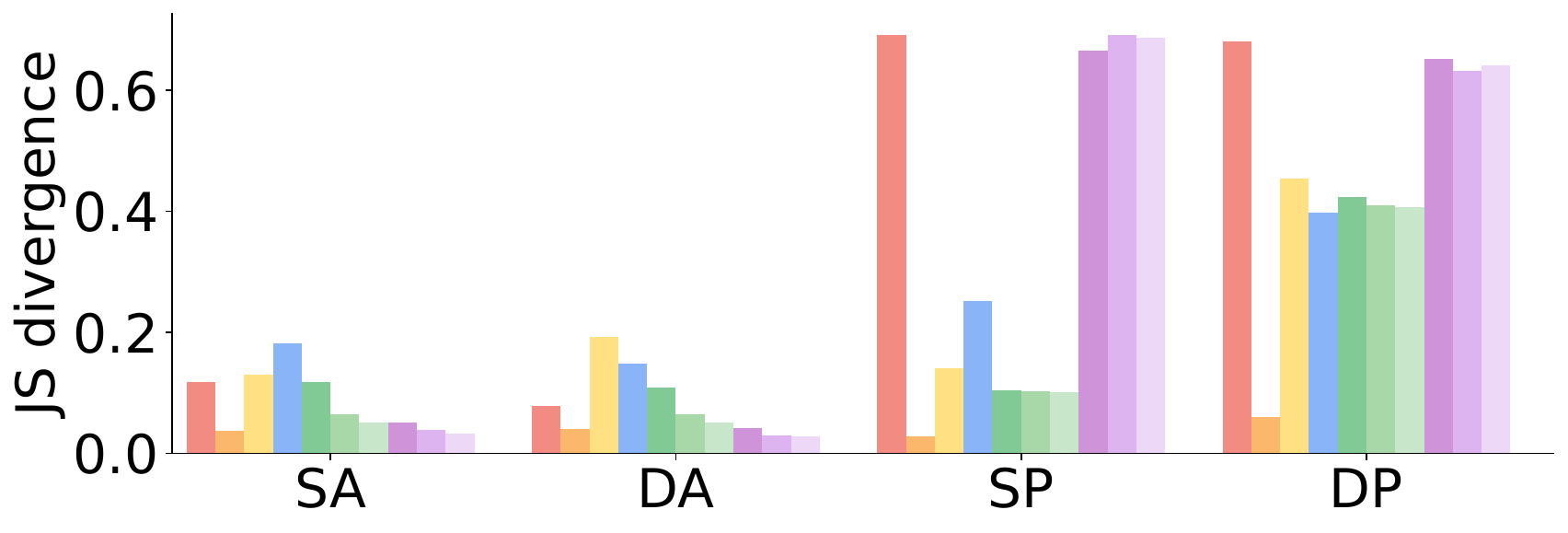}  
    \caption{JSD, CAIDA}
    \label{}
\end{subfigure}
\begin{subfigure}{.49\linewidth}
    \centering
    \includegraphics[width=.95\linewidth]{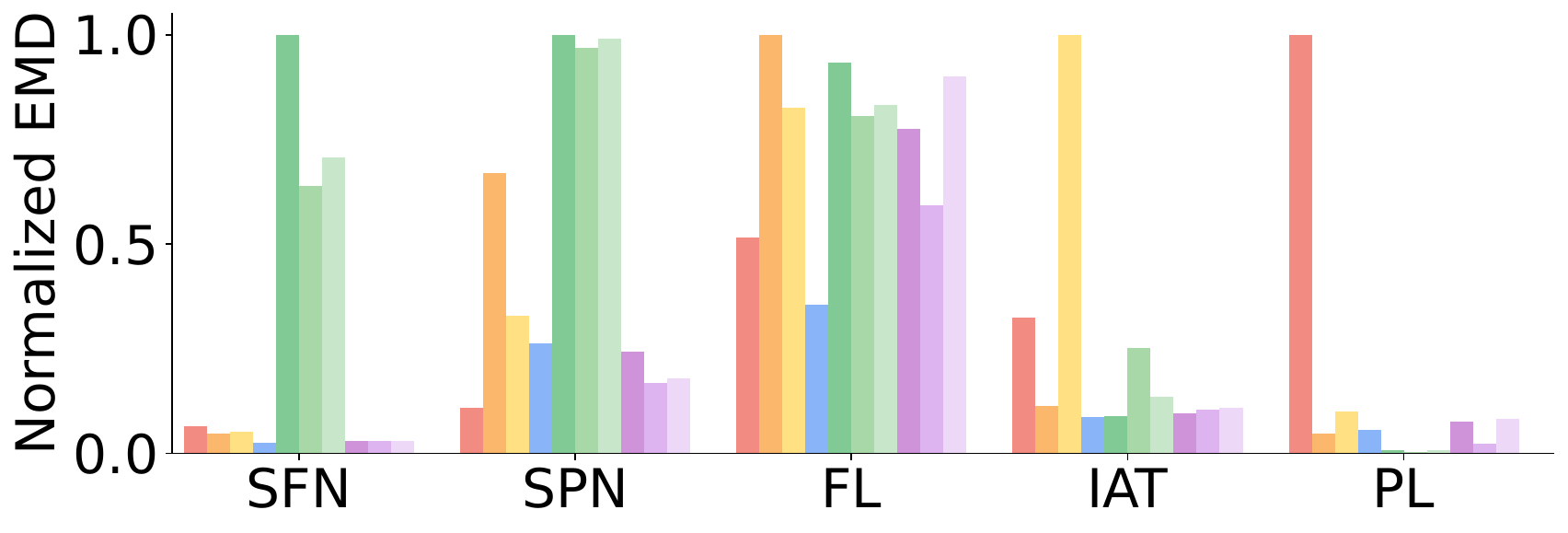}  
    \caption{EMD, CAIDA}
    \label{}
\end{subfigure}

\begin{subfigure}{.49\linewidth}
    \centering
    \includegraphics[width=.95\linewidth]{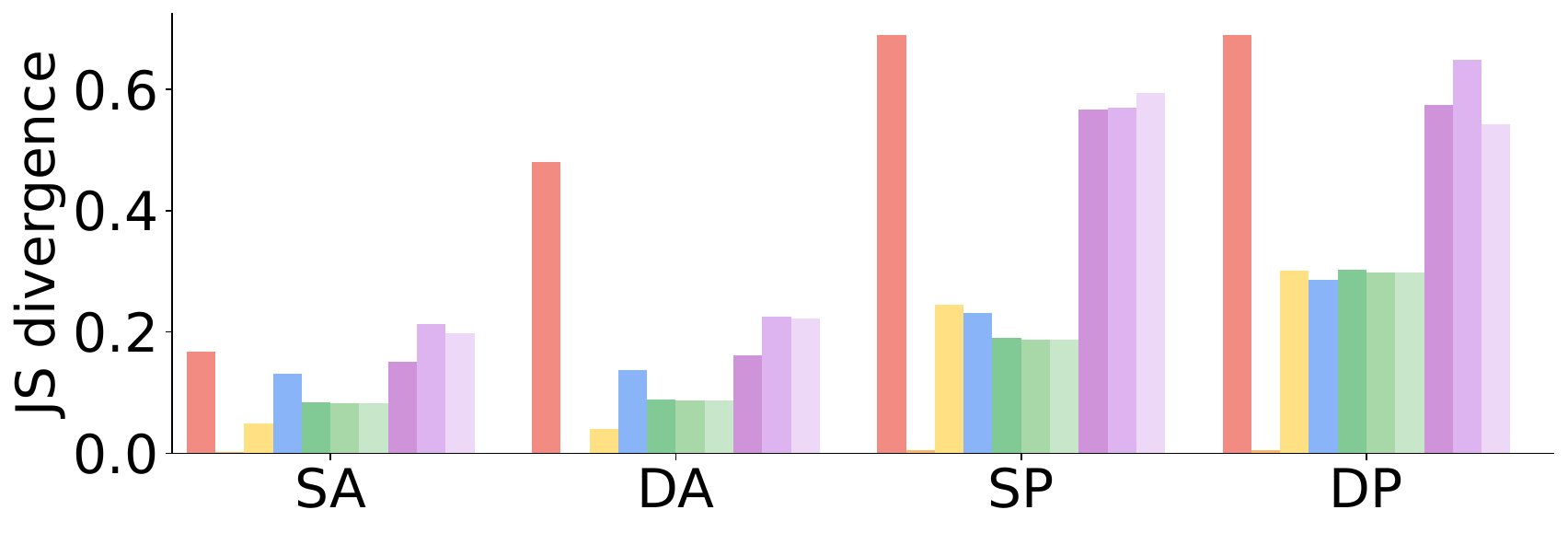}  
    \caption{JSD, DC}
    \label{}
\end{subfigure}
\begin{subfigure}{.49\linewidth}
    \centering
    \includegraphics[width=.95\linewidth]{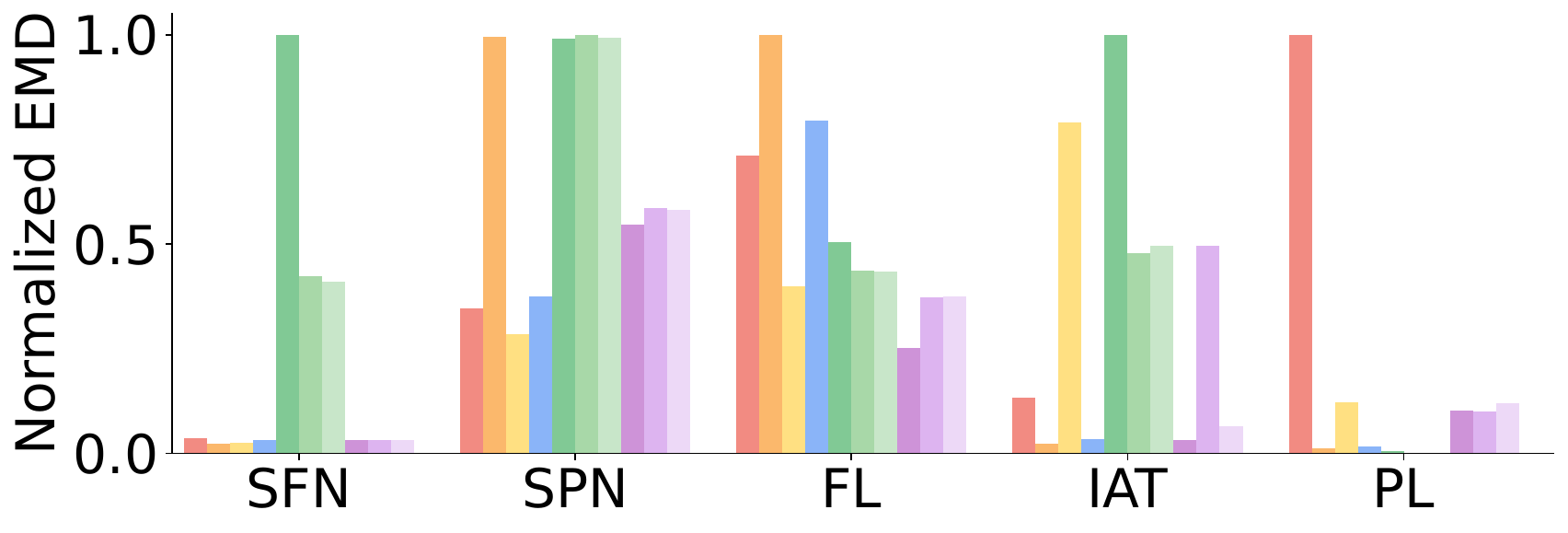}  
    \caption{EMD, DC}
    \label{}
\end{subfigure}

\begin{subfigure}{.49\linewidth}
    \centering
    \includegraphics[width=.95\linewidth]{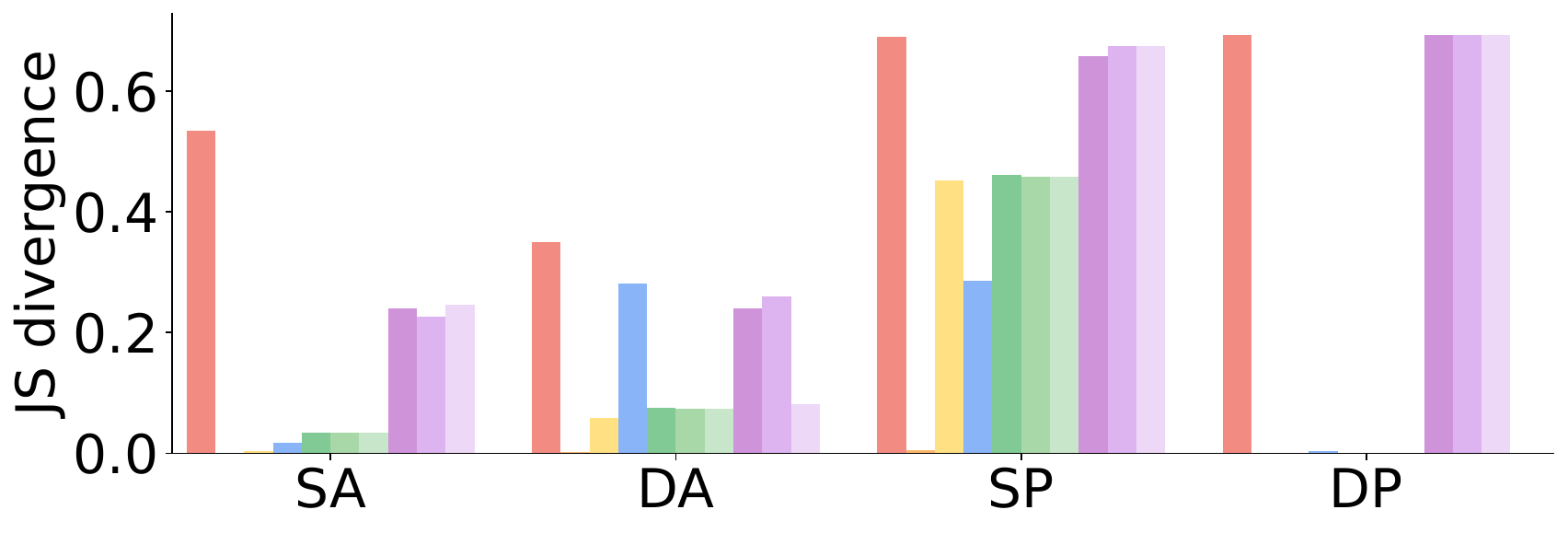}  
    \caption{JSD, WFP}
    \label{}
\end{subfigure}
\begin{subfigure}{.49\linewidth}
    \centering
    \includegraphics[width=.95\linewidth]{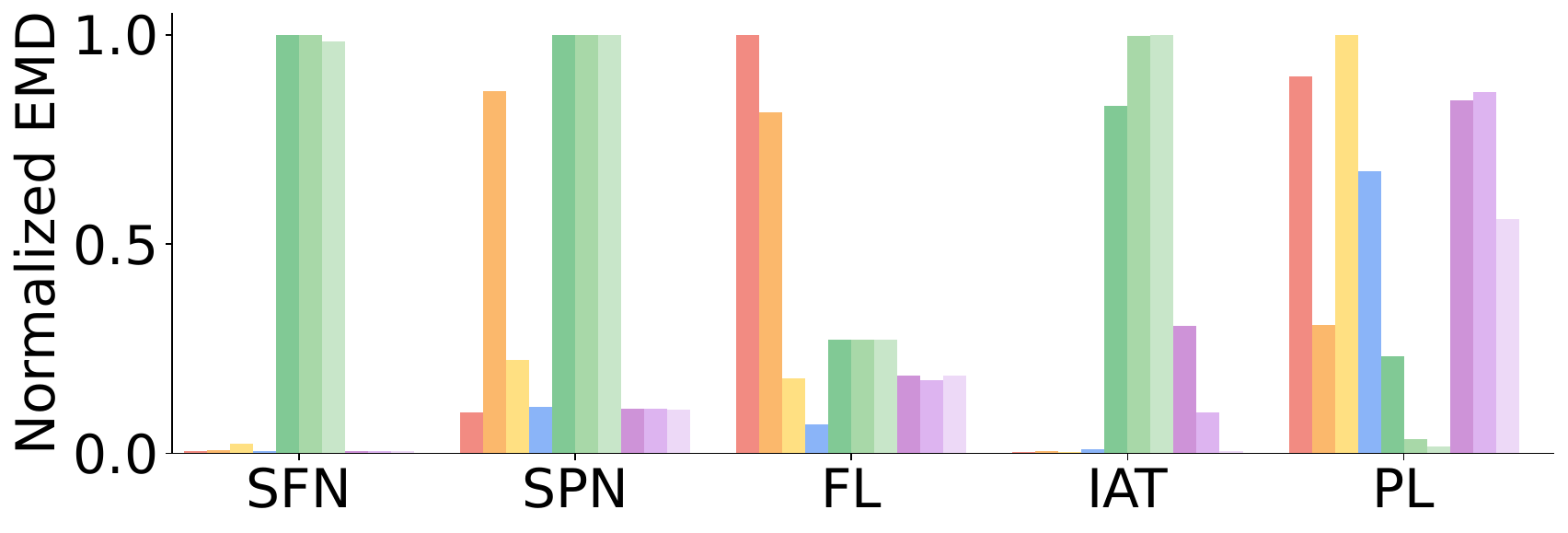}  
    \caption{EMD, WFP}
    \label{}
\end{subfigure}

\begin{subfigure}{.49\linewidth}
    \centering
    \includegraphics[width=.95\linewidth]{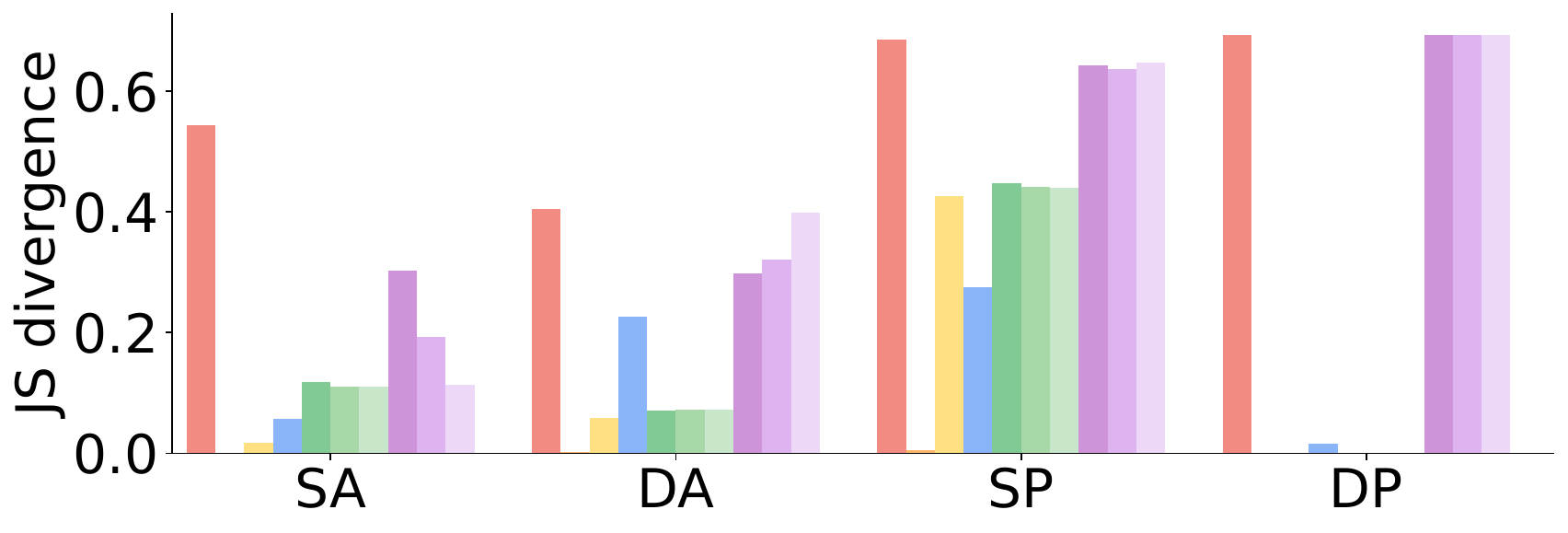}  
    \caption{JSD, BFP}
    \label{subfig:fidelity_web_jsd}
\end{subfigure}
\begin{subfigure}{.49\linewidth}
    \centering
    \includegraphics[width=.95\linewidth]{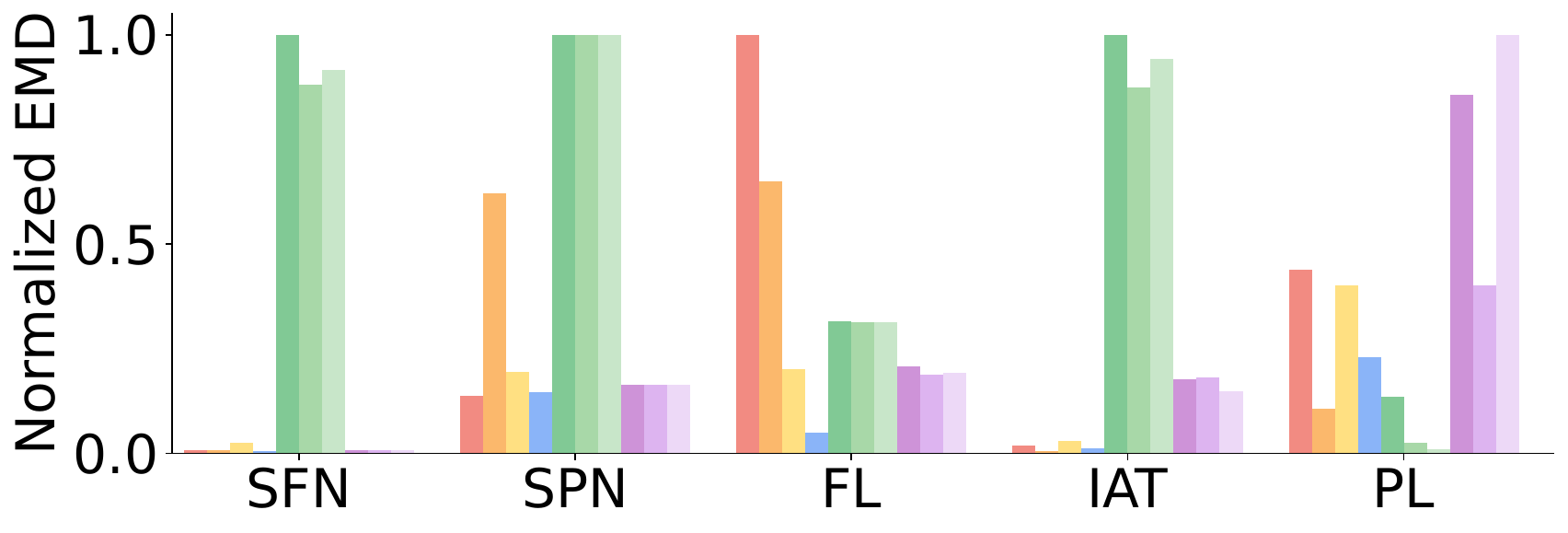}  
    \caption{EMD, BFP}
    \label{subfig:fidelity_web_emd}
\end{subfigure}

\caption{Fidelity report of \generators for CAIDA, MAWI, DC, WFP and BFP datasets. We use JS Divergences to evaluate the the Source IP (SA), Destination IP (DA), Source Port (SP) and Destination Port (DP) distribution similarity. we use normalized EMD for source-level flow number, source-level packet number, flow length, inter-arrival time and packet length distribution similarity. Generated synthetic traces are of reasonably high fidelity.}
\label{fig:apdx:general_fidelity}
\vspace{-0.1cm}
\end{figure*}

Fig.~\ref{fig:apdx:general_fidelity} plots the fidelity of synthetic data for the dataset MAWI, CAIDA, DC, BFP and WFP.
Synthetic data by \generators all show reasonably well fidelity.

\section{Testbed}
\label{apdx:testbed}
For consistent evaluation, we evaluate \sys using the instance of C240g5 in CloudLab Wisconsin cluster~\cite{cloudlab}. It has two Intel Xeon Silver 4114 10-core CPUs at 2.20 GHz, 192GB ECC DDR4-2666 Memory, and 1 Nvidia 12GB P100 GPU.
We train \generators using the university research computing cluster which has 2.8 GHz Intel Ice Lake CPU and Nvidia 80G H100.  

\section{\sys source coverage and Precision}
\label{apdx:sec:prec_cov}
\begin{figure*}[t]
\centering
\begin{subfigure}{.69\linewidth}
    \centering
    \includegraphics[width=.99\linewidth]{figs_ccs/syn_cov_prec_legend.pdf}  
    \label{}
\end{subfigure}

\begin{subfigure}{.245\linewidth}
    \centering
    \includegraphics[width=.99\linewidth]{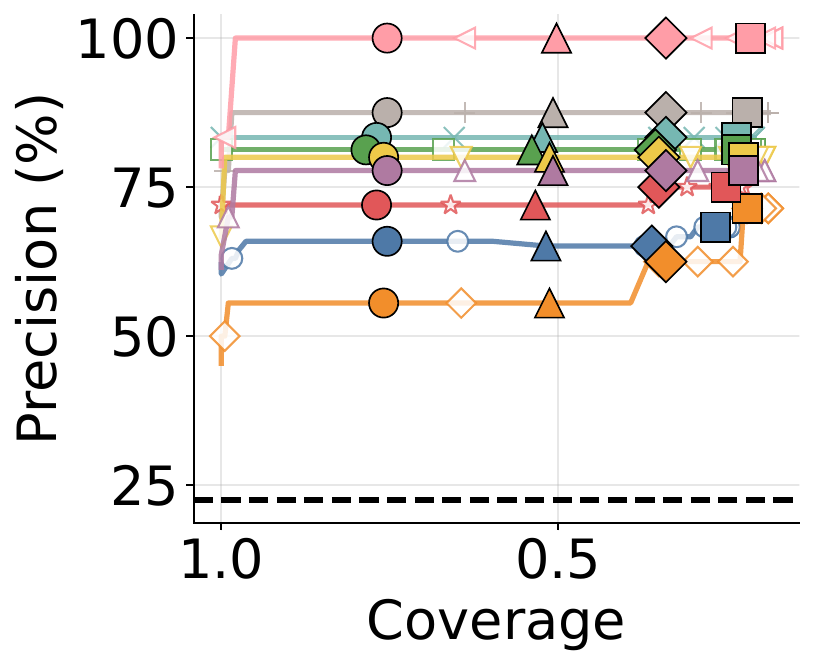}  
    \caption{MAWI}
    \label{}
\end{subfigure}
\begin{subfigure}{.245\linewidth}
    \centering
    \includegraphics[width=.99\linewidth]{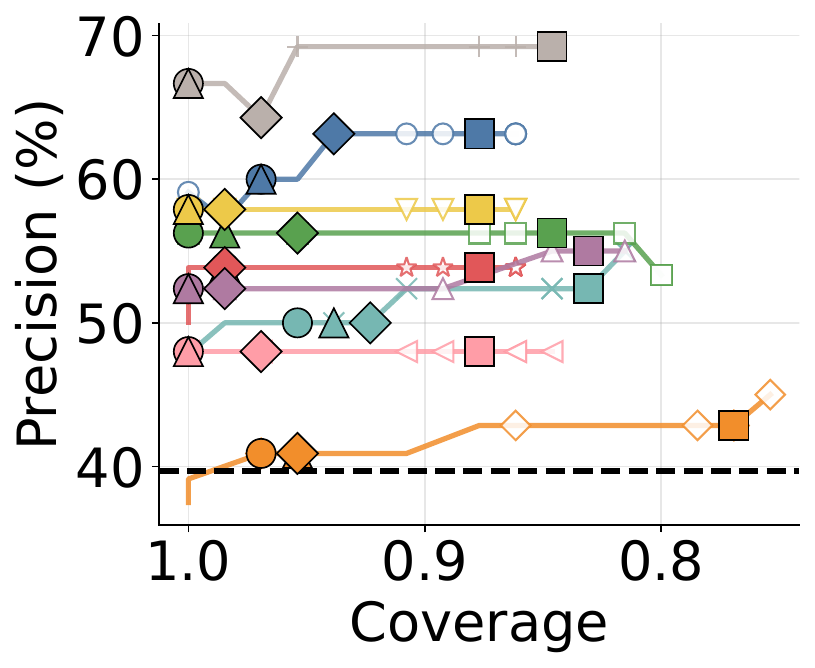}  
    \caption{BFP}
    \label{}
\end{subfigure}
\begin{subfigure}{.245\linewidth}
    \centering
    \includegraphics[width=.99\linewidth]{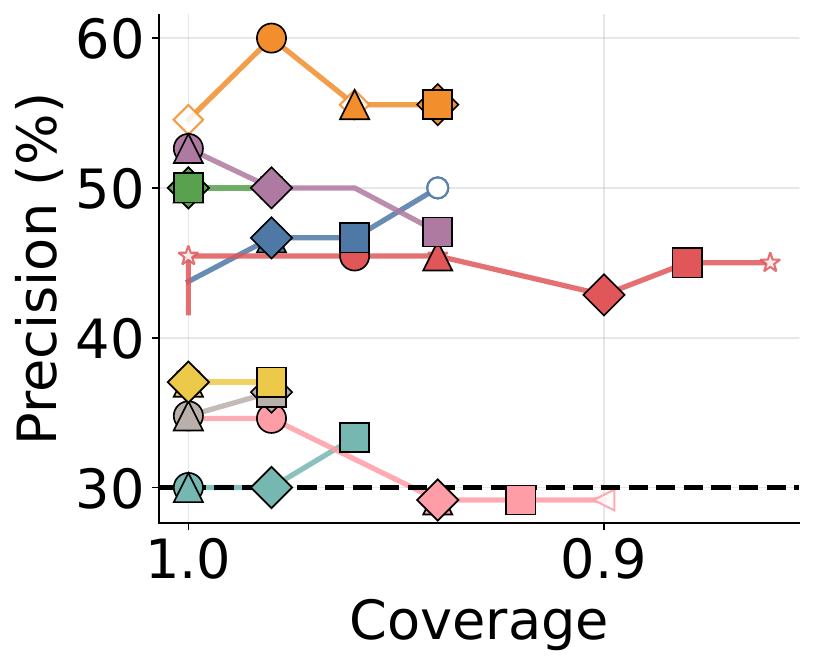}  
    \caption{WFP}
    \label{}
\end{subfigure}

\caption{Precision and source coverage at different confidence levels for MAWI, BFP and WFP.
}
\label{apdx:fig:prec_cov}
\end{figure*}

Fig.~\ref{apdx:fig:prec_cov} shows \sys's source coverage and Precision for MAWI, BFP and WFP dataset. When tuning the confidence from 10\% to 99.9999\%, the Precision generally increases with a drop of source coverage. Different with CAIDA and DC, the increase for MAWI and BFP is more flat, representing a more uniform vulnerability among sources. The WFP shows \generator-dependent pattern, meaning that the source vulnerability also depends on concrete \generator used for data generation.

\section{\sys against \generators generating mode data}
\label{apdx:sec:more_data}

\begin{figure}[t]
\centering
\begin{subfigure}{.99\linewidth}
    \centering
    \includegraphics[width=.99\linewidth]{figs_ccs/hctp_legend.pdf}  
    \label{}
\end{subfigure}

\begin{subfigure}{.45\linewidth}
    \centering
    \includegraphics[width=.95\linewidth]{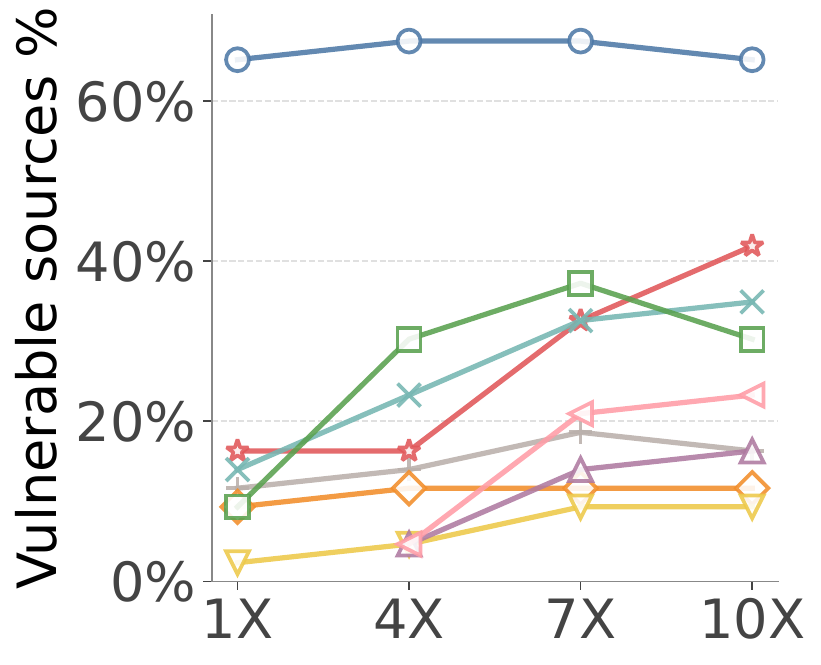}  
    \caption{MAWI}
    \label{}
\end{subfigure}
\begin{subfigure}{.45\linewidth}
    \centering
    \includegraphics[width=.95\linewidth]{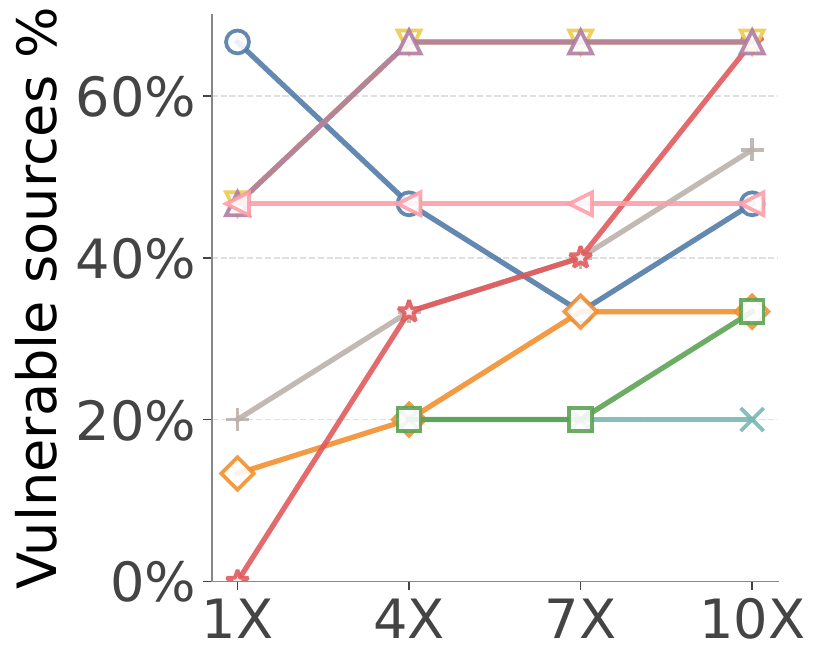}  
    \caption{WFP}
    \label{}
\end{subfigure}
\caption{
\generators sharing more data will increase \sys effectiveness to identify more vulnerable sources.
}
\label{apdx:fig:more_data}
\end{figure}

Fig.~\ref{apdx:fig:more_data} shows the fraction of vulnerable sources identified by \sys when \generators sharing more data for MAWI and WFP dataset. 

\section{Source-level DP}
\label{apdx:sec:source-dp}
Table~\ref{tab:netshare_user_dp} shows the statistics about the flow length for NetShare and its source-level (DP) extension. The fidelity of source-level DP-based NetShare suffers from significant degradation.

\begin{table}[t]
\centering
\begin{tabular}{c|c|c|c}
\hline
flow length & max & mean & stddev \\ \hline
raw & 3444 & 4.78 & 35.60 \\ \hline
NetShare (flow, original) & 778 & 3.35 & 10.42 \\ \hline
NetShare (source) & 41 & 1.04 & 0.35 \\ \hline
NetShare (source, $\epsilon\approx$7) & 1 & 1 & 0 \\ \hline
\end{tabular}
\caption{Compared to the original flow-level Non-DP-Protected (DP-Protected) NetShare, source-level NetShare fails to generate long flows. The longest generated flow drops from 778(808) to 41(1) after changing the granularity from per network flow to per source.  The standard deviation of the flow length degrades from 3.35(6.98) to 1.04(1).}
\label{tab:netshare_user_dp}
\end{table}

\section{Privacy and Fidelity after applying obfuscation}
\label{apdx:sec:trace_obfuscation}
\begin{figure*}[t]
\centering
\begin{subfigure}{.3\linewidth}
    \centering
    \includegraphics[width=.7\linewidth]{figs_ccs_defense/defense_obfuscation_before_after_legend_caida.pdf}  
\end{subfigure}

\begin{subfigure}{.49\linewidth}
    \centering
    \includegraphics[width=.9\linewidth]{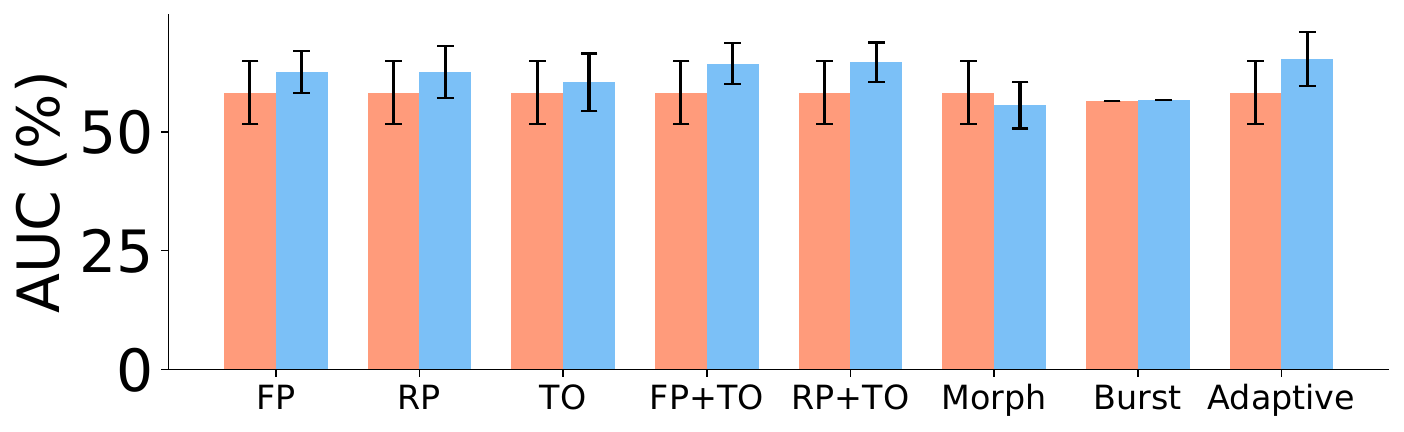}  
    \caption{AUC (MAWI)}
    \label{}
\end{subfigure}
\begin{subfigure}{.49\linewidth}
    \centering
    \includegraphics[width=.9\linewidth]{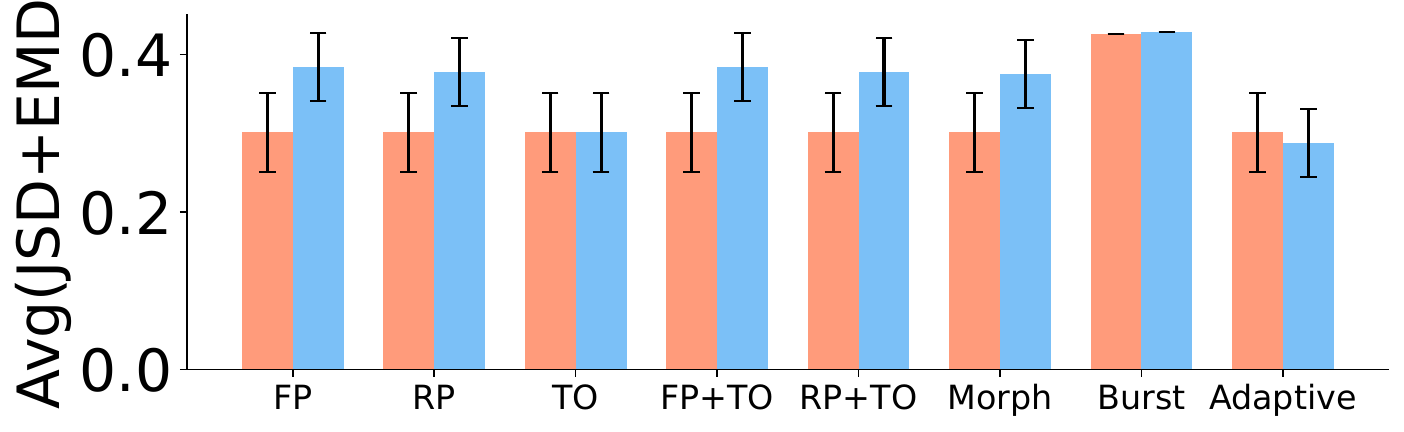}  
    \caption{Avg(JSD+EMD) (MAWI)}
    \label{}
\end{subfigure}

\begin{subfigure}{.49\linewidth}
    \centering
    \includegraphics[width=.9\linewidth]{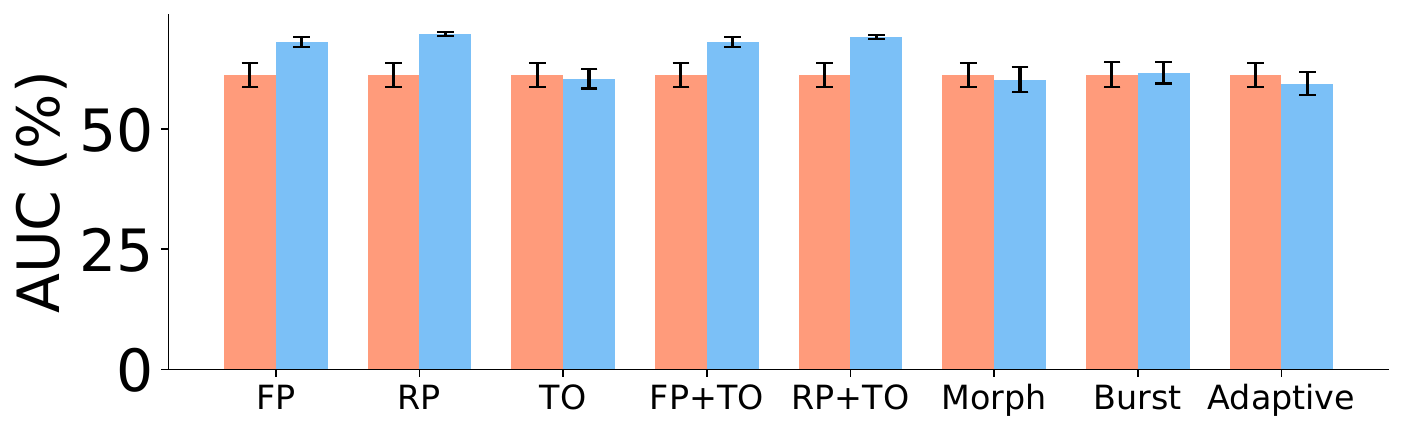}  
    \caption{AUC (DC)}
    \label{}
\end{subfigure}
\begin{subfigure}{.49\linewidth}
    \centering
    \includegraphics[width=.9\linewidth]{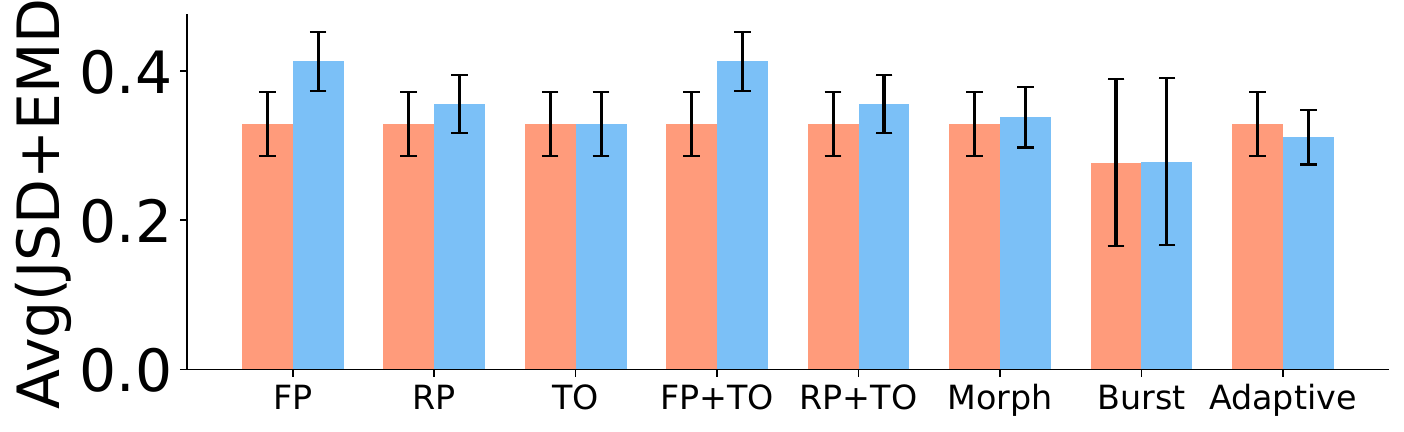}  
    \caption{Avg(JSD+EMD) (DC)}
    \label{}
\end{subfigure}

\begin{subfigure}{.49\linewidth}
    \centering
    \includegraphics[width=.9\linewidth]{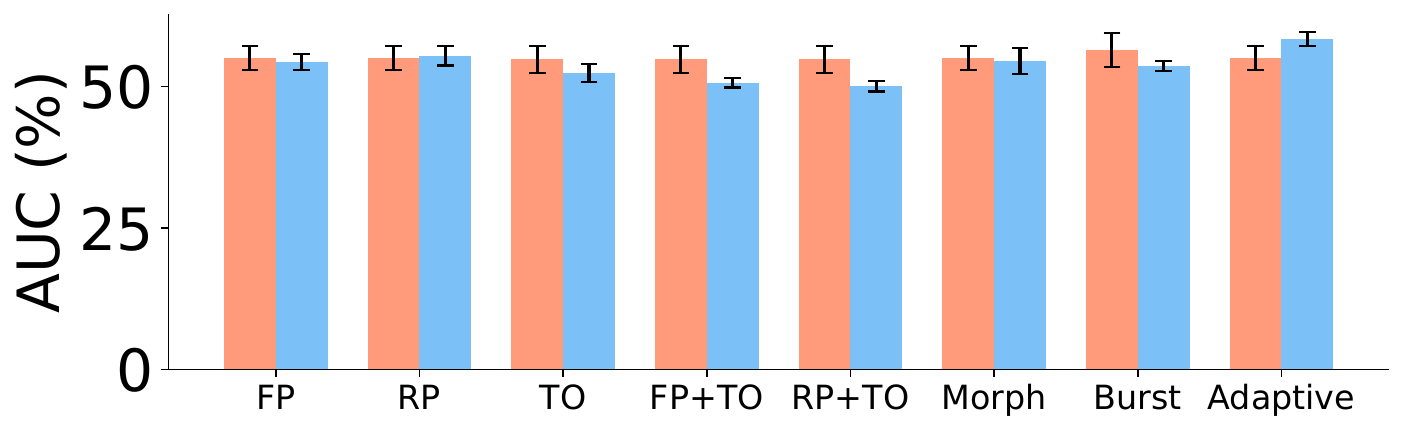}  
    \caption{AUC (BFP)}
    \label{}
\end{subfigure}
\begin{subfigure}{.49\linewidth}
    \centering
    \includegraphics[width=.9\linewidth]{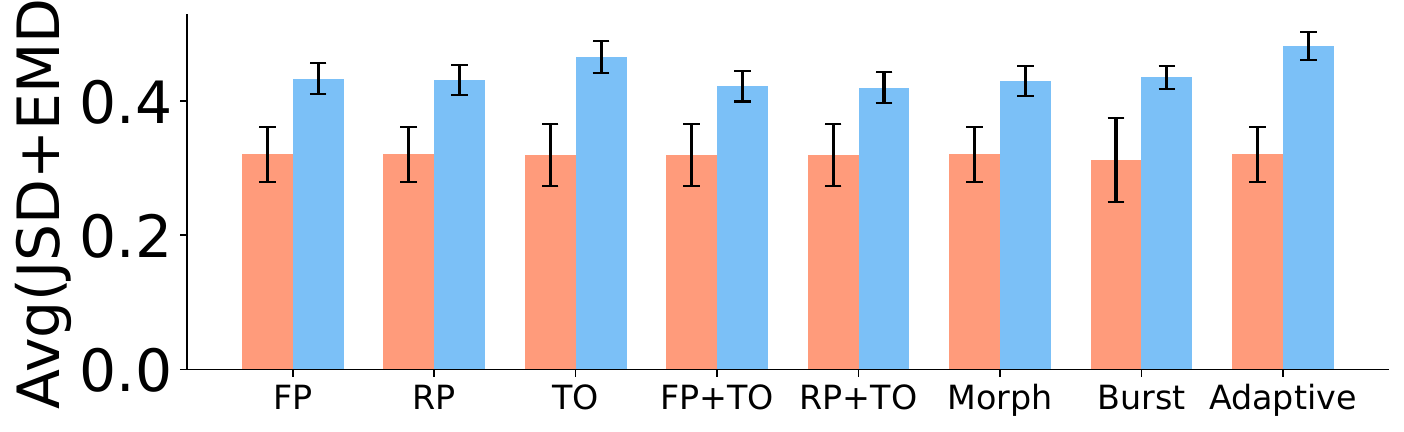}  
    \caption{Avg(JSD+EMD) (BFP)}
    \label{}
\end{subfigure}

\begin{subfigure}{.49\linewidth}
    \centering
    \includegraphics[width=.9\linewidth]{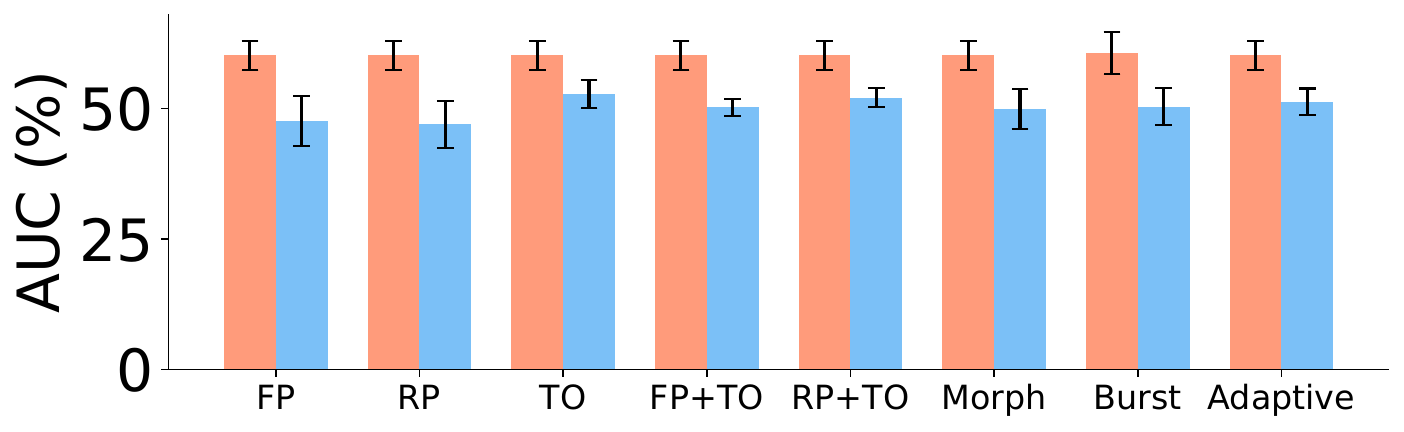}  
    \caption{AUC (WFP)}
    \label{}
\end{subfigure}
\begin{subfigure}{.49\linewidth}
    \centering
    \includegraphics[width=.9\linewidth]{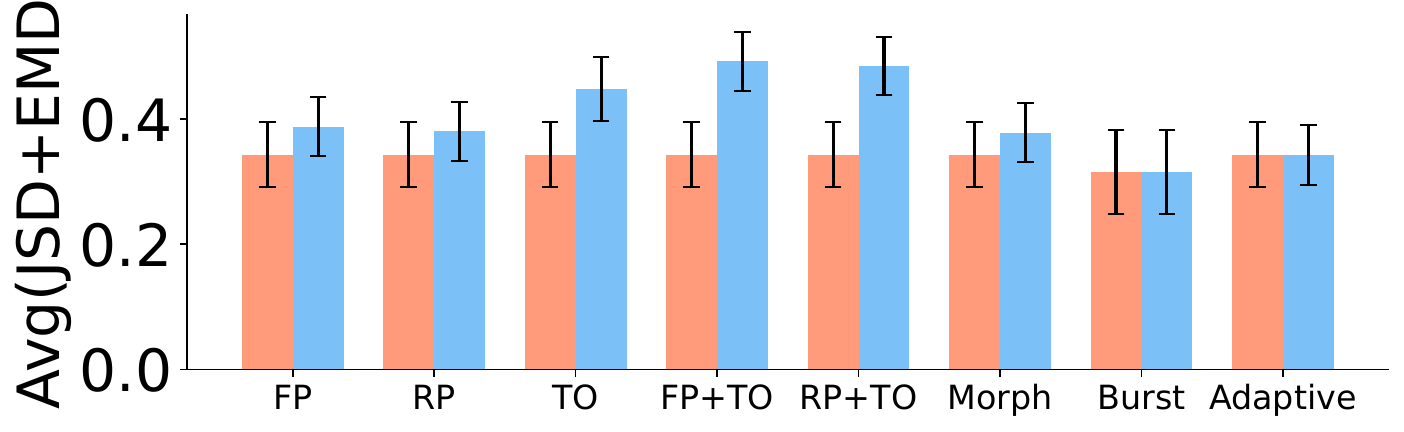}  
    \caption{Avg(JSD+EMD) (WFP)}
    \label{}
\end{subfigure}
\caption{Privacy and fidelity change after applying obfuscation for MAWI, DC, BFP and WFP dataset. Similarly, the trace obfuscation approaches cannot improve data privacy while suffering from universal fidelity drop.}
\label{apdx:fig:trace_obfuscation}
\end{figure*}

Fig.~\ref{apdx:fig:trace_obfuscation} shows the change of privacy and fidelity after applying the obfuscation approaches for MAWI, DC, BFP and WFP dataset. Unfortunately, these approaches fail to improve privacy while consistently degrading the fidelity.

\section{Details of User-level WF}
\label{apdx:awf}

The dataset constructed for training \sys includes roughly 40 different websites. When given a test traffic chunk for prediction, \sys's traffic encoder maps it to representation and finds the closet 10 representation and their corresponding websites from $T$. The predicted website for the test sample is based on the majority vote of these closet representations' label. However, if the distance for test sample with its closet representation in $T$ is larger than the threshold, \sys simply considers it as "unknown", meaning that the test sample's website is not included in $T$. 

After training \sys, we build user-level testing dataset $U$. Similar to Multi-VA and WEB, the selected website traffic to construct $U$ is collected at a different vantage point with the training set of \sys. We randomly select three different websites for each user from a collection of roughly 30 websites. Most of these websites are also collected in $T$ while some of them are not. We simulate around 60 users.

When predicting websites from $U$ and $P(U)$, we include random guess as a baseline to evaluate \sys's effectiveness and the privacy gained by synthetic data. Suppose the number of websites in $T$ is $N_T$, the number of users in $U$ is $N_U$ and the number of websites per user is 3, the random guess of error is calculated as 
\[
\text{Error}
= 1 - \frac{1}{3}\left(
    \frac{3 N_U}{\binom{N_T}{3}}
    + \frac{2 N_U \binom{3}{2}}{\binom{N_T}{3}}
    + \frac{1 N_U \binom{3}{1}}{\binom{N_T}{3}}
\right)
\]

\end{document}